\documentclass[twocolumn,tighten]{aastex62}
\usepackage{amsmath}

\newcommand\msun{\rm M_{\odot}}

\newcommand\msunyr{\rm M_{\odot}\,yr^{-1}}
\newcommand\be{\begin{equation}}
\newcommand\en{\end{equation}}

\begin{document}

\title{A Global 3-D Simulation of Magnetospheric Accretion: I.  Magnetically Disrupted Discs and Surface Accretion}

\shorttitle{Magnetospheric Accretion}
\shortauthors{Zhu et al.}

\correspondingauthor{Zhaohuan Zhu}
\email{zhaohuan.zhu@unlv.edu}

\author[0000-0003-3616-6822]{Zhaohuan Zhu}
\affiliation{Department of Physics and Astronomy, University of Nevada, Las Vegas, 4505 S.~Maryland Pkwy, Las Vegas, NV, 89154, USA}
\affiliation{Nevada Center for Astrophysics, University of Nevada, Las Vegas, 4505 S.~Maryland Pkwy, Las Vegas, NV, 89154, USA}

\author[0000-0001-5603-1832]{James M. Stone}
\affiliation{Institute for Advanced Study, 1 Einstein Drive, Princeton, NJ, 08540, USA}

\author[0000-0002-3950-5386]{Nuria Calvet}
\affiliation{Department of Astronomy, University of Michigan, 1085 South University Ave., Ann Arbor, MI 48109, USA}

\begin{abstract}

We present a 3-D ideal MHD simulation of magnetospheric accretion onto a non-rotating star.  The accretion process unfolds with intricate 3-D structures driven by various mechanisms. First, the disc develops filaments at the magnetospheric truncation radius ($R_T$) due to magnetic interchange instability. These filaments penetrate deep into the magnetosphere, form multiple accretion columns, and eventually impact the star at $\sim$30$^o$ from the poles at nearly the free-fall speed. 
Over 50\% (90\%) of accretion occurs on just 5\% (20\%) of the stellar surface. Second, the disc region outside $R_T$ develops large-scale magnetically dominated bubbles, again due to  magnetic interchange instability. These bubbles orbit at a sub-Keplerian speed, persisting for a few orbits while leading to asymmetric mass ejection. The disc outflow is overall weak because of mostly closed field lines. Third, magnetically-supported surface accretion regions appear above the disc, resembling a magnetized disc threaded by net vertical fields, a departure from traditional magnetospheric accretion models. Stellar fields are efficiently transported into the disc region due to above instabilities, contrasting with the ``X-wind'' model.
The accretion rate onto the star remains relatively steady with a 23\% standard deviation. The periodogram reveals variability occurring at around 0.2 times the Keplerian frequency at $R_T$, linked to the large-scale magnetic bubbles.  
The ratio of the spin-up torque to $\dot{M}(GM_*R_T)^{1/2}$ is around 0.8. 
Finally, after scaling the simulation, we investigate planet migration in the inner protoplanetary disc.
The disc driven migration 
is slow in the MHD turbulent disc beyond $R_T$, while aerodynamic drag plays a significant role in migration within $R_T$.

\end{abstract}


\section{Introduction}
Magnetospheric accretion plays a key role in many astrophysical systems, from  neutron stars (e.g. \citealt{Lewin2006}) to T-Tauri stars \citep{Hartmann2016} and even young planets \citep{Wagner2018,Haffert2019,Zhou2021} \footnote{The problem generator and input file for the simulation presented in this work can be found at \url{https://github.com/zhuzh1983/magnetospheric2023}}. 
For neutron stars, it could be related to the spin-up/spin-down of accreting X-ray pulsars \citep{Bildsten1997}, QPOs in low-mass X-ray binaries \citep{vanderklis2006}, ultraluminous X-ray sources \citep{King2023},  and relativistic jets/outflows \citep{Fender2004}, the latter of which are crucial for studying Gamma-ray bursts and neutron star mergers. For T-Tauri stars, magnetospheric accretion is directly detected via accretion shocks at the surface of the star \citep{CalvetGullbring1998,Ingleby2011} and atomic line
s produced within the magnetosphere \citep{Hartmann1994, Muzerolle1998, Muzerolle2001}, which serve as the main ways to constrain the disc accretion rates \citep{Rigliaco2012,Manara2013,Espaillat2022}. Magnetospheric accretion also affects the structure of protoplanetary discs within 1 au, which is crucial for studying the formation of close-in exoplanets  \citep{Lee2017, Liu2017}. For young planets, magnetospheric accretion may be responsible for the detected $H_{\alpha}$ lines around young giant planets (e.g. \citealt{Zhu2015, Thanathibodee2019, Marleau2022}, but see \citealt{Aoyama2018,Szulagyi2020}), and emission from magnetospheric accretion could reveal young planets in protoplanetary discs. 

Magnetospheric accretion is a complex process primarily driven by the influence of the stellar magnetic fields. The magnetic field strength decreases sharply as distance from the star increases.  In close proximity to the star, stellar fields are so strong that the flow is in the force-free regime. Moving outwards to the magnetospheric truncation radius, the flow and magnetic fields are dynamically balanced and the flow is in the strong field regime. Further out into the disc, the disc thermal pressure is far higher than the magnetic pressure so that the disc enters the weak field regime, where disc instabilities (e.g. magneto-rotational instability, MRI) could play an essential role in disc accretion.

Several outstanding theoretical questions persist regarding magnetosheric accretion, as reviewed by \citealt{Lai2014}): 1) how do the stellar magnetic fields connect with the disc? 2) Can accretion occur steadily throughout the region? 3) What controls the stellar spin? 4) How are outflows launched? and many others. To address these questions, numerous models have been proposed. In a seminal paper, \cite{Ghosh1979II} supposed that stellar fields can permeate a substantial radial region of the disc, reaching a steady state when field dragging is balanced by dissipation. This model suggests the existence of a broad transition zone (a detailed description of this model is given in \S \ref{sec:magstr}). However, alternative models propose that the disc is a good conductor so that the field lines are pinched inwards at the boundary (\citealt{Arons1986}, ''X-wind'' model from \citealt{Shu1994}). Furthermore, some models suggest that the accretion is not steady (e.g. \citealt{Aly1990, Lovelace1995, Uzdensky2002}). In these scenarios, magnetic fields connecting the star and the disc undergo winding, causing the flux tube to expand with building magnetic pressure -  so called ``field inflation''. Subsequently, reconnection events take place, leading to the expulsion of outer field lines, resulting in mass ejection.  Meanwhile, the inner reconnected field lines continue to wind up,  perpetuating this cyclic process.

Magnetospheric accretion could also be linked to the launching of jets and winds, expanding the scope beyond the conventional extended disc winds\citep{BlandfordPayne1982,WardleKoenigl1993,Ferreira1995,Casse2000} and stellar winds \citep{Sauty1994,Hartmann1980}.
Various models have been suggested, including accretion-powered stellar winds \citep{Matt2005}, a wide-angle ``X-wind'' \citep{Shu1994}, or even unsteady magnetospheric winds due to reconnection \citep{Ferreira2000,Hayashi1996,Matt2002}. 

More detailed insights have been unveiled through direct numerical simulations (see review by \citealt{Romanova2015}). Earlier simulations employed axisymmetric 2-D configurations \citep{Goodson1997,miller1997,Fendt2000}. These simulations have already revealed that the accretion structure depends sensitively on the initial stellar and disc field configurations (e.g. parallel or anti-parallel). Both accretion and stellar winds \citep{Zanni2013} could also vary strongly with time in these simulations. Later, 3-D simulations have been carried out \citep{Romanova2003,Romanova2004}. Given the difficulty in simulating the polar region using the conventional spherical-polar coordinate system in 3-D, \cite{Romanova2003} adopt the ``cubed sphere'' grid. To simulate the accretion disc, an $\alpha$ viscosity is adopted in the disc region. Tilted dipole fields \citep{Romanova2003}, fast rotators \citep{Romanova2004b}, titled rotating stars with titled dipole fields \citep{Romanova2020} have all been studied. These works show that the accretion structure depends on the dipole tilt angle. Furthermore, warping of the disc, interchange instability, and Rayleigh-Taylor instability \citep{Kulkarni2008} can all play important roles for accretion with titled dipole fields. However, due to the use of an $\alpha$ viscosity, the self-consistent treatment of  disc accretion driven by the MRI is not achieved in these models. 

MHD simulations including both the magnetosphere and MRI turbulence 
have been carried out with both axisymmetric 2-D and fully 3-D simulations \citep{Romanova2011,Romanova2012}. The 3-D simulations again use the ``cubed sphere'' configuration. Both boundary layer accretion with weak fields and magnetospheric accretion with strong fields have been explored. These simulations reveal significant accretion variability.  They show that the stress in the accretion disc is consistent with local MRI simulations. On the other hand, the global flow structure has not been thoroughly examined. 

{ Recent global MHD  simulations reveal that, for discs threaded by external vertical magnetic fields (without including the stellar fields), the flow structure is dramatically different from that in local MRI simulations. The disc accretion structure sensitively depends on the strength of net vertical magnetic fields.} When the net field is relatively weak (initial plasma $\beta\sim 10^3-10^4$ at the disc midplane), most accretion occurs in the magnetically supported disc surface region that can extend vertically to $z\sim R$ \citep{Beckwith2009,ZhuStone2018,Takasao2018,Mishra2020M,Jacquemin2021}. A quasi-static global field geometry is established when the flux transport by the fast inflow at the surface is balanced by the slow vertical turbulent diffusion. When strong vertical fields thread the disc around black-holes (BH), the disc flow can enter the regime of magnetically arrested discs (MAD, \citealt{Narayan2003,Igumenshchev2003}), { where the accumulated poloidal fields disrupt the accretion flow to become discrete blobs/streams and the blobs/streams fight their way towards the BH through magnetic interchanges and reconnections. With poloidal magnetic fields acting like a wire lowering  material down to the BH, the MAD state leads to the efficient release of the rest mass energy} and
strong quasi-periodic outflows \citep{Tchekhovskoy2011}. The variability may be related to low-frequency QPOs, variability in AGN, and GRB outflows \citep{Proga2006}.  Considering that the disc-threading stellar dipole fields weaken sharply with distance to the star (from force-free, strong fields, to weak fields), some aspects of the MAD state and 
the newly discovered surface accretion mode can be applied to magnetospheric accretion.

Furthermore, owing to the high computational cost in 3-D simulations, most previous 3-D magnetospheric accretion simulations cannot follow the disc evolution over the viscous timescale, and thus focused on studying the magnetosphere itself.
{ However, magnetopsheric accretion plays a pivotal role in shaping the disc's long-term evolution. An unresolved challenge within accretion disc theory is the uncertainty of the inner boundary conditions, except for accretion onto black holes. For example, zero torque or zero mass flux inner conditions lead to different disc evolution paths \citep{Lynden-Bell1974}.  The proper inner boundary condition can only be understood if we incorporate the central object in the disc simulation.}
To study the interplay between the magnetosphere and the disc, encompassing all three different MHD regimes, we carry out high-resolution long-timescale global simulations that incorporate both the magnetized star and the disc. As a first step, we adopt the simplest setup which only includes a non-rotating star with a dipole field surrounded by an accretion disc. The disc is threaded  by the stellar field only (without the external field) to avoid the complex interaction between the stellar and external fields. Since the star is non-rotating, the corotation radius is at infinity. Such a simple setup allows us to model a relatively clean problem as the foundation for future more realistic simulations. Nevertheless, this setup can  be directly applied to slow rotators in many astrophysical systems.
On the other hand, this setup does not allow us to study accretion around fast rotators, which could launch powerful jets and winds \citep{miller1997,Lovelace1999,Romanova2018} that make the star to spin down \citep{Matt2005}. While we were preparing this manuscript, \cite{Takasao2022} published the results of 3-D ideal MHD simulations studying magnetospheric accretion onto stars with different spin rates. Our work and \cite{Takasao2022} share some similarities, but also bear significant differences. By exploring stars with different spin rates, \cite{Takasao2022} can address how wind launching and spin-up/down torque is affected by the stellar spin. On the other hand, our simulation adopts a Cartesian grid with mesh-refinement, which significantly reduces the computational cost. This allows our simulation to include a relatively large magnetosphere generated from the 1kG stellar dipole (compared with 100 G in \citealt{Takasao2022}) and study the disc evolution on much longer timescales (2000 vs 400 innermost orbits). Thus, besides the processes within the magnetosphere, we can study the dynamics of the disc over a large dynamical range and explore how the disc and the magnetosphere interact with each other in a quasi-steady accretion stage.  

In \S2, we lay out the theoretical framework for magnetospheric accretion, and introduce the key physical quantities in the classical \cite{Ghosh1979II} model. Our numerical model is presented in \S 3. We present results in \S 4 from the inner magnetosphere to the outer disc. After discussions in \S5, we conclude the paper in \S 6.

\section{Theoretical Framework}
\subsection{The Classical Model}
\label{sec:magstr}

\begin{figure*}[t!]
\includegraphics[trim=0mm 5mm 0mm 30mm, clip, width=6.5in]{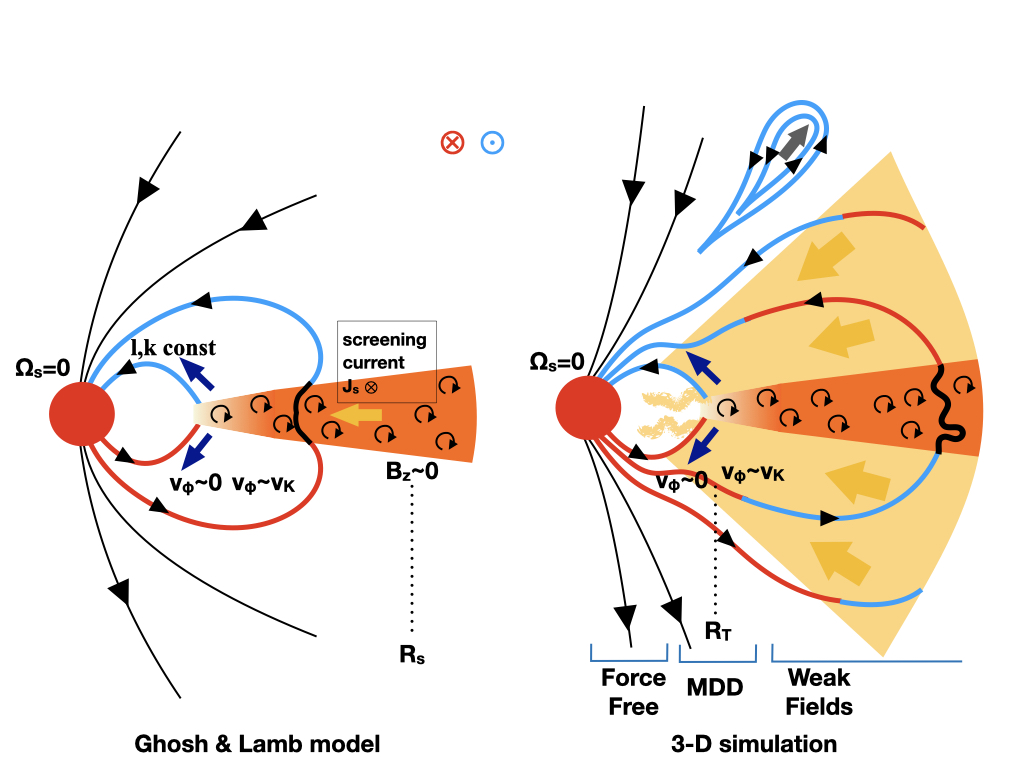}
\figcaption{The flow and magnetic field structure in the  magnetospheric accretion model of \cite{Ghosh1979II}  (the left panel)
and in our 3-D simulations (the right panel). The solid curves represent the magnetic field lines. The red and blue colors
represent $B$ fields with positive and negative $B_{\phi}$ components. The yellow and purple arrows
 label the flow with positive and negative $v_{\phi}$ components respectively. The positive $\phi$ component indicates the same direction as disc rotation. The screening current in the traditional model  has a positive
 $\phi$ component, which leads to the vanishing stellar $B_z$ beyond $R_s$. { $R_s$ is the screening radius where  $B_z=0$}.  The most noticeable difference
 between the two models include the highly magnetized surface layer which inflows at supersonic speeds. Keplerian shear in this layer
 drags the radial magnetic field lines to develop significant $B_{\phi}$ components inside the layer. { Other differences include the filaments and density voids  in the magnetically disrupted disc (MDD)}, and the non-steady outflow which is also illustrated in the right panel.
 \label{fig:copy003}}
\end{figure*}

The magnetospheric accretion model by \cite{Ghosh1979II}  is summarized in the left panel of Figure \ref{fig:copy003}. In close proximity to the star, magnetic fields
are so strong that the plasma is forced to corotate with the star. For an  axisymmetric and steady flow, the plasma structure there can be solved via four conserved quantities which are constant along magnetic field lines  \citep{Ghosh1977}.
These four constants \citep{Mestel1961,WeberDavis1967} are also widely used in disc wind studies \citep{BlandfordPayne1982}. 
When $\bf{v}$ and $\bf{B}$ are separated into the poloidal  and toroidal components ($\mathbf{v}=\mathbf{v_{p}}+\Omega R \mathbf{e_{\phi}}$ and $\mathbf{B}=\mathbf{B_{p}}+B_{\phi}\mathbf{e_{\phi}}$), 
the induction equation implies that $\mathbf{v_{p}}$ and $\mathbf{B_{p}}$ are in the same direction, and the first
constant is
\begin{equation}
k=\frac{\rho \mathbf{v_{p}}}{\mathbf{B_{p}}}\,, \label{eq:kintegral}
\end{equation}
which is the mass loading parameter.
In the azimuthal direction, we have the second constant 
\begin{equation}
\omega_s=\Omega-\frac{k B_{\phi}}{\rho R}\,.\label{eq:omegaintegral}
\end{equation}
Note that $\omega_s$ should be equal to the angular frequency of the star, denoted as $\Omega_s$. Otherwise, there is a strong shear
at the stellar surface where $v_p$ decreases to 0. 
Equations \ref{eq:kintegral} and \ref{eq:omegaintegral} imply that, in the rotating frame with the angular frequency $\omega_s$, the flow's velocity is in the same direction as the magnetic field lines, or in simpler terms, the material flows along the field lines.
We can define the pitch angle of the magnetic field lines as tan $\psi\equiv B_p/B_\phi$.
Using the angular momentum equation, we have the third constant 
\begin{equation}
l=R(v_{\phi}-\frac{B_{\phi}}{k})\,, \label{eq:lintegral}
\end{equation}
which is the specific angular momentum of the flow. 
In a barotropic fluid, Bernoulli's equation can be used to derive the fourth constant
\begin{align}
e&=\frac{1}{2}v^2+\Phi+h+\frac{B^2}{\rho}-\frac{{\bf B}\cdot{\bf v}}{k}\nonumber\\
&=\frac{1}{2}v^2+\Phi+h+\frac{B_{\phi}B_{\phi}}{\rho}-\frac{B_{\phi}v_{\phi}}{k}\,,\label{eq:eintegral}
\end{align}
where $h$ is $\int_0^p dp/\rho$. For a strongly magnetized plasma, 
$h$ is much smaller than other terms. Within the magnetosphere, the poloidal
velocity and gravitational term dominate in Equation \ref{eq:eintegral}, so that the poloidal velocity is nearly equal
to the free-fall velocity. With both  $B_p$ (dipole fields) and $v_p$ known, the flow and magnetic field structure within the magnetosphere can be derived using
these constants with given $\omega_s$ and $l$. One main result regarding this inner magnetosphere \citep{Ghosh1977} is that, in the case of slow rotators, matter inside the
Alfv\'en surface ($r_A$) rotates in the opposite direction from $l$. The  reason is that, when $\omega_s$ has the same sign but is much smaller than $l/r_A^2$, 
the spiral field lines have a forward pitch since field lines are dragged forward by the fast disc rotation. Matter that falls inwards along these spiral fields has a backward azimuthal velocity. These magnetic fields could also spin up the star through the magnetic stress. 

Meanwhile, the values of $l$ along different field lines depend on the condition in the transition zone (inside the screening radius $R_s$ in Figure \ref{fig:copy003}) where  the corotating flow (with $\bf{v_p}$ following $\bf{B_p}$) changes to the Keplerian disc flow. Since the flow in the transition zone is neither steady nor axisymmetric, the full angular momentum equation is needed to describe this region:
\begin{equation}
\frac{\partial R\rho v_{\phi}}{\partial t}+\nabla\cdot\left(R\left(\rho{\bf v}v_\phi-{\bf B}B_{\phi}+P{\bf e_{\phi}}\right)\right)=0\,,\label{eq:ame}
\end{equation}
where $P$ is the total pressure. If using spherical-polar coordinates,
$R$ is replaced with $r\times$sin$\theta$.

\cite{Ghosh1979II} built a simple disc model for the transition zone, assuming the accretion can reach a steady state with an effective electrical conductivity ($\sigma_{eff}$) in this region. 
In a steady state, we have
\begin{equation}
\nabla\times{\bf B}=\frac{4\pi {\bf J}}{c}\,,
\end{equation}
with the Ohm's law ${\bf J}=\sigma_{eff}({\bf E}+({\bf J}\times{\bf B})/c)$. If $B_{\phi}$ changes over the disc scale height H, we have
\begin{equation}
\frac{B_{\phi}}{H}=\frac{(\Omega-\Omega_s)rB_z}{\eta} \,,\label{eq:eta}
\end{equation}
where $H\equiv c_s/\Omega$ is the disc scale height and $\eta\equiv c^2/(4\pi \sigma_{eff})$ is the resistivity. The resistivity has to satisfy Equation \ref{eq:eta} for a steady state, meaning that the slippage of field lines  balances the azimuthal shear. 
The azimuthal current screens the background stellar fields. 
\cite{Ghosh1979II} did not specify the source of the resistivity. But if we consider turbulent resistivity $\eta\sim \nu\sim\alpha H c_{s}$ and $\Omega_s=0$, we have $B_{\phi}/B_{z}=R/(\alpha H)$. If $\alpha<R/H$, $B_{\phi}$ is then larger than $B_z$, and the magnetic pressure
from the toroidal fields can drive field inflation until fields open up \citep{Aly1990, Lovelace1995, Uzdensky2002}. These later works suggest that a steady state may not be possible with the turbulent resistivity.

The transition zone is separated into the inner transition zone
(also called the boundary layer) where the azimuthal velocity changes dramatically due to the magnetic stress and the outer transition zone where the disc is Keplerian but the residual stellar magnetic fields affect the disc accretion. The boundary between these two transition zones is denoted as $R_b$ and is a sizable fraction of the Alfv\'en radius for spherical accretion
\begin{equation}
r_{A}=r_{*}\left(\frac{B_*^4 r_*^5}{2GM_*\dot{M}^2}\right)^{1/7} {\rm\;\;\; in\;\; C.G.S.}\,, \label{eq:RT}
\end{equation}
where $r_A$ is sometimes called the magnetospheric truncation radius $R_T$ \citep{Hartmann2016}. In this paper, we define $R_T$ as the radius where the averaged azimuthal velocity drops to half of the local Keplerian velocity ($v_K$). { As will be shown later, the azimuthal velocity changes dramatically around the magnetospheric truncation radius. Thus, choosing other velocities between 0 and $v_K$ for the definition of $R_T$ barely affects $R_T$. } In Section \ref{sec:RT}, we will compare our measured $R_T$ in the simulation with various definitions of the magnetospheric truncation radius, and show that the measured $R_T$ is quite close to $r_A$ in Equation \ref{eq:RT}. 
The outer transition zone ends at $R_s$, where the screening current reduces the background stellar magnetic fields to zero. In the \cite{Ghosh1979II} model, $R_s$ could be tens to hundreds of times larger than $R_b$. The broad outer transition zone in their model plays a key role in the coupling between the disc and the star, especially for fast rotators. If the torque on the star is written as
\begin{equation}
T_*=n (GM_* R_b)^{1/2}\dot{M}\,,\label{eq:Tstar}
\end{equation}
\cite{Ghosh1979III} derived  n$\approx$1.4 for a non-rotator. The $\alpha$ accretion disc \citep{Shakura1973} exists beyond the transition zone, and continuously provides mass inwards. Since we focus on the simplest setup with a non-rotator, we will not discuss the rich physical processes associated with moderate and fast rotators \citep{Ghosh1979III,Lovelace1999,Romanova2003b,Romanova2004b}.

\subsection{Angular Momentum Transport and Disc Evolution}
Our first-principle 3-D simulation reveals that the flow structure is more complicated than that assumed in the traditional Ghosh \& Lamb model. To understand the flow structure in the 3-D simulation, we need to study how angular momentum is transported among different regions. { For the disc region that reaches a steady state, the second term in Equation \ref{eq:ame} becomes zero. In other words, if we define $S$ as a surface surrounding the star which is also along the $\phi$ direction (so $S$ could be cylinders or spheres surrounding the star), we have
\begin{equation}
\int R\left(\rho{\bf v}v_\phi-{\bf B}B_{\phi}\right) \cdot d{\bf S} = const\, \label{eq:torqueconst}
\end{equation}
along the radial direction. 
If we assume $S$ is the cylinder surface around the star, we have
\begin{equation}
    R \langle v_{\phi} \rangle \dot{M} + R \int (\rho v_{R}(v_{\phi}-\langle v_{\phi} \rangle ) - B_{R} B_{\phi}) dS = const\,\label{eq:torqueconst2}
\end{equation}
at different $R$. The symbol $\langle \rangle$ denotes averaging over the $\phi$ direction, and we have assumed that $\langle v_{\phi}\rangle $ is constant along the cylinder. The first and second terms within the integral are Reynolds and Maxwell stresses. Thus, for the steady state, the accretion rate is directly determined by the total stress and the constant. The constant represents the torque between the star and the disc. Considering that $\langle v_{\phi}\rangle\sim v_{K}\sim R^{-1/2}$ and $\dot{M}$ is constant with $R$, the first term on the left-hand side increases with $R$. Thus, at large distances, the constant on the right hand side becomes negligible, and the $\dot{M}$  term is balanced by the stress term. The accretion structure there is solely determined by the stress values. On the other hand, the constant is crucial for the disc's structure at the inner disc edge. We will discuss the constant in detail in \S \ref{sec:torque}}

We can also derive the disc's accretion rate even if the disc has not reached the steady state. We can average the angular momentum equation (Equation \ref{eq:ame}) in the azimuthal direction to derive\begin{align}
\frac{\partial R\langle\rho v_{\phi}\rangle}{\partial t}=&-\frac{1}{R}\frac{\partial}{\partial R}\left(R^2\langle \rho v_{R}v_{\phi}-B_{R}B_{\phi}\rangle\right)\nonumber\\
&-\frac{\partial}{\partial z}\left(R\langle\rho  v_{z}v_{\phi}-B_{z}B_{\phi}\rangle\right)\,,\label{eq:amephi}
\end{align}  
If we integrate Equation \ref{eq:amephi} vertically in the disc region and use the mass conservation equation, we have
\small
\begin{align}
\frac{\partial \int R \langle \rho \delta v_{\phi}\rangle dz}{\partial t}&=-\frac{1}{R}\frac{\partial}{\partial R}\left( R^2 \int\left(\langle \rho v_{R}\delta v_{\phi}\rangle-\langle B_{R}B_{\phi}\rangle\right)dz\right) \nonumber\\
&-\frac{\dot{M}_{acc}}{2\pi R}\frac{\partial R v_{k}}{\partial R}-R\left(\langle \rho v_{z} \delta v_{\phi}\rangle-\langle B_{z}B_{\phi}\rangle\right)\bigg |_{z_{min}}^{z_{max}}\,,\label{eq:angcyl2}
\end{align}
\normalsize
where $\delta v_{\phi}\equiv v_{\phi}-v_{k}$\footnote{$v_{k}$ is assumed to be
constant along $z$. Without this assumption, there will be an additional term related to $\dot{M}_{loss}\partial v_{k}/\partial z$.}. The equation connects the disc's radial mass accretion rate
($\dot{M}_{acc}= 2\pi R\int \rho v_{R}dz$) to the $R\phi$ stress within the disc and $z\phi$ stress at the disc surface. 
 Equation \ref{eq:angcyl2} is widely used in accretion disc studies (e.g., \citealt{Turner2014a}). However, it can also be used to study flows in the magnetosphere where the $v_z$ term describes the vertical flow lifted out of the disc plane.  The terms
$\langle \rho v_{R}\delta v_{\phi}\rangle$ and $-\langle B_{R}B_{\phi}\rangle$ are the radial Reynolds and Maxwell stresses, and the corresponding $\alpha$ parameters can be defined as
\begin{equation}
\alpha_{Rey}=\langle \rho v_{R}\delta v_{\phi}\rangle/\langle p\rangle\quad {\rm and}\quad \alpha_{Max}=-\langle B_{R}B_{\phi}\rangle/\langle p\rangle \,.
\end{equation}
Stresses and $\alpha$ parameters in spherical-polar coordinates can be defined in similar ways.
If we define
the vertically integrated $\alpha$ parameter as
\begin{equation}
\alpha_{int}=\frac{\int T_{R\phi}dz}{\Sigma c_{s}^2}\,,\label{eq:alphaint}
\end{equation}
where $T_{R\phi}$ is the sum of both radial Reynolds and Maxwell stresses,
Equation \ref{eq:angcyl2} can be written as
\small
\begin{align}
&\dot{M}_{acc}=-\frac{2\pi}{\partial R v_{k}/\partial R}\times\nonumber\\
&\left(\frac{\partial}{\partial R}\left( R^2 \alpha_{int}\Sigma c_{s}^2\right)+R^2\left(\langle \rho v_{z} \delta v_{\phi}\rangle-\langle B_{z}B_{\phi}\rangle\right)\bigg |_{z_{min}}^{z_{max}}\right)\,,\label{eq:mdot}
\end{align}
\normalsize
for a steady state. { It is the differential form of Equation \ref{eq:torqueconst2}.}
Equation \ref{eq:mdot} suggests that both the internal stress (the radial gradient of the $r$-$\phi$ stress) and the surface stress can lead to accretion. To understand the flow structure, We will
measure these stresses and $\alpha$ values directly from our simulations.

\section{Method}

\begin{figure*}[t!]
\includegraphics[trim=0mm 0mm 0mm 0mm, clip, width=7.in]{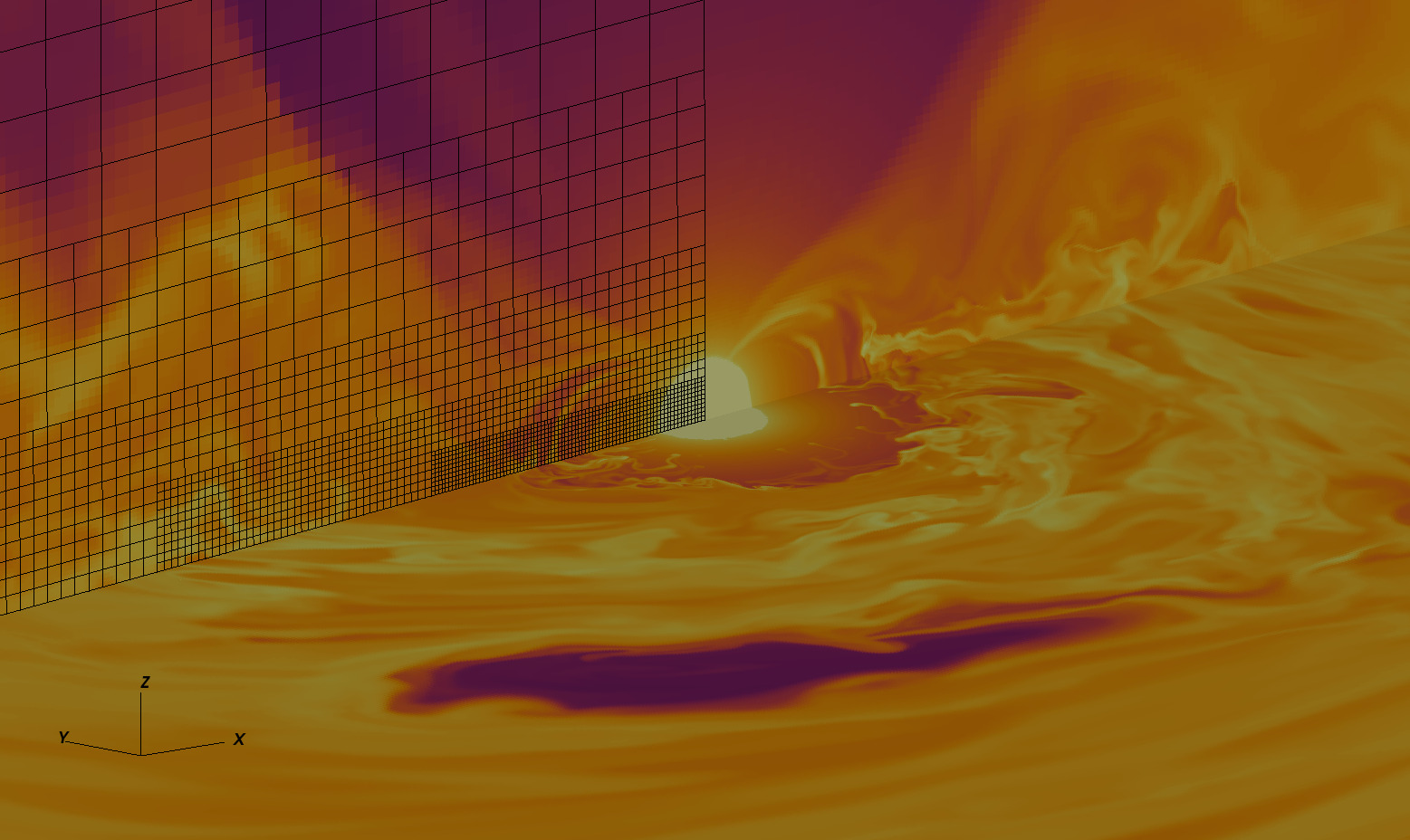}
\figcaption{ Color contours of density at the disc midplane (the $x$-$y$ plane) and in a vertical slice (the $x$-$z$ plane) at time 11 $T_0$. 
The adopted mesh structure is also shown on the left side of the $x$-$z$ plane. The movie can be downloaded at \url{https://figshare.com/articles/media/Magnetospheric_Accretion/24103623}.
  \label{fig:setup}}
\end{figure*}

We solve the magnetohydrodynamic (MHD) equations  in the ideal MHD limit using Athena++ \citep{Stone2020}.
Athena++ is a grid-based code using a higher-order Godunov scheme for MHD and 
 constrained transport (CT) to conserve the divergence-free property for magnetic fields. 
Compared with its predecessor Athena \citep{GardinerStone2008,Stone2008},  Athena++ is highly optimized for 
speed and uses a flexible grid structure that enables mesh refinement, allowing global
numerical simulations spanning a large radial range. 

We adopt a Cartesian coordinate system ($x$, $y$, $z$) with mesh-refinement to include both the central magnetized star and the accretion disc. 
Although the spherical-polar coordinate system is normally used for accretion disc studies, its grid cells become highly distorted at the grid poles in this case. 
Although Athena++ includes a special polar boundary condition to treat these grid cells \citep{ZhuStone2018}, they still introduce large numerical errors and significantly
limit the evolution timestep. Considering that most accretion onto the star occurs close to the star's magnetic poles, accurately simulating the flow in these regions
is crucial. Thus, the Cartesian coordinate system is better suited for magnetospheric accretion studies. 
After the simulation is completed, we transform all quantities into spherical-polar coordinates and cylindrical coordinates to simplify the data analysis. 
In this paper, we use ($R$, $\phi$, $z$) to denote positions in cylindrical coordinates and
($r$, $\theta$, $\phi$) to denote positions in spherical polar coordinates.  In both coordinate systems, 
$\phi$ represents the azimuthal direction (the direction of disc rotation). 

Our simulation domain expands from -64$R_0$ to 64$R_0$ in each $x$, $y$, and $z$ direction, with 320 grid cells in each direction at the root level. $R_0$ is the code length unit, which is quite close to the magnetospheric truncation radius $R_T$ at the end of the simulation. { The stellar radius, denoted as $r_{in}$, is chosen as 0.1 $R_0$, so that the magnetospheric truncation radius is roughly 10 times larger than the stellar radius. }
Static mesh-refinement has been adopted with the fourth level (cells with 2$^4$ times shorter length) at [-8, 8]$R_0\times$[-8, 8]$R_0\times$[-1, 1]$R_0$ for $x\times y\times z$. 
Within this fourth level domain, one additional higher level is used for every factor of 2 smaller domain until the seventh level at [-$R_0$, $R_0$]$\times$[-$R_0$, $R_0$]$\times$[-0.125$R_0$, 0.125$R_0$]. Color contours of density and the grid structure for the innermost several levels are shown in  Figure \ref{fig:setup}.
If the disc's aspect ratio is 0.1, the disc scale height is resolved by 16 to 32 grid cells at the disc region between $R=0.5 R_0$ to $R=8 R_0$. At the finest level, the cell size is 0.003125 $R_0$. { Piecewise linear method is used in the spatial reconstruction, while Van Leer integrator is adopted for the time integration. }

\subsection{Disc Setup}

The initial density profile at the disc midplane (the $x-y$ plane) is
\begin{equation}
\rho_{d}(R,z=0)=\rho_{0}\left(\frac{R}{R_{0}}\right)^p\,.\label{eq:rho0mid}
\end{equation}
In our code unit, we set $\rho_0\equiv\rho_{d}(R=R_{0},z=0)$=1, and the time unit $1/\Omega(R=R_0)=T_0/2\pi$.  $T_{0}$ is the orbital period $T(R)$ at $R_0$. Thus, $T_0$ is also close to the orbital period at the magnetospheric truncation radius.
The temperature is assumed to be constant on cylinders
\begin{equation}
c_s^2(R,z)=c_s^2(R=R_{0},z=0)\left(\frac{R}{R_{0}}\right)^q\,.\label{eq:qvalue}
\end{equation}
where $c_{s}=\sqrt{p/\rho}$ is the isothermal sound speed.
We choose $p=-2.25$ and $q=-1/2$ so that the disc surface density $\Sigma\propto R^{-1}$.  For the disc temperature, we set $(H/R)_{R=R_{0}}$=0.1, where $H=c_{s}/\Omega_{K}$. We note that this disc is thicker than a typical protoplanetary disc at the magnetospheric truncation radius. Simulations for thinner discs are more computational expensive and will be presented in the future. We have used the adiabatic equation of state with $\gamma$=1.4, while an almost instantaneous cooling is applied for each grid cell at every timestep (the thermal relaxation time is 10$^{-5}$ local orbital time using the cooling treatment in \citealt{Zhu2015c}).

The density and velocity in the $R$-$z$ plane are set to be
\small
\begin{equation}
\rho_{d}(R,z)=\rho_{d}(R,z=0) {\rm exp}\left[\frac{GM_*}{c_{s}^2}\left(\frac{1}{\sqrt{R^2+z^2}}-\frac{1}{R}\right)\right]\,,\label{eq:rho0}
\end{equation}
\normalsize
and
\small
\begin{equation}
v_{\phi,d}(R,z)=v_{K}\left[(p+q)\left(\frac{c_{s}}{v_{\phi,K}}\right)^2+1+q-\frac{qR}{\sqrt{R^2+z^2}}\right]^{1/2}\,,\label{eq:vphi}
\end{equation}
\normalsize
with $v_{K}=\sqrt{GM_{*}/R}$ (e.g. \citealt{Nelson2013}). To avoid the density and velocity becoming infinite at $R=0$, we use $R=\max(R,r_{in})$ on the right-hand side of Equations \ref{eq:rho0mid}-\ref{eq:vphi}. 

\subsection{Star Setup}

The gravitational acceleration from the central star is set as
\begin{equation}
a_r=
\begin{cases}
-\frac{GM_*}{r^2}\frac{(r-r_{in})^2}{(r-r_{in})^2+r_{sm}^2}\;\;\;&at\;r>r_{in}\\
0\;\;\;&at\;r<r_{in}\,,
\end{cases}
\label{eq:ar}
\end{equation}
where both the stellar radius $r_{in}$ and the smoothing length $r_{sm}$ are set to be 0.1 $R_0$. 
{  At the region with $R<r_{in}$, $c_s(R,z)$ is set as a constant that is equal to $c_s(r_{in},z)$ in Equation \ref{eq:qvalue}.}

The density and pressure of the star at $r\le r_{in}$ are set to be constants as $\rho(r)=\rho_{*,in}$  and $P(r)=\rho_{*,in}c_s^2(R=r_{in})$ with a large $\rho_{*,in}$ value of $10^7\rho_{0}$. We decrease the disc density by a factor of $exp(-(r-r_{mag})^2/r_{sm}^2)$ beyond $r_{in}$ but within the initial magnetosphere radius $r_{mag}=0.5 R_0$. Then, we add the density of the star that is in pressure equilibrium against the stellar gravity (Equation \ref{eq:ar}). We first solve
\begin{equation}
\frac{dP (r)}{dr}=a_r(r)\frac{P(r)}{c_s^2(R=r_{in})}
\end{equation}
from $r_{in}$ to $r$, and then set $\rho(r)=P(r)/c_s^2(R=r_{in})$. 
The velocity structure of the star and the magnetosphere at $r\leq r_{mag}$ is
\begin{equation}
v_\phi(r)=\Omega_* R+v_{\phi,d}e^{-(r-r_{mag})^2/r_{sm}^2}\,.  \\
\end{equation}
In this work, we set $\Omega_*$=0 as appropriate for a non-rotating star. 
To maintain the star's structure, we fix
the density, velocity, and pressure to be the initial values at $r<r_{in}$.

The stellar magnetic field is assumed to be a dipole. { Although the magnetic fields of a real star also consist of higher multipole components, the strength of the open fields and thus the magnetospheric truncation radius are mainly determined by the dipole component \citep{Johnstone2014}. }
To maintain $\nabla\cdot\bf{B}=0$, we use the
vector potential $\bf{A}$ to initialize magnetic fields (${\bf B}=\nabla\times{\bf A}$) :
\begin{equation}
{\bf A}=\frac{{\bf \overline{m}}\times {\bf r}}{r_c^3}\,,
\end{equation}
where $r_c=max(r, r_{in})$ to avoid the singularity at r=0.  The magnetic moment ${\bf m}$ is thus $4\pi{\bf \overline{m}}$. Note that the vacuum permeability constant is assumed to be 1
in Athena++, and thus the magnetic pressure is simply $B^2/2$ in code units. We choose ${\bf {\overline m}}$=-0.0447${\bf e_z}$ so that the initial
plasma $\beta=2P/B^2$ at $R=R_0$ is 10. { The fields within the star evolve through numerical diffusion. But at the disc surface $r_{in}$, the midplane field strength only decreases by 5\% at the end of the simulation.}

To avoid small timesteps within the highly magnetized magnetosphere, we employ a density floor that varies with position 
\begin{equation}
\rho_{fl}=\rho_{fl,0}\left(\frac{r}{R_0}\right)^{p}+\rho_{flm,0}\left(\frac{r}{R_0}\right)^{pm} \,,\label{eq:floor}
\end{equation}
where $\rho_{fl,0}=10^{-5}\rho_0$, $\rho_{flm,0}$=1.33$\times 10^{-5}\rho_0$, and $pm=-5.5$. 
When $\rho_{fl}$ gets smaller than $10^{-9}\rho_0$, we choose $10^{-9}\rho_0$ as the density floor. { Since the smoothing length (0.1 $R_0$) is resolved by 32 cells at the finest level, the star maintains hydrostatic equilibrium quite well in the absence of magnetic fields. However, the adoption of the high density floor due to the strong stellar magnetic fields leads to inflow onto the star. The density floor is chosen low enough that this inflow is much weaker than the magnetospheric accretion from the disc. }

At both the $x$ and $y$ boundaries, the flow and magnetic fields are fixed at the initial values during the whole simulation.
In the $z$ directions, we adopt outflow boundary conditions. In the rest of the paper, we drop the code unit (e.g. $\rho_0$, $R_0$, $T_0$)
after the physical quantities for simplification. 

\section{Results}
\begin{figure}[t!]
\centering
\includegraphics[trim=2mm 10mm 0mm 10mm, clip, width=3.4in]{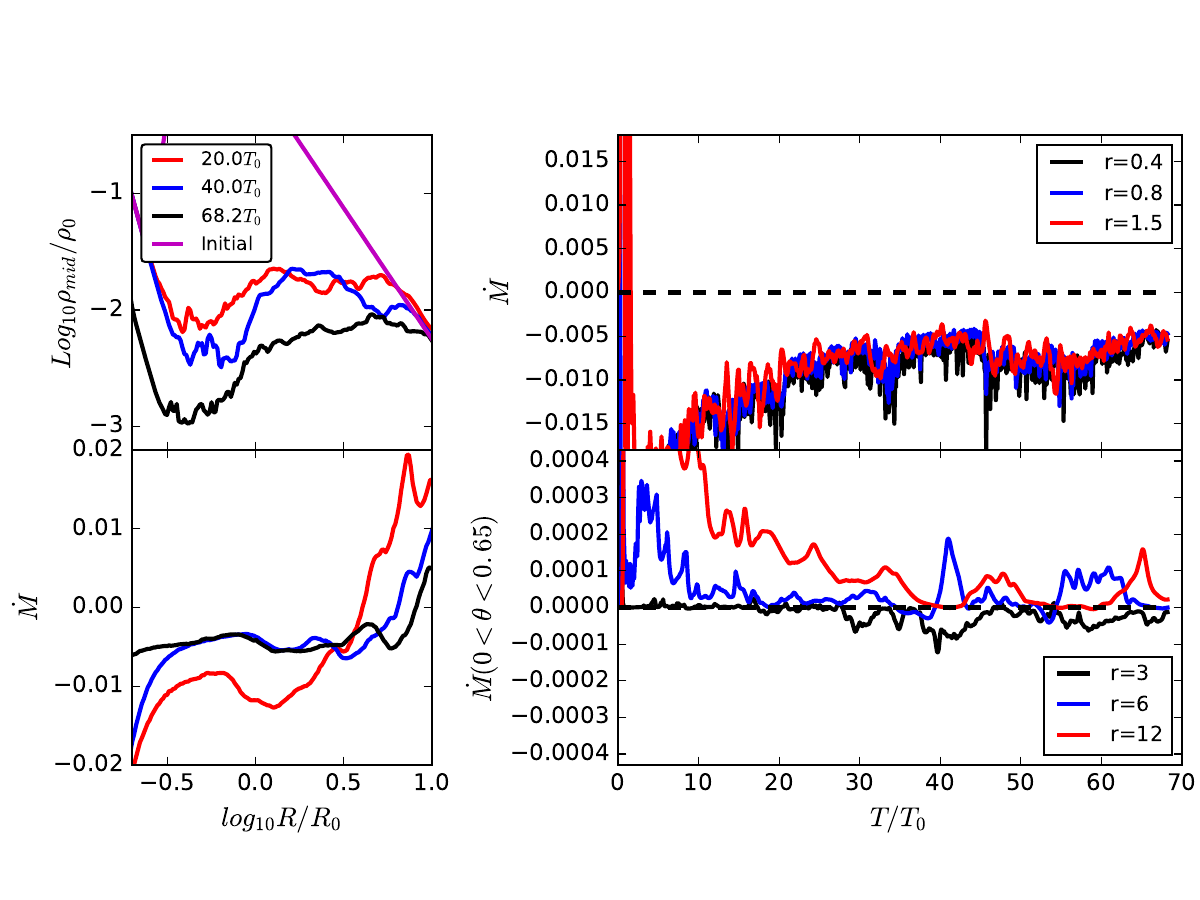}
\figcaption{Left panels: the disc midplane density and  mass accretion rate along the R direction at three different times. Quantities are averaged over both the azimuthal direction and time (from each time span at 18 to 20 $T_0$, 38 to 40 $T_0$, and 66.2 to 68.2 $T_0$, we average over 21 snapshots). Right panels: the  mass accretion rate at r=0.4, 0.8, 1.5 with time (upper panel), and the mass outflow rate at the disc atmosphere (integrated within $0<\theta<0.65$) at r=3, 6, 12 with time (lower panel). \label{fig:steadystate}}

\end{figure}

We run the simulation for 65 orbits at $R_0$, and subsequently, we continue the simulation for another 3.2 orbits albeit with a 10 times smaller $\rho_{flm,0}$ for the density floor. The total time is equivalent to 2157 Keplerian orbits at the stellar surface $R_{in}=0.1$. The disc has settled to a quasi-steady state, as shown in the right panels of Figure \ref{fig:steadystate} where
the disc's mass accretion rate and one-sided outflow rate are plotted against time. 
The lower left panel of Figure \ref{fig:steadystate} also shows that the disc accretion  has reached a steady state within $R\sim 6$ at the end of the simulation. The upper left panel of Figure \ref{fig:steadystate} shows a sharp drop in density within the magnetospheric truncation radius, $R_T\sim$1.  Beyond this radius, the density at the  disc midplane remains relatively constant. In the following subsections, we will present our findings for different regions in the order of the regions' proximity to the star, starting from the magnetosphere region.

\subsection{Magnetosphere and Instabilities}
\label{sec:instability}

\begin{figure*}[t!]
\centering
\includegraphics[trim=0mm 35mm 15mm 25mm, clip, width=6.in]{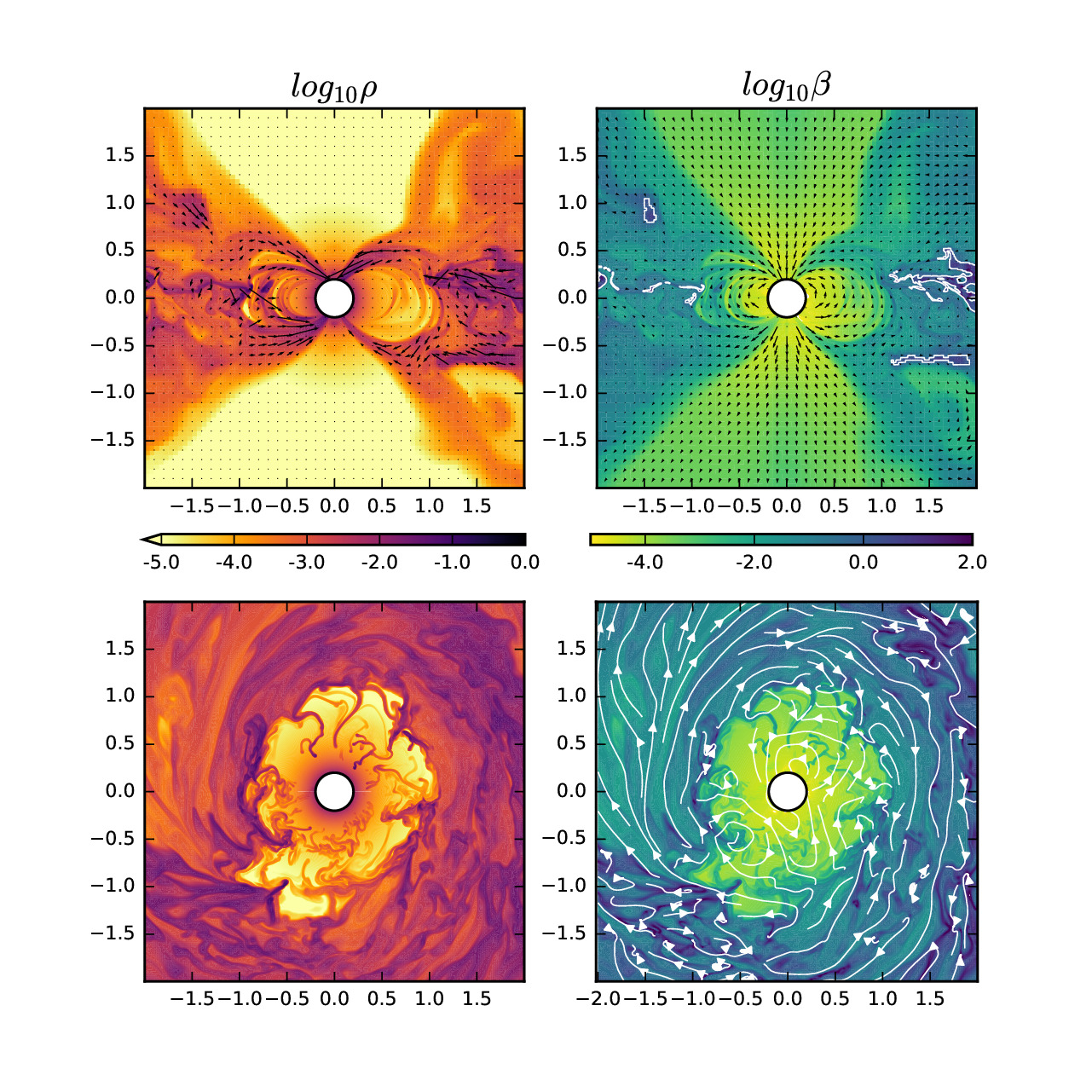}
\figcaption{Poloidal and midplane slices (upper and lower panels) for density and plasma $\beta$ (left and right panels) at the end of the simulation. The $\beta=1$ contour is also shown as white contours in the upper right panel. Vectors of the momentum and magnetic field are overplotted on the upper $\rho$ and $\beta$ panels respectively. The magnetic streamlines are overplotted in the lower right panel. \label{fig:mid2d}} 
\end{figure*}

The flow within the magnetospheric truncation radius is highly dynamic. Figure \ref{fig:mid2d} shows the contours for the density and magnetic fields at the end of the simulation. In the upper panels of Figure \ref{fig:mid2d}, the poloidal plane reveals a distinct contrast in flow structure within and beyond $R_0 \sim R_T$. The disc region outside of $R_0$ is turbulent, with certain denser regions exhibiting $\beta$ values $\sim$ 100. In contrast, the region within $R_0$ appears to be laminar in the poloidal plane and $\beta$ is less than $10^{-2}$. It seems that the material cannot penetrate into the magnetosphere and has to follow the field lines falling onto the star close to the polar directions. However,  a different picture emerges from the midplane slices at the lower panels of Figure \ref{fig:mid2d}. At the edge of the magnetosphere around $R_0$, the disc material becomes filamentary and develops ``fingers'' that penetrate into the magnetosphere. Due to their higher density, these filaments have higher $\beta$ values than the rest of the magnetosphere. As they move in, they are lifted and move along the dipole magnetic field lines. In the upper panels of Figure \ref{fig:mid2d}, we can see the poloidal cut of these intruding filaments deep inside the magnetosphere. Closer to the star, fewer filaments persist at the midplane (lower panels), as some filaments have been lifted and subsequently accrete onto the star. 

\begin{figure}[t!]
\includegraphics[trim=0mm 0mm 5mm 5mm, clip, width=3.5in]{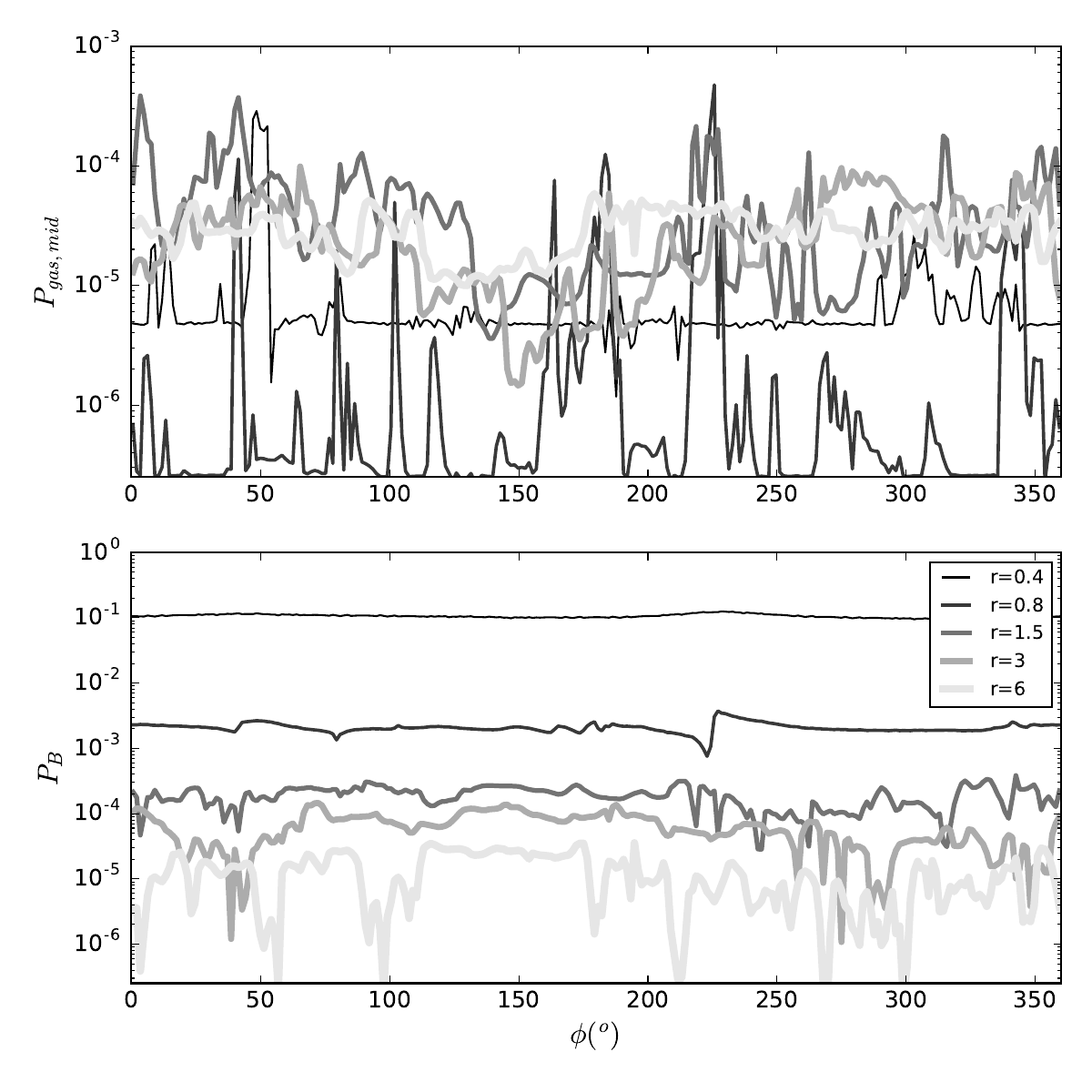}
\figcaption{The midplane gas and magnetic pressure at different radii along the azimuthal direction at the end of the simulation.  \label{fig:mid1dazi}} 
\end{figure}

To show the development of the filaments, we plot the midplane gas pressure and magnetic pressure at different radii along the azimuthal direction in Figure \ref{fig:mid1dazi}. At the outer edge (e.g. $r=6$),  the gas pressure is within the same order of magnitude as the magnetic pressure.  Both pressures fluctuate within one order of magnitude with slight anti-correlation (higher $P_{gas}$ corresponds to lower $P_{B}$). As $r$ becomes smaller, magnetic fields become stronger while the gas pressure decreases due to the magnetospheric truncation. This results in a smoother profile of the magnetic pressure but a significant increase in density fluctuations. At $r=0.8$, the density can fluctuate more than 3 orders of magnitude and the gas pressure is near the magnetic pressure only at the highest density peaks. Since the lowest density region has reached the density floor, the real pressure fluctuations are more significant than what is depicted in the plot. Very close to the star (e.g. $r=0.4$), the magnetic pressure substantially exceeds the gas pressure, even within the densest filaments. With a higher density floor imposed in this region, the density fluctuations within the filaments are less accurately captured.

\begin{figure*}[t!]
\centering
\includegraphics[trim=0mm 10mm 5mm 5mm, clip, width=6.in]{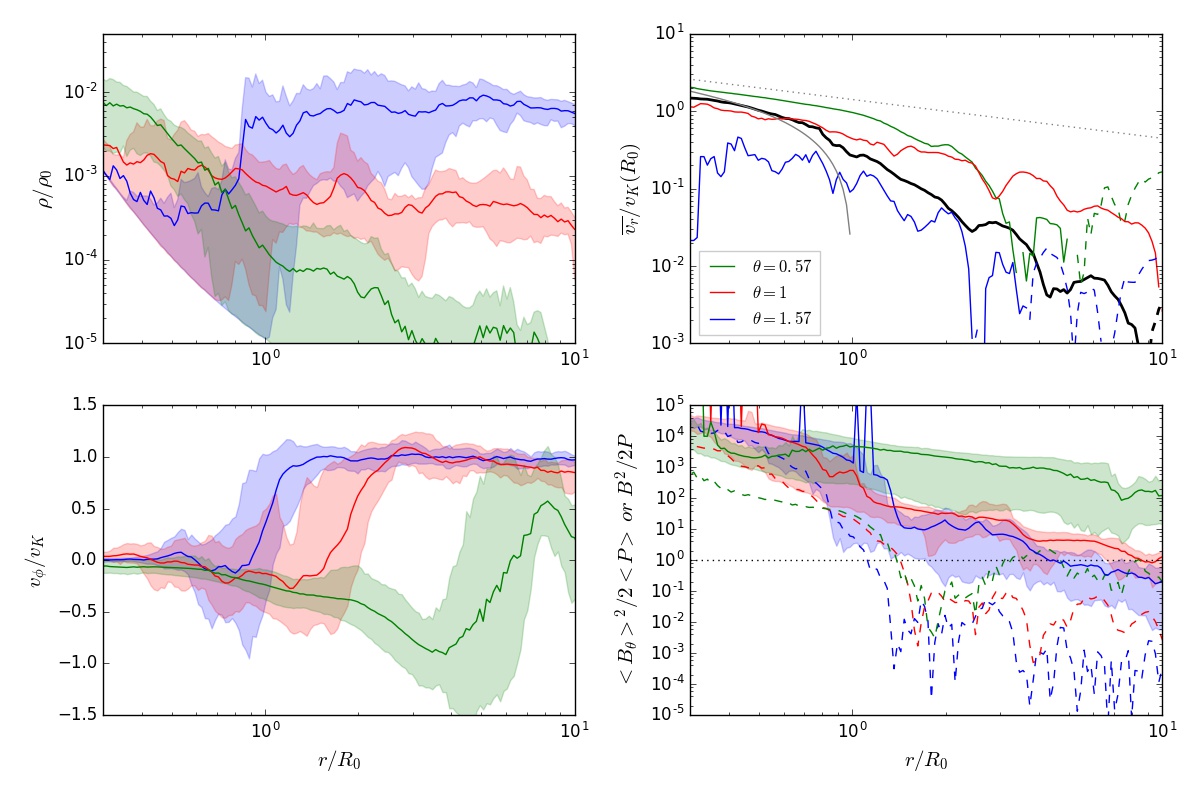}
\figcaption{ The azimuthally averaged density, radial and azimuthal velocities, and 1/$\beta$ along different $\theta$ directions (different colors) at the end of the simulation. The shaded area shows the quantities within 10\% and 90\% 
of all the data along the azimuthal direction. In the $\overline{v_r}$ panel, the thick black solid curve  uses the spherically integrated
mass flux divided by the spherically integrated density. The dashed curves in this panel represent negative numbers. The thin solid curve is the free-fall velocity starting at $R_0$, and the dotted line is the free-fall velocity starting from infinity. In the lower right panel, the dashed curves
are calculated with the azimuthally averaged $B_{\theta}$ and pressure, while the solid curves  are the azimuthally averaged value of 1/$\beta$ from each cell.
\label{fig:mid1dfluc}}
\end{figure*}

The density fluctuation is also shown in the upper left panel of Figure \ref{fig:mid1dfluc}. At the disc midplane and along $\theta=1$, the amplitude of the density fluctuation (the shaded region) increases towards the inner disc where the disc magnetic field is stronger. The lower density boundary at $r\lesssim1$ is  the density floor. Although the density at the midplane has a sharp drop at $r\sim$1, the density profiles at $\theta=1$ and 0.57 are relatively smooth. This suggests that material high above the disc midplane smoothly accretes into the magnetosphere and onto the star, more similar to the spherical accretion onto a magnetized star. Such smooth transition from the disc to the magnetosphere at higher altitudes is also apparent in Figure \ref{fig:mid2d}. Magnetospheric accretion in our simulation  seems to be a mixture of the traditional thin disc magnetospheric accretion at the midplane and the spherical accretion above the midplane { (more discussion in \S \ref{sec:surface})}.

The ``fingers'' penetrating into the magnetosphere is due to a  type of Rayleigh-Taylor (RT) instability that involves magnetically supported material, called the ``interchange instability'' \citep{Kruskal1954,Newcomb1961}. \cite{Arons1976} suggested that this instability can occur at the disc-magnetosphere boundary, which is confirmed later by numerical simulations \citep{Kulkarni2008,Romanova2008,Blinova2016}. 
\cite{Spruit1995} 
has derived the general condition for the instability taking the velocity shear into account, which agrees well with numerical simulations \citep{Romanova2008,Takasao2022}. For our simulation with a non-rotator, the instability is expected at $R_T$. Within the magnetosphere, the flow couples strongly with stellar magnetic fields and its azimuthal velocity reduces to zero (the lower left panel of Figure \ref{fig:mid1dfluc}). With such a small azimuthal velocity, the magnetosphere  can be considered a hydrostatic  fluid supported by magnetic pressure against gravity.  This magnetized fluid is unstable when $-d\rho/dz<\rho^2 g/\gamma P$ \citep{Newcomb1961}, where the gravity is towards negative $z$. This condition is identical to the condition of the Rayleigh-Taylor (RT) instability and independent of the field strength.  Since the density always increases with $r$ (opposite to the direction of the gravity) at the boundary between the magnetosphere and the disc due to the magnetospheric truncation, the instability condition is satisfied and the instability develops. In the nonlinear regime of the instability, the material becomes filamentary and sinks to the star \citep{Stone2007}. We caution that, although the instability grows fast in our simulation with a non-rotating star, it will be suppressed for a rotating star \citep{Blinova2016}.

Slightly different from previous simulations, our high-resolution simulation reveals that the filaments have substructures. A larger filament can split into multiple smaller sub-filaments when it moves in, implying a highly dynamic magnetosphere. When the filaments move closer to the star, the magnetic pressure and stress keep increasing  and more filamentary material starts to climb vertically along the dipole fields and onto the star. Eventually, most material accretes onto the star at high latitudes.

Once the material begins to follow the field lines, it undergoes a free-fall motion driven by the stellar gravity (Equation \ref{eq:eintegral}). To verify the free-fall motion, we plot the averaged radial velocity at different $\theta$ angles in the upper right panel of Figure \ref{fig:mid1dfluc}. The velocity is calculated by dividing the azimuthally averaged radial momentum by the azimuthally averaged density. The thick black solid curve  uses the spherically integrated mass flux divided by the spherically integrated density. The thin solid curve is the free-fall velocity starting from $R_T\sim R_{0}$:
\begin{equation}
v_{ff}=\left(\frac{2GM_*}{r}\right)^{1/2}\left(1-\frac{r}{R_{0}}\right)^{1/2}\,,
\end{equation}
while the dotted line is the free-fall velocity starting from infinity. 
Within the magnetosphere, the accretion generally follows the free-fall velocity except for the disc midplane region. At the disc midplane, magnetic fields are in the vertical direction so that the radial inflow is only possible via the interchange instability.

\subsection{Transition between the Magnetosphere and the Disc}
\label{sec:RT}
In the classical model, the magnetosphere and the disc are sharply separated at the magnetospheric truncation radius. Although our simulation supports this sharp transition at the disc midplane, the transition is more gradual and less obvious in the disc atmosphere, especially based on the $\rho$ and $v_r$ structure.  The upper right $\overline{v_r}$ panel of Figure \ref{fig:mid1dfluc} shows that, at $\theta=1$, the inflow velocity is still 10\% of the free-fall velocity even at $r\sim$8, and the accretion process smoothly transitions from the disc surface accretion to the magnetospheric infall at smaller $r$ (more discussion on the disc surface accretion in Section \ref{sec:surface}).
At $\theta=0.57$, we see mass outflows far away at large $r$ (Section \ref{sec:asym}). 

\begin{figure*}[t!]
\includegraphics[trim=0mm 5mm 35mm 55mm, clip, width=6.7in]{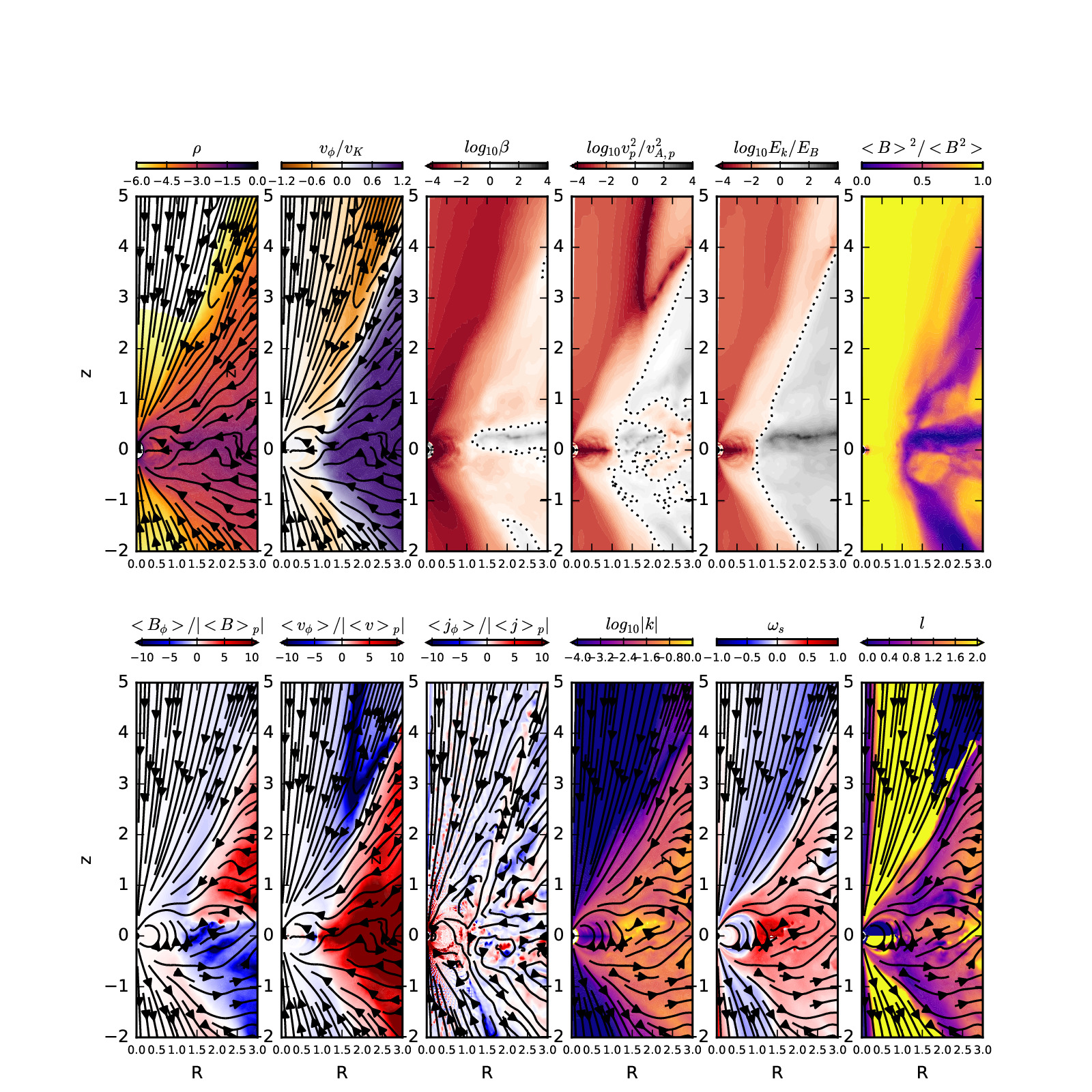}
\figcaption{Various quantities related to magnetic fields at the end of the simulation. The streamlines of poloidal velocity are shown in the $\rho$ and $v_{\phi}$ panels, while the streamlines of poloidal magnetic field are shown in $B_{\phi}$ and $k$, $\omega_s$, $l$ panels. The streamlines  of electric current are shown in the j panel. The dotted curves in the upper middle 3 panels indicate the regions where the quantity equals zero. Plotted quantities have been averaged over the azimuthal direction except the ones with $\langle\rangle$ which indicates the azimuthal averaging for specific quantities. For streamlines and $k$, $\omega_s$, $l$ constants, the primitive variables have been averaged over the azimuthal direction.).  \label{fig:magconst}} 
\end{figure*}

The distinction between the two regions becomes more pronounced when examining the structure of $v_{\phi}$ and $B_{\phi}$ in the R-z plane (the lower left two columns of Figure \ref{fig:magconst}). $v_{\phi}$ changes from negative values at small $R$ to positive values in the disc, which is also shown in  the lower left panel of Figure \ref{fig:mid1dfluc}. The deviation from the Keplerian rotation suggests that the region is more magnetically and less rotationally supported. $B_{\phi}$ also changes sign where $v_{\phi}$ changes sign (the reason will be discussed in Section \ref{sec:surface}). Furthermore, magnetic fields are mostly poloidal around the star while toroidal in the disc. The different field geometries in these two regions are also shown in the lower right panel of Figure \ref{fig:mid1dfluc} (the $1/\beta$ panel). Within the disc region ($R\gtrsim$1), the dominance of azimuthal fields is evident from the substantial difference between $\langle B^2/2P \rangle$ and $\langle B_{\theta}\rangle^2/2\langle P \rangle$. Conversely, at $R\lesssim$1, these two quantities closely align, signifying the dominance of poloidal fields. Magnetic fields are  mostly axisymmetric within the magnetosphere ($\langle B \rangle^2/\langle B^2\rangle\sim$1 in the upper right panel of Figure \ref{fig:magconst}). Using azimuthally averaged quantities, we calculate the conserved constants along magnetic field lines ($k$, $l$, $\omega_s$ from Equations \ref{eq:kintegral} to \ref{eq:lintegral}), which are plotted in Figure \ref{fig:magconst}. These quantities are still roughly constant along the streamlines within the magnetosphere, although the density is highly filamentary as shown above. We will discuss the conserved constants in detail in Appendix \ref{sec:constantsofintegral}.

We have tried various methods to quantitatively define the transition radius between the magnetosphere and the disc. 
At the disc midplane, we define the transition radius as where $\Omega=0.5 \Omega_K$, and consider it as the magnetospheric truncation radius. We measure this radius as 1.02 $R_0$ at the end of the simulation. 
The  magnetospheric truncation radius at the disc midplane has been defined in various ways in the literature. But, for a slow-rotator, we find that they all provide similar values. 

The mostly widely used $R_T$ definition (Equation \ref{eq:RT}) is derived from
$v_r^2/v_{A,p}^2=1$, or more specifically
\begin{equation}
\rho (R_T) v_{r}^2(R_T)=\frac{B_p^2(R_T)}{4\pi} {\rm\;\;\; in\;\; C.G.S.}\,,
\end{equation}
where $v_r=\sqrt{2GM_*/r}$ and $B_p$ follows a dipole stellar field.
This equation can be interpreted in several ways, including ram pressure balancing magnetic pressure, free-fall radial speed equal to the Alfv\'en speed, or magnetic energy density equal to radial kinetic energy density. In our unit system, we have
\begin{equation}
R_T=\left(\frac{\overline{m}^4 \left(4\pi\right)^2}{2GM_*\dot{M}^2}\right)^{1/7}\,.\label{eq:RTcode}
\end{equation}
With our adopted initial dipole fields ($\bf{B_0}$) and the measured $\dot{M}=0.005$ at the end of the simulation, we can calculate $R_T=1.44$. Considering that the stellar dipole field is $\sim$0.6 $B_0$ around $R_0$ at the end of the simulation (Section \ref{sec:field}), $R_T$ calculated with 0.6 $B_0$ is 1.07 $R_0$, which is remarkably close to our measured $1.02 R_0$. Such a good agreement is due to: 1) the surface radial velocity is indeed close to the free-fall velocity (Figure \ref{fig:mid1dfluc}); 2) the dipole field has a very strong radial dependence ($P_{B}\propto r^{-6}$) so that $R_T$ has a weak dependence on parameters (Equation \ref{eq:RTcode}). { Although our derivation is limited to non-rotators, \cite{Takasao2022} find that Equation \ref{eq:RTcode} could also apply to fast rotators. }

The other ways to define $R_T$ include the radius where $\beta$=1 \citep{Pringle1972,Bessolaz2008,Kulkarni2013}, and the radius where the total kinetic energy ($E_k$) equals the magnetic energy ($E_B$)\citep{Lamb1973}. 
In the $\beta$, $v_p^2/v_{A,p}^2$, $E_k/E_B$  panels of Figure \ref{fig:magconst}, we see that all three definitions (black dotted curves) give similar $R_T$ at the disc midplane.  

However, when considering regions above the disc midplane, the transition radius between the magnetosphere and the disc exhibits significant variation depending on the method employed. The $\beta$ panel shows that the magnetic pressure dominates over the thermal pressure in most regions except for the disc midplane and high above the surface where $B_{\phi}$ changes sign. The $v_p^2/v_{A,p}^2$ panel shows that the disc surface (with a high infall velocity and weak fields) has super-Alfv\'enic speed but with some sub-Alfv\'enic patches. Among these three diagnostics, $E_k/E_B=1$ provides the clearest separation between the disc and highly magnetized regions. This $E_k/E_B=1$ boundary also corresponds to the boundary where $B_\phi$ and $v_{\phi}$ change sign in the disc's atmosphere. Therefore, we consider $E_k/E_B=1$ as the boundary that effectively separates the magnetosphere region from the disc region.

\subsection{Magnetically Disrupted Disc and Outflow}
\label{sec:asym}
\begin{figure*}[t!]
\includegraphics[trim=0mm 30mm 0mm 20mm, clip, width=6.5in]{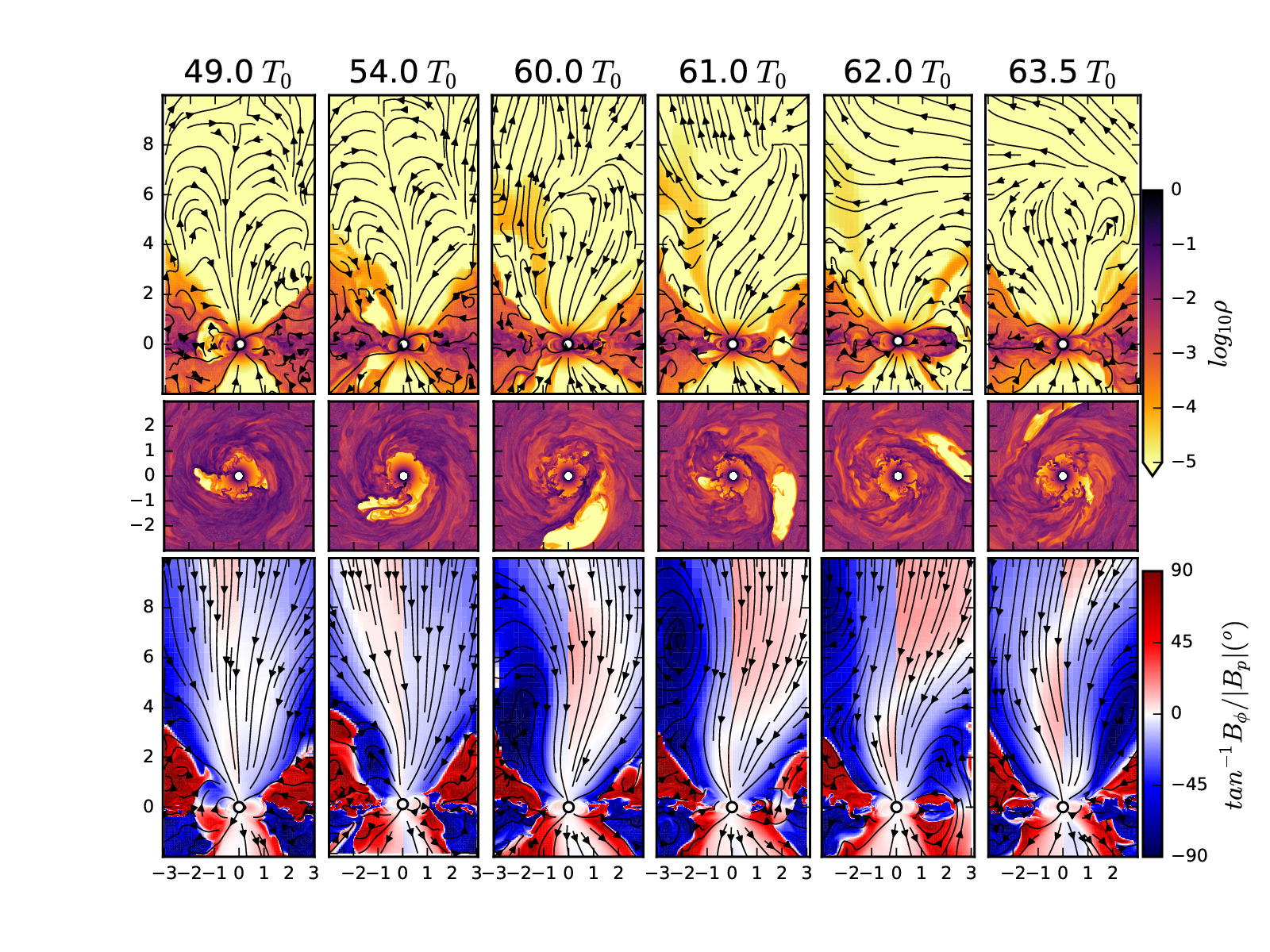}
\figcaption{The density (upper and middle panels) and $B_{\phi}$ (lower panels) structure at different times, highlighting the density void in the disc. The vectors in the upper and lower panels are the velocity and magnetic field vectors respectively.  While the upper and lower panels show the poloidal cut, the middle panels show the midplane cut. The movie can be downloaded at \url{https://figshare.com/articles/media/Magnetospheric_Accretion/24103623}. \label{fig:outflow}} 
\end{figure*}

Although the disc is pressure supported, the strong magnetization makes this region highly dynamic,   occasionally leading to magnetic disruptions in some regions. Magnetic reconnection and interchange instability could sometimes reorganize magnetic fields around the truncation radius, leading to a large-scale density void (the middle panels of Figure \ref{fig:outflow}). { Before the density void forms, $B_{\phi}$ changes sign 3 times when transitioning from one side of the disc to the other (bottom panels).} { While the density void is forming,} the magnetic fields on both sides of the disc directly connect with each other (49 $T_0$ panel in the bottom row). { The void starts at the magnetospheric truncation radius and expands outwards. The outward motion slows down around 2-3 $R_T$ and the void} orbits around the central star at sub-Keplerian speed. As shown in Figure \ref{fig:outflow}, it takes 6 $T_0$ for the density void to finish one orbit, while the Keplerian orbital timescale at $R=2$ is only 2.8 $T_0$. This suggests that the asymmetric density structure is magnetically connected to either the region at larger scales with a slower Keplerian speed or { the region within the magnetosphere which also rotates slowly.}  The magnetic field lines are shown in the bottom rows of Figure \ref{fig:outflow}, where the density void is connected with the strong azimuthal magnetic fields at the disc surface and these magnetic fields rise up and outwards with time. { When the magnetic bubble rises, it  opens field lines and drives outflow.}

These density voids and magnetic islands are remarkably similar to those in models of magnetically arrested discs (MAD, \citealt{Tchekhovskoy2011}). { The density voids in MAD discs are similarly associated with flux tubes which move outwards until the circularization radius, and eventually dissipate after several orbits \citep{Porth2021}.  These voids appear quasi-periodically, regulating the spin-up/down of the central blackhole and outflow rates, which may be associated with the flares of Sgr A*. Similarly, these voids due to the strong disc fields have also been invoked to explain the giant flares in protostars \citep{Takasao2019}. }  On the other hand, it is essential to highlight that magnetospheric accretion differs from the MAD state around BHs or discs threaded by external vertical fields.
In magnetospheric accretion, mass in the disc accretes to the central star following the stellar field lines after mass is loaded into these field lines through turbulence or reconnection processes. In contrast, in the MAD state or discs with net vertical fields, there is no field line connecting the disc and the central object. Hence, we choose to designate our observed disc state, characterized by the presence of magnetically buoyant bubbles during the magnetospheric accretion, as the magnetically disrupted disc (MDD) rather than categorizing it as the MAD state.

In our simulations, it is also evident that the appearance of  magnetic islands is related to subsequent disc mass ejection/outflow events. As shown in Figure \ref{fig:outflow}, the magnetic islands are magnetically connected to the disc surface, generating strong negative $B_\phi$ at $z>0$. The magnetized ``bubble'' at the disc surface moves out (bottom panels), and pushes material outwards (top panels), leading to asymmetric non-steady outflow. Such an outflow event is also shown in the $\dot{M}(0<\theta<0.65)$ panel of Figure \ref{fig:steadystate} as the bumps on the blue and red curves. At $r=6$, the outflow rate  starts to rise at t=55 $T_0$ and drops down at t=63 $T_0$, which corresponds to the time interval in Figure \ref{fig:outflow}. It takes time for the ejection to move to larger distances. At $r=12$, the outflow rate rises at t=65 $T_0$. 

On the other hand, in our simulation that includes a non-rotating central star, the outflow rate is small compared with the accretion rate even during these mass ejection events. Even including outflow from both sides of the disc, the outflow rate during these mass ejection events ($\dot{M}\sim 0.0002$ based on Figure \ref{fig:steadystate}) is less than $5\%$ of the disc accretion rate ($\dot{M}\sim 0.005$). Such a small outflow rate is due to the magnetic field structure in the low density region surrounding the non-rotating star (Figure \ref{fig:magconst}). At one side of the disc (e.g. $z>0$ at $R=2$), all magnetic field lines are pointing towards the star. The material at the disc surface is channeled towards the star directly, and cannot move across the field lines to be loaded to the low density region high above the disc. Furthermore, even if the material can slip into the low density region, the material will simply fall to the star following the field lines. This occurs due to the magnetic fields being connected to the non-rotating star, and as a result, the flow lacks any rotation and  centrifugal support.
Due to this latter reason, it is not even clear if the mass ejection seen in Figure \ref{fig:outflow} can eventually escape the system. 
In contrast, we expect significantly higher outflow rates for rotating stars. Mass loaded into the stellar open fields, either by some steady diffusion or by the magnetic ``bubbles'', could be accelerated by twisted stellar magnetic fields \citep{Matt2005}. However, the outflow in this case depends on the density structure around the star (e.g. from the stellar wind) which is highly uncertain.

\subsection{ Surface Accretion and Disc Evolution}
\label{sec:surface}
\begin{figure}[t!]
\centering
\includegraphics[trim=5mm 5mm 0mm 5mm, clip, width=3.5in]{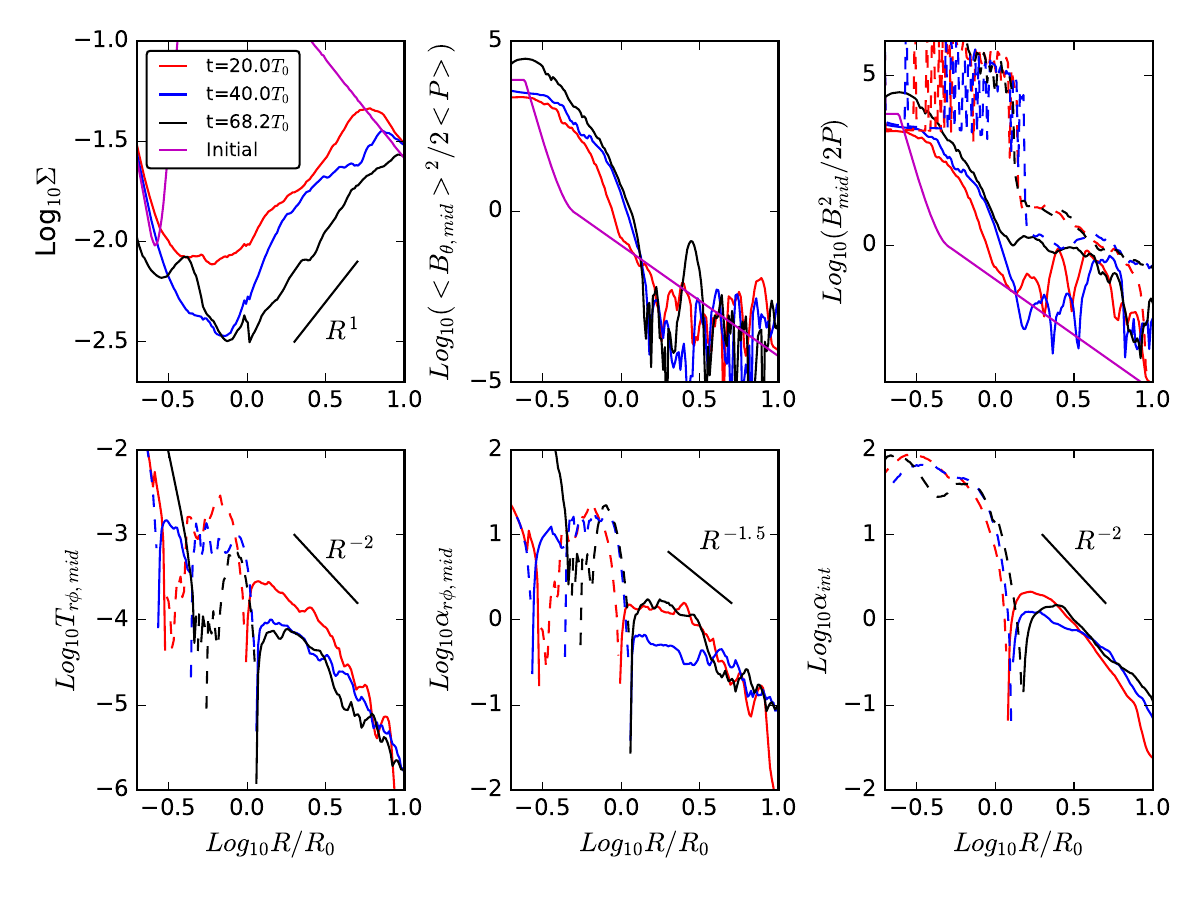}
\figcaption{The disc surface density, $\langle B_{\theta,mid}\rangle^2/2\langle P\rangle$, $B_{mid}^2/2 P$ (upper panels),  midplane r-$\phi$ stress, midplane $\alpha_{r\phi}$, 
vertically integrated $\alpha_{int}$ (lower panels) along the R direction at different times. The dashed curves in the lower panels represent  negative values. In the upper right panel, the solid curves are $\langle B_{mid}\rangle^2/2\langle P\rangle=(\langle B_{r,mid}\rangle^2+\langle B_{\theta,mid}\rangle^2+\langle B_{\phi,mid}\rangle^2)/2\langle P\rangle$, while the dashed curves are  $\langle B_{mid}^2/2P \rangle$. All plotted quantities are averaged over both the azimuthal direction and time (from each time span from 18 to 20 $T_0$, 38 to 40 $T_0$, and 66.2 to 68.2 $T_0$, 21 snapshots are averaged). The quantities with $\langle\rangle$ are azimuthally averaged first, and then the plotted quantities are averaged over time.  \label{fig:onedradial}}
\end{figure}

Our grid setup enables us to evolve the disc for a long timescale (2157 Keplerian orbits at the stellar radius) with a reasonable computational cost. The disc outside the magnetosphere has reached a steady state over a large dynamical range ($R\lesssim 6 R_T$). To study the steady disc accretion, we plot the radial profiles of surface density, stress, $\alpha$, and magnetic fields in Figure \ref{fig:onedradial}.  
The $r$-$\phi$ stress changes sign within the magnetospheric truncation radius, mainly due to the reversal of $B_{\phi}$. Beyond $R_0\sim R_T$, the midplane $r$-$\phi$ stress decreases as $R^{-2}$, so that the midplane $\alpha_{r\phi}$ changes as $R^{-1.5}$. The vertically integrated $\alpha_{int}$ (Equation \ref{eq:alphaint}) changes as $R^{-2}$ and $\Sigma$ changes as $R$. With these profiles, $\dot{M}$ is constant with $R$ (Equation \ref{eq:mdot}) if we ignore the stresses at the disc surface (which will be justified in Appendix B). We note that these profiles are significantly different from those in \cite{ZhuStone2018} where the  disc is threaded by net vertical magnetic fields with a constant initial $\beta$. However, the disc accretion rates in both cases are constant radially. The disc in \cite{ZhuStone2018} has $\Sigma\propto R^{-0.6}$ and $\alpha_{int}\propto R^{-0.4}$, which also leads to a constant $\dot{M}$ along $R$. This suggests that an MHD turbulent disc with net vertical magnetic fields (either from the star or molecular cloud) can reach  different steady states depending on the initial magnetic field distribution and field transport within the disc. The disc evolution is inconveniently affected by the global magnetic field structure that is difficult to be constrained for real astrophysical systems. The magnetic fields adjust themselves quickly and affect the disc structure at a very short timescale, as indicated by \cite{ZhuStone2018}. 

{ While we lack a comprehensive theory to explain the observed differences in the smaller $\alpha$ slope and the higher $\Sigma$ slope in this study compared to those in \cite{ZhuStone2018}, we can offer a preliminary explanation based on stress considerations. To reach a steady state, the vertically integrated stress needs to follow $R^{-1.5}$, independent from the external field configurations, which means $\alpha_{int}\Sigma\propto R^{-1}$ with our temperature profile. Since the net vertical fields in this work decrease much faster outwards compared with those in \cite{ZhuStone2018}, the midplane $\beta$ increases faster outwards even with the same surface density profile, which leads to a faster decrease of $\alpha_{int}$ moving outwards (indicating a smaller slope or a more negative slope). Since $\alpha_{int}\Sigma$ needs to maintain the same $R^{-1}$ slope, $\Sigma$ needs to increase faster outwards (indicating a higher slope), which drives an even steeper $\beta$ and a smaller $\alpha$ slope. Eventually, a balance is achieved with a small $\alpha_{int}$ slope and a high $\Sigma$ slope. To derive the exact slope values, we need to understand how field is transported and amplified in the disc. Such an analytical model has not been constructed. We hope that our simulations here could shed light on how to construct such a model in future.}

We could also estimate the $\alpha$ value by equating the viscous timescale at $R\sim$6 to the simulation time.
The derived $\alpha$ value is on the order of unity, similar to $\alpha_{int}$ derived above. This high $\alpha$ value has important implications for planet formation theory (Section \ref{sec:planet}).

\begin{figure}[t!]
\includegraphics[trim=12mm 0mm 0mm 40mm, clip, width=3.7in]{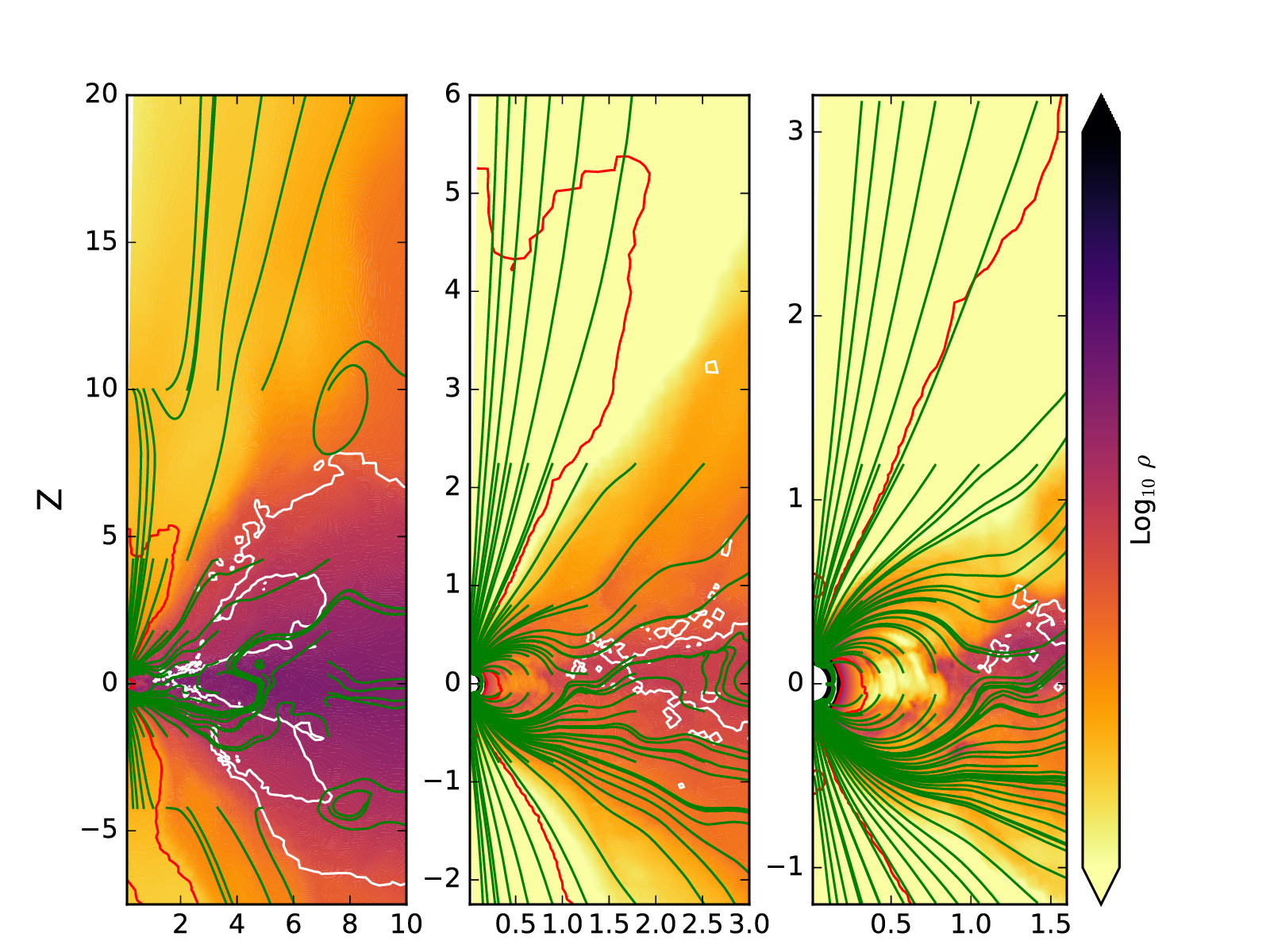}
\figcaption{Azimuthally averaged density at different scales at $t$ = 68.2$T_0$. The green curves are the velocity streamlines calculated with azimuthally averaged
velocities. The white curves label where the azimuthally averaged $\beta$ = 1. The red curves label where the azimuthally averaged density is only larger than the density floor by
10\%, indicating that the majority of grids have reached the density floor there. Only the pole region has reached the density floor.  
The range of the colorbar is [-8,1], [-5,1],[-3.5,-1] in the left, middle, and right panels.
\label{fig:twodrho}}
\end{figure}

\begin{figure}[t!]
\includegraphics[trim=12mm 0mm 0mm 40mm, clip, width=3.7in]{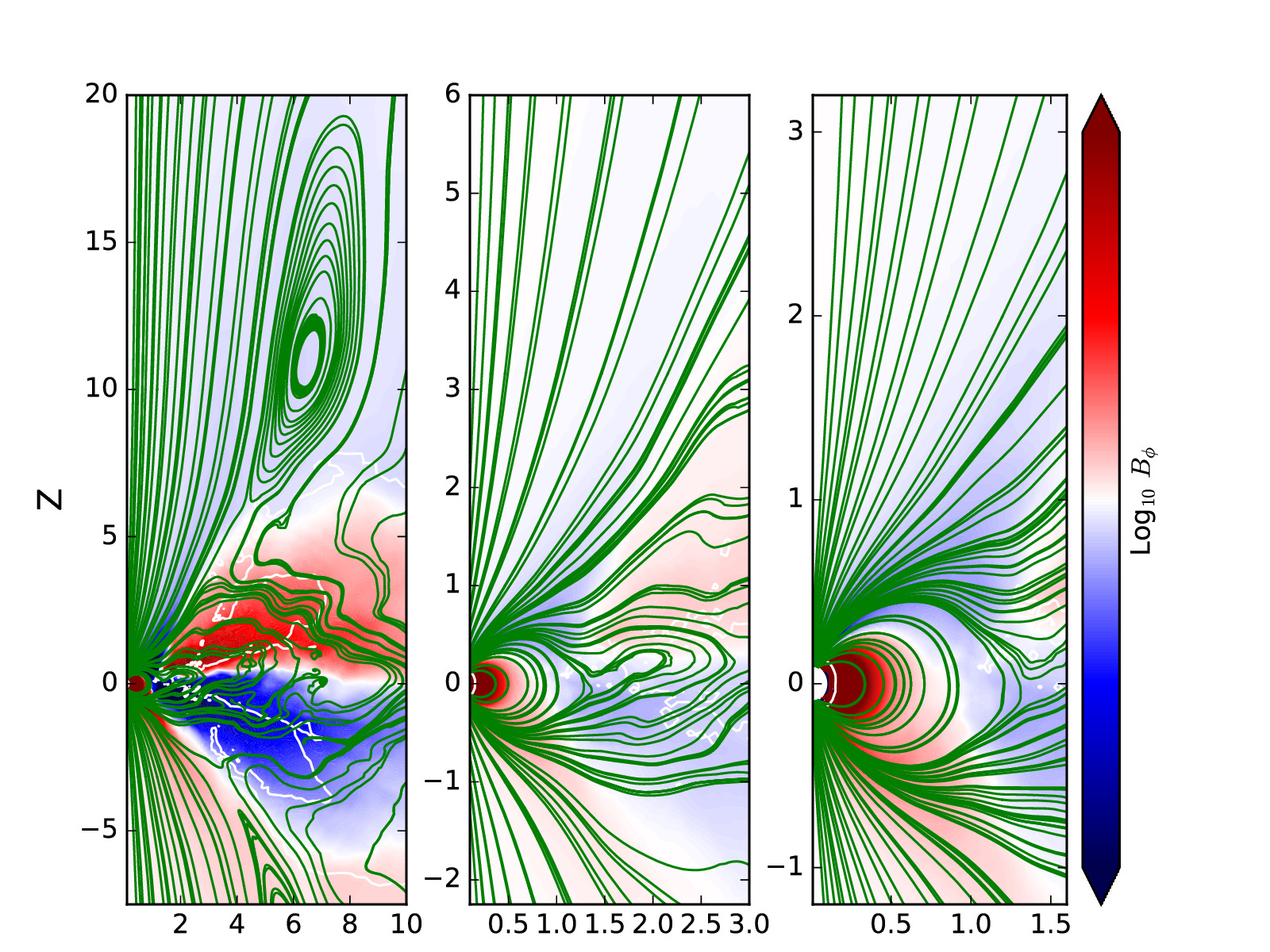}
\figcaption{Similar to Figure \ref{fig:twodrho}  but for $B_{\phi}$, and the green curves are the magnetic streamlines. The range of the colorbar is [-0.01,0.01], [-0.1,0.1], [-0.1,0.1] in the left, middle, and right panels. \label{fig:twoBrho}}
\end{figure}

The most surprising feature in our simulation is the vertically extended surface accretion region at $R\gtrsim R_T$, as shown in the middle panel of Figure \ref{fig:twodrho}. This region is  magnetically supported (Figure \ref{fig:twoBrho}), and is remarkably similar to the surface accretion region in the MRI turbulent discs with net vertical magnetic fields (e.g., \citealt{ ZhuStone2018, Mishra2020M, Jacquemin2021}).  The region extends to
$z\sim r$ and the flow in the region moves inwards supersonically. It is supported by magnetic pressure and located beyond the $\langle \beta\rangle=1$ surface.  The strong magnetic fields 
are generated by the azimuthal stretching of the radial magnetic fields from the Keplerian shear. The resulting large $B_r B_{\phi}$ stress from net magnetic fields drives the surface accretion, while, at the disc midplane where $\langle\beta\rangle\gtrsim 1$, the disc is turbulent due to MRI. { The magnetically dominated surface is similar to that in the magnetic elevated disc model \citep{Begelman2023} and the failed disc wind in accretion discs around weakly magnetized stars \citep{Takasao2018}.}

\begin{figure*}[t!]
\includegraphics[trim=0mm 0mm 0mm 5mm, clip, width=7.in]{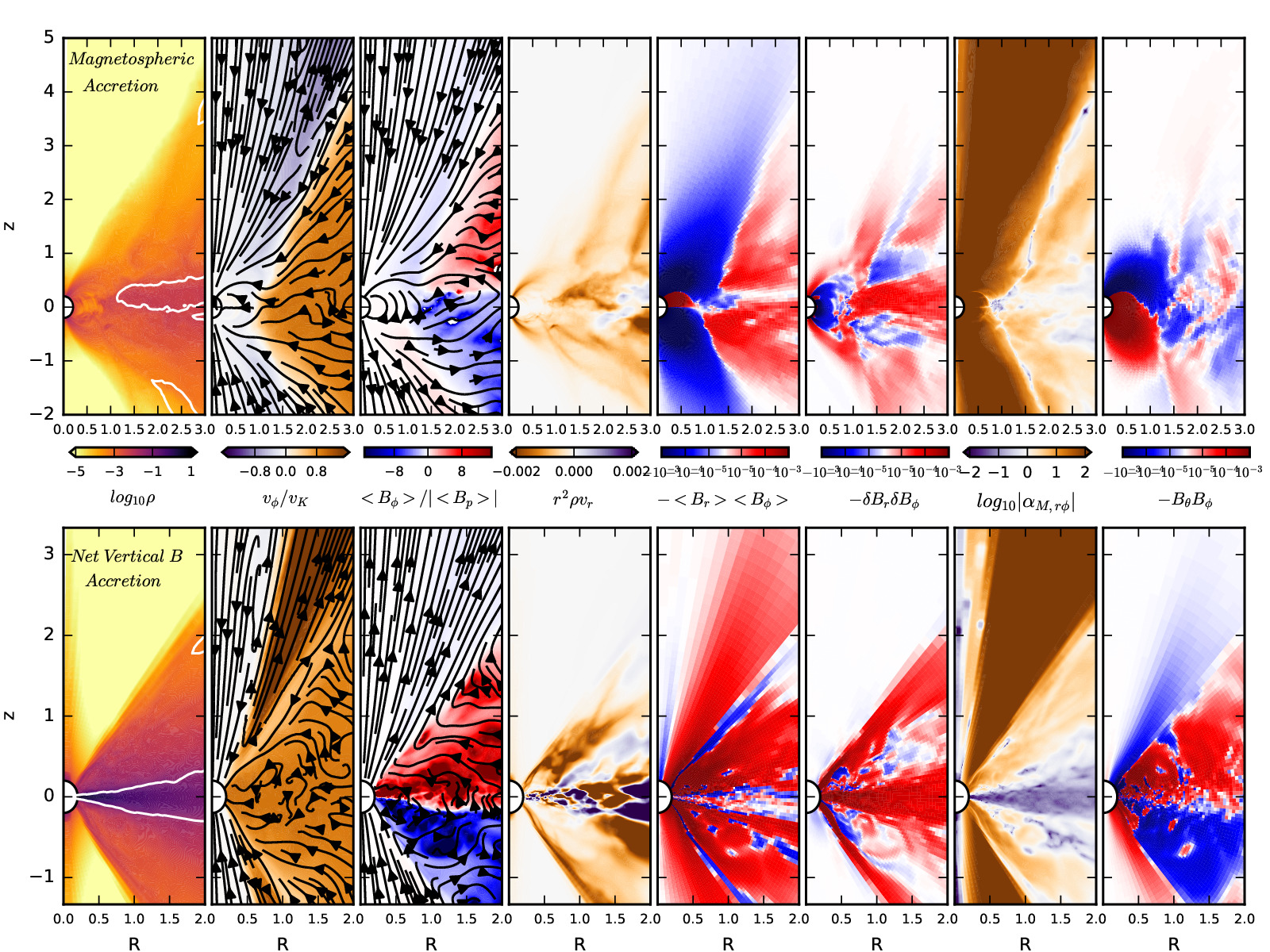}
\figcaption{Disc structure from this work (upper panels), compared to the structure of a disc threaded by net vertical magnetic fields (lower panels, from \citealt{ZhuStone2018}). Both simulations are shown at the final output, and all plotted quantities are azimuthally averaged. The white contours in the leftmost panels (the density panel) represent where $\langle \beta\rangle=1$.   The streamlines in the $\langle v_{\phi}\rangle$ panels show the poloidal velocity, while the streamlines in the $\langle B_{\phi}\rangle$ panels show the poloidal magnetic fields. 
\label{fig:compare}}
\end{figure*}

To see the similarities and differences between accretion discs threaded by dipole versus net vertical magnetic fields, Figure \ref{fig:compare} compares this magnetospheric accretion simulation against the simulation in 
\cite{ZhuStone2018} where the disc is threaded by net vertical fields with an initial $\beta_0=10^3$. The magnetically-dominated accreting surface is evident in both simulations. Significant
radial inflow at the disc surface can be seen in the $\langle r^2\rho v_{r}\rangle$ panels. It is driven by the high $r$-$\phi$ stress (the $\langle B_r\rangle\langle B_{\phi}\rangle$ panel) that is produced by stretching the radial fields azimuthally. The  radial fields in magnetospheric accretion come from the stellar dipole fields after reconnection events (the $\langle B_{\phi}\rangle$ panel). On the other hand, the radial fields in the net vertical field simulations come from the surface accretion itself, which drags the fields at the surface inwards tilting the vertical fields into the horizontal direction. The magnetic fields also connect the midplane with the disc atmosphere vertically in both cases. The $B_\theta B_\phi$ stress  acts like magnetic braking, which removes angular momentum from the surface to the midplane (the $\langle B_{\theta} B_{\phi}\rangle$ panels). Thus, the  $B_\theta B_\phi$ stress increases surface accretion further while it slows down the midplane's radial inflow (or even makes it move outwards). On the other hand, the $B_\theta B_\phi$ stress within the disc can only redistribute angular momentum vertically within the disc, and it cannot lead to the overall disc accretion if we integrate the disc accretion rate vertically throughout the disc. The overall accretion is led by the $r$-$\phi$ stress integrated vertically within the disc and the $\theta$-$\phi$ stress at the disc surface (Equation \ref{eq:mdot}). The simulations in \cite{ZhuStone2018} show that the $r$-$\phi$ stresses play a more important role than the $\theta$-$\phi$ stresses at the disc surface, which is also the case for magnetospheric accretion presented here (more details in Appendix \ref{sec:diskaccretion}).

{ \cite{Jacquemin2021} discover that, for discs threaded by net vertical fields, the surface accretion region consists of two parts: the laminar region at lower $z$ where the net fields ($\langle B_r\rangle\langle B_{\phi}\rangle$) dominate the angular momentum transport and the turbulent region at higher $z$ where the turbulent fields ($\delta B_r \delta B_{\phi}$) dominate the transport. This is also shown in our Figure \ref{fig:compare}. They identify that the turbulent region at $z\sim R$ is due to MRI when the net azimuthal fields become weaker than the net vertical fields. For our magnetospheric accretion simulation, we also detect the turbulent surface accretion region above the laminar region. Turbulent stress is also observed in the vicinity of the magnetosphere, which could be attributed to the interchange instability. However, it's worth noting that the turbulent stress is significantly weaker than the laminar stress in that region.}

The most noticeable differences between these two models are mainly at the magnetosphere and the lowest density region at $z>r$ . In the upper panels of Figure \ref{fig:compare}, the stellar dipole fields in the low-density region are pointing toward the star. Since the star is non-rotating, the stellar field lines can magnetically break the low-density region, leading to accretion onto the star instead of launching outflows. As discussed in Section \ref{sec:magstr}, for an axisymmetric steady flow around a non-rotating star, the gas velocity and the magnetic field lines are along the same direction ($B_\phi/B_r\sim v_\phi/v_r$). High above the disc at $z>0$,  $B_\phi$, $B_r$, $v_{\phi}$, and $v_{r}$ are all negative ($\langle v_\phi\rangle$ and $\langle B_\phi\rangle$ panels). Thus, the flow falling onto the star is in the opposite azimuthal direction from the direction of the Keplerian rotation, similar to the traditional picture in Section \ref{sec:magstr}. In contrast, for discs threaded by net vertical magnetic fields shown in the lower panels, the net vertical field lines at the disc surface are tilted away from the star. The material in the disc surface connects to the higher and outer low density region magnetically. Since the disc rotates faster than the higher and outer low density region, the magneto-centrifugal wind is launched. 

\begin{figure*}[t!]
\includegraphics[trim=0mm 30mm 0mm 40mm, clip, width=6.in]{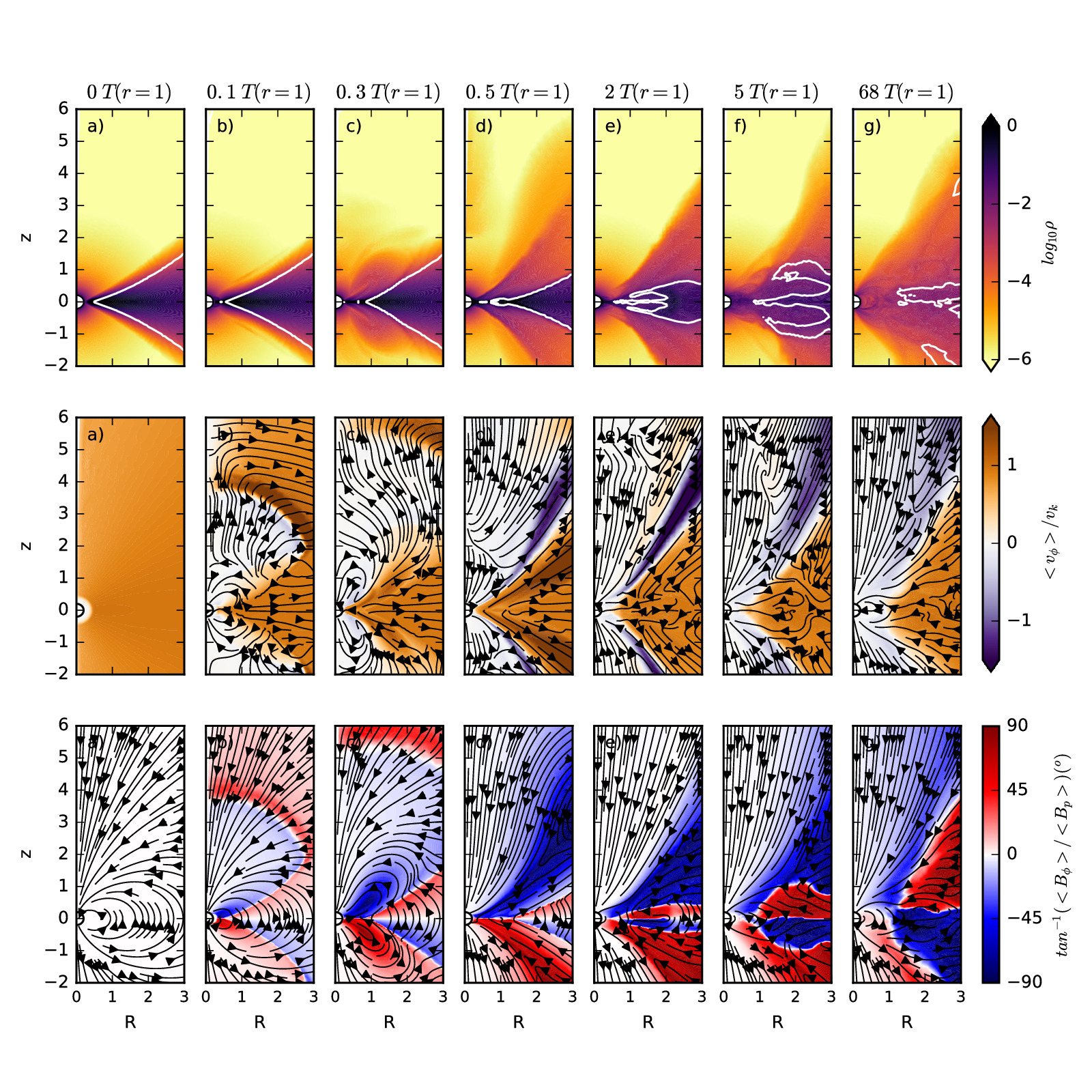}
\includegraphics[trim=-10mm 82mm 0mm 60mm, clip, width=5.5in]{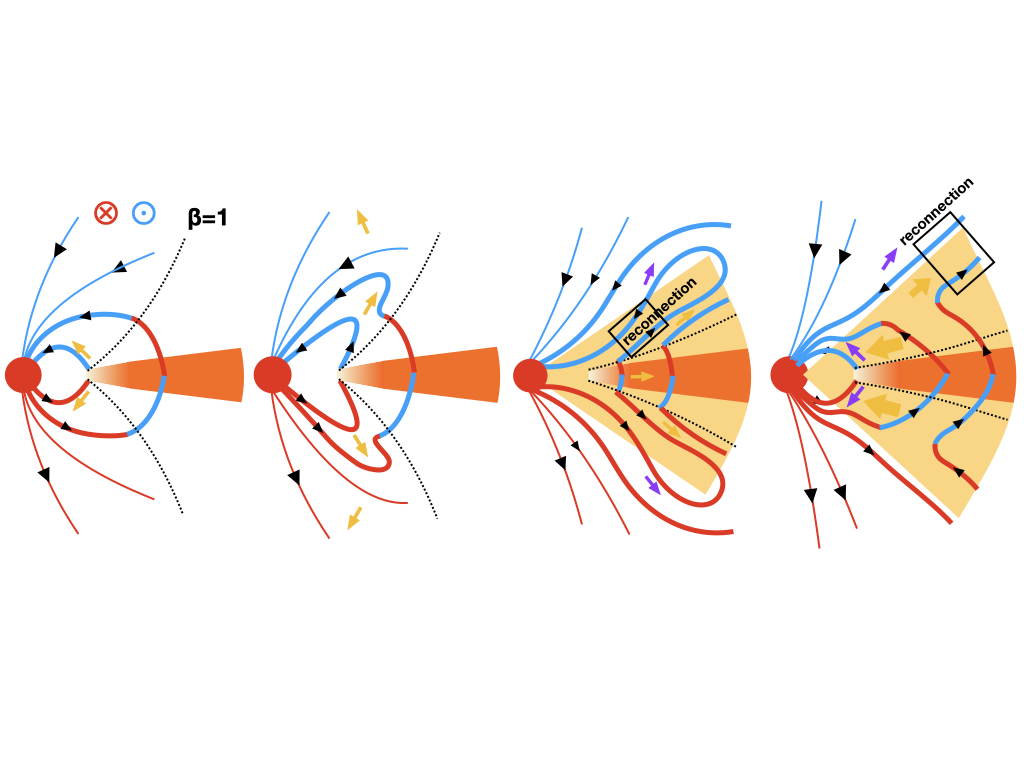}
\figcaption{Top three rows: the disc density, velocity, and magnetic structure at different times during the simulation (left to right panels). The white contours in the top panels label where  $\overline\beta\equiv2\langle P\rangle/(\langle B_{r}\rangle^2+\langle B_{\theta}\rangle^2+\langle B_{\phi}\rangle^2)=1$. The streamlines in the panels of the second row represent the poloidal velocity, while the streamlines in the panels of the third row represent the poloidal magnetic field. The bottom row: the schematic diagrams showing how magnetic fields and flow structure change with time. The red and blue curves are magnetic field lines with positive and negative $B_{\phi}$ components. The arrows show the flow in the poloidal direction, and their colors (yellow or purple) represent positive and negative $v_{\phi}$ in the region. 
\label{fig:evolve}}
\end{figure*}

Such quasi-steady structure is not established instantaneously, it is important to understand how the disc evolves to such a state from the initial condition. Such evolution may have implications for outbursting discs.  Figure \ref{fig:evolve} shows the density, velocity, and magnetic structure at different times. In the initial condition, we set up a disc that is in hydrostatic equilibrium with the stellar gravity threaded by a dipole magnetic field. Due to the azimuthal shear, the poloidal fields are quickly stretched to produce toroidal components. The second panel in the third row shows that, inside the matter dominated disc region ($\overline\beta\geq 1$), the faster rotation at the inner disc drags the magnetic field with negative $B_{R}$ at $z>0$ to develop a positive $B_{\phi}$ component. On the other hand, in the magnetically dominated region ($\overline\beta< 1$) close to the star, the flow's rotation  is slowed down by magnetic fields from the non-rotating star (the second panel in the second row). Since the Keplerian rotating disc with $\overline\beta\geq 1$ is outside the magnetosphere with $\overline\beta<1$,  magnetic fields with negative $B_{R}$ in the magnetosphere are stretched to develop the negative $B_{\phi}$ component. The strong shear quickly amplifies $B_{\phi}$, especially at the $\overline\beta\sim1$ region. 

The increasing magnetic pressure starts to push material outwards (the third and fourth panels in the second row), forming an outwardly moving magnetic bubble that stretches the magnetic fields in the radial direction (the third row). Such strong magnetic fields in the atmosphere push the $\overline\beta\sim1$ curve further into the disc region (the fourth panel in the first row). After the magnetic bubble moves outwards,  the field lines open up, similar to the ``X-wind model'' \citep{Shu1994}. Later, magnetic field lines at the base reconnect so that once again they connect the disc region to the star (the fifth and sixth panels in the third row). This is very similar to the unsteady field inflation model in \cite{LyndenBell1994, Lovelace1995, Uzdensky2002}. { However, unlike these previous studies which are built upon $R$-$z$ 2-D models,  after this initial relaxation stage, our disc generates magnetically supported and turbulent surface regions, which expand with time and allow the disc to accrete quasi-steadily. The difference on the steadiness of accretion is due to the operation of various 3-D instabilities in our simulation, including the magnetic interchange instability around $R_T$ and the MRI in the outer disc. These instabilities allow field lines to diffuse across disc material, acting like a large anomalous resistivity. Thus, open fields lines return to the dipole configuration due to the anomalous resistivity. 
With a large resistivity, the slippage of field lines  balances the azimuthal shear, allowing a quasi-steady accretion \citep{Ghosh1979II}. }
 The $B_{\phi}$ in the disc atmosphere increases steadily due to the Keplerian shear until the field growth is balanced by turbulent diffusion. This whole process is self-similar at different radii, and eventually the poloidal field structure looks similar to the initial dipole structure except with a very strong azimuthal component. 
Even at the end of the simulation, the largest scale (e.g. the leftmost panels in Figures \ref{fig:twodrho} and \ref{fig:twoBrho}) is still far from steady state. Instead, the flow and magnetic structure there look like the structure at $R\sim$1 when $t=5 T_0$ (Figure \ref{fig:evolve}), demonstrating that the flow and magnetic structure expand self-similarly with time. 

\section{Discussion}

\subsection{Comparison with Observations: Filling Factor and Variability}

\begin{figure*}[t!]
\includegraphics[trim=0mm 55mm 35mm 75mm, clip, width=5.8in]{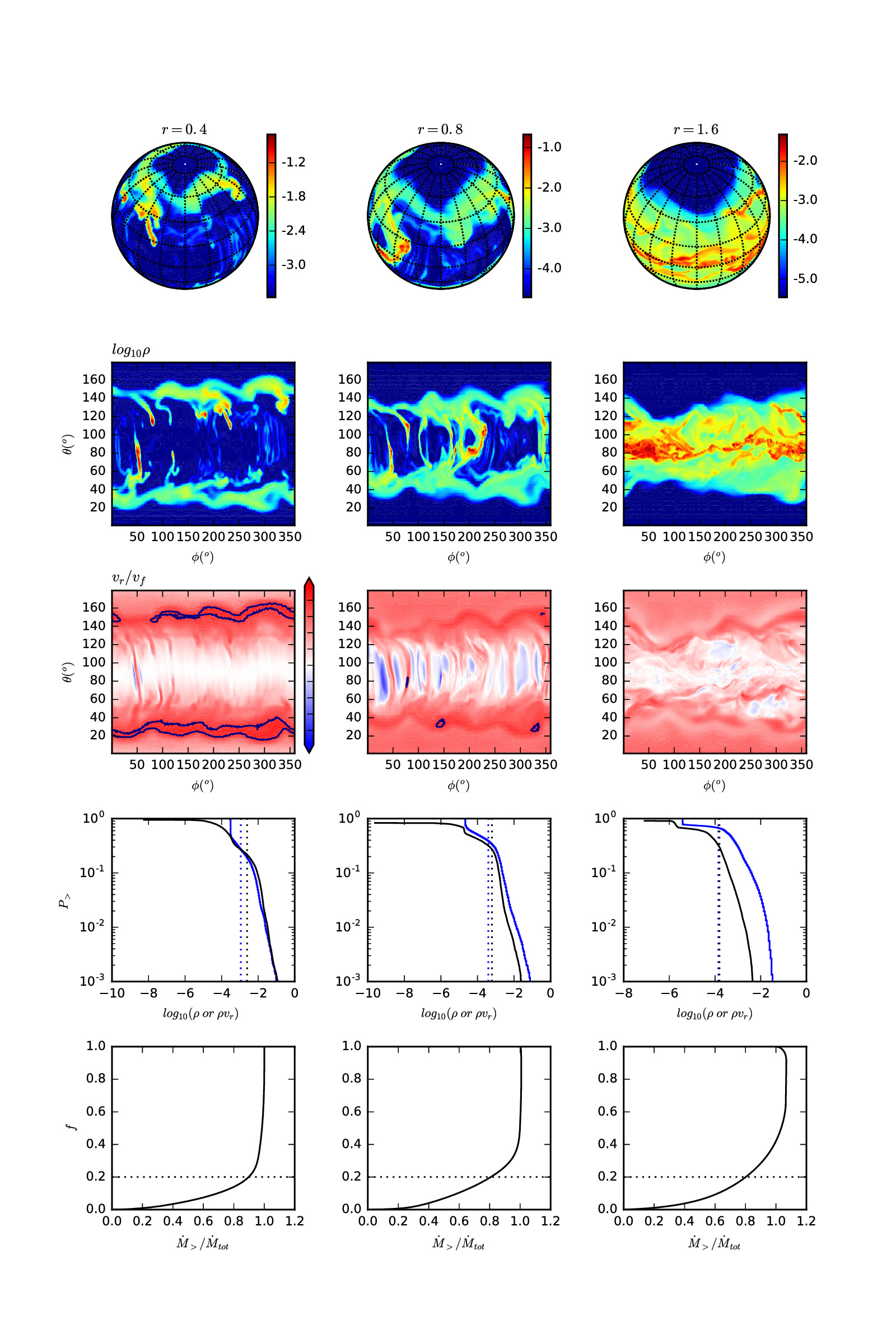}
    \figcaption{Various quantities at $r=0.4$, 0.8, and 1.6 (left, middle, and right columns). The upper two rows show $log_{10}\rho$ mapped to a sphere (the top row) and in the $\phi$-$\theta$ plane (the second row).  The third row shows the radial velocity normalized to the free-fall velocity at those three different radii. The colorbar extends from -1 to 1, and the black contour represents the value of 0.8.   The fourth row shows the fraction of the area on the sphere with density ($\rho$, blue curves) or radial mass flux ($\rho v_r$, black curves) larger than the given value at x-axis. The vertical black and blue lines label where the radial mass flux and density should be for spherically symmetric accretion with the measured accretion rate (0.005) and the free-fall velocity. The bottom panels show the fraction of the sphere ($f$) where the integrated accretion rate is $\dot{M}_>$. To derive $\dot{M}_>$, we integrate the mass flux from the highest flux region to the lowest flux region. 
\label{fig:sphere}}
\end{figure*}

Observable accretion signatures of classical T-Tauri stars are produced within the magnetosphere \citep{Hartmann2016}. The accretion shock at the surface of the star produces the excess emission at ultraviolet, which is the most robust tracer for estimating the disc's accretion rate. Atomic lines are produced over a large volume (probably covering the whole magnetosphere) and the line shapes can be used to constrain the flow structure within the magnetosphere. Detailed radiative transfer modeling reveals that: 1) the maximum infall velocity onto the star is roughly consistent with free-fall velocity \citep{Hartmann1994,Muzerolle1998,Muzerolle2001,Kurosawa2011}; 2) the infall is at moderate latitudes from the disc plane but not at the poles \citep{Bonnell1998,Muzerolle1998}; 3) the covering factor of the accretion columns at the stellar surface (called ``filling factor'') is 0.001 to 0.1 \citep{CalvetGullbring1998};  4) the outflow rate (with 100 km/s velocity) is correlated with disc accretion rate \citep{Hartigan1995, Rigliaco2013, Natta2014}.  

We can compare the flow structure in our simulated magnetosphere with above observed properties. Physical quantities on the $\theta$-$\phi$ plane at different radii are shown in Figure \ref{fig:sphere}. 
The top two rows clearly demonstrate that the material is lifted from the disc (the $r=1.6$ panel) to higher altitudes within the magnetosphere ($r$=0.4 and 0.8 panels).  The $r$=0.4 panel suggests that most material will eventually fall onto the star at $\theta\sim$ 30$^o$ and 150$^o$. { We caution that the position of the hot spot also depends on the tilt of the dipole fields and/or the multipole field components \citep{Long2008}. Our discussion here is based on our simulation with the aligned dipole fields.}

Figure \ref{fig:sphere} also shows that the filamentary features stretch from north to south, and fewer filaments penetrate into $r=0.4$ at the midplane compared with filaments at $r=0.8$. The infalling material also accelerates as it falls, reaching free fall speed at $r\sim$0.4. As discussed in Section \ref{sec:instability}, Figure \ref{fig:mid1dfluc} shows that the infall speed at moderate latitudes is almost the free-fall speed. We notice that there is more than one accretion hot spot and column, also shown in Figure \ref{fig:setup}. There could be many layers of magnetosphere with an onion-like structure. Recent observations by \cite{Thanathibodee2019b} find that some systems do have multiple geometrically isolated accretion flows, which seems to be consistent with our simulation.

Considering that accretion is concentrated at high altitudes and within several accretion columns, we calculate the fraction of the sphere where most accretion occurs (the ``filling'' factor). The fourth row in Figure \ref{fig:sphere} shows the fraction of the area with density ($\rho$) or radial mass flux ($\rho v_r$) higher than a given value. The vertical dotted lines label the density and radial mass flux for spherically symmetric accretion (labeled as $\rho_{sph}$ and $\rho_{sph}v_{ff}$) that is calculated using the measured accretion rate (0.005) and the free-fall velocity. At all three radii, roughly 30\% of the area has a mass flux higher than $\rho_{sph}v_{ff}$. On the other hand, the density of the accretion columns in the outer disc is significantly higher than the $\rho_{sph}$, since the radial velocity there is significantly lower than the free-fall velocity.
At $r=1.6$, 70\% of the area has a density higher than $\rho_{sph}$.  

One important parameter in the magnetospheric accretion model is the filling factor, $f$ \citep{Hartmann2016} defined as the surface covering fraction of the accretion columns. With a small $f$, most accretion occurs within a small patch on the stellar surface. Since all accretion energy is released in such a small region, a small $f$ produces a high temperature hot spot. For young stars, this produces UV excess emission  over the photospheric SED, a distinct feature indicating magnetospheric accretion. On the other hand, $f$ is less well defined in our simulations, since accretion occurs across a wide range of densities and mass fluxes over the sphere, rather than in discrete patches. Thus, we define a filling factor function, $f(\dot{M})$, as the surface covering fraction for regions where the integrated mass flux is $\dot{M}$. We integrate the mass flux from the patch with the highest mass flux to the patch with the lowest mass flux. More specifically, to calculate this function, we first calculate the distribution function for $M_r\equiv \rho v_{r}$ as the probability of finding accreting columns within a certain range of $M_r$ values across all 4$\pi$ direction at $r$,
\begin{equation}
P(M_{r,1}<M_r<M_{r,2})=\int_{M_{r,1}}^{M_{r,2}} P(M_r) dM_r\,,
\end{equation}
where $P$ is calculated by dividing the solid angle corresponding to the selected $M_r$ range by the total solid angle of 4$\pi$ steradians.
Essentially, we rearrange all patches on the stellar surface according to its mass flux.
Then, we integrate $P$ from the patch with the highest mass flux $M_{r,max}$ all the way down to $M_{r,\dot{M}}$ to derive $f(\dot{M}_>)$:
\begin{equation}
f(\dot{M}_>)=-\int_{M_{r,max}}^{M_{r,\dot{M}}}P(M_r )d M_r\,,
\end{equation}
where $\dot{M}_>$ is the integrated flux for top accreting patches
\begin{equation}
\dot{M}_>=\int_{M_{r,max}}^{M_{r,\dot{M}}}\rho v_r P(M_r)4\pi r^2 dM_r\,.
\end{equation}
The filling factor $f(\dot{M}_>)$ is shown in the bottom panels of Figure \ref{fig:sphere}. From the right to left panels, the filling factor is smaller at the inner radius. At $r=0.8$,  90\% of the accretion occurs within 30\% of the area.
At $r=0.4$, 90\% of accretion occurs within 20\% of the area, and 50\% of accretion
is within 5\% of the area.Thus, we estimate that the filling factor is $\sim$5\%-20\%.   { We caution that, at smaller radii, the filling factor could be even smaller.}

\begin{figure}[t!]
\includegraphics[trim=15mm 20mm 15mm 5mm, clip, width=3.5in]{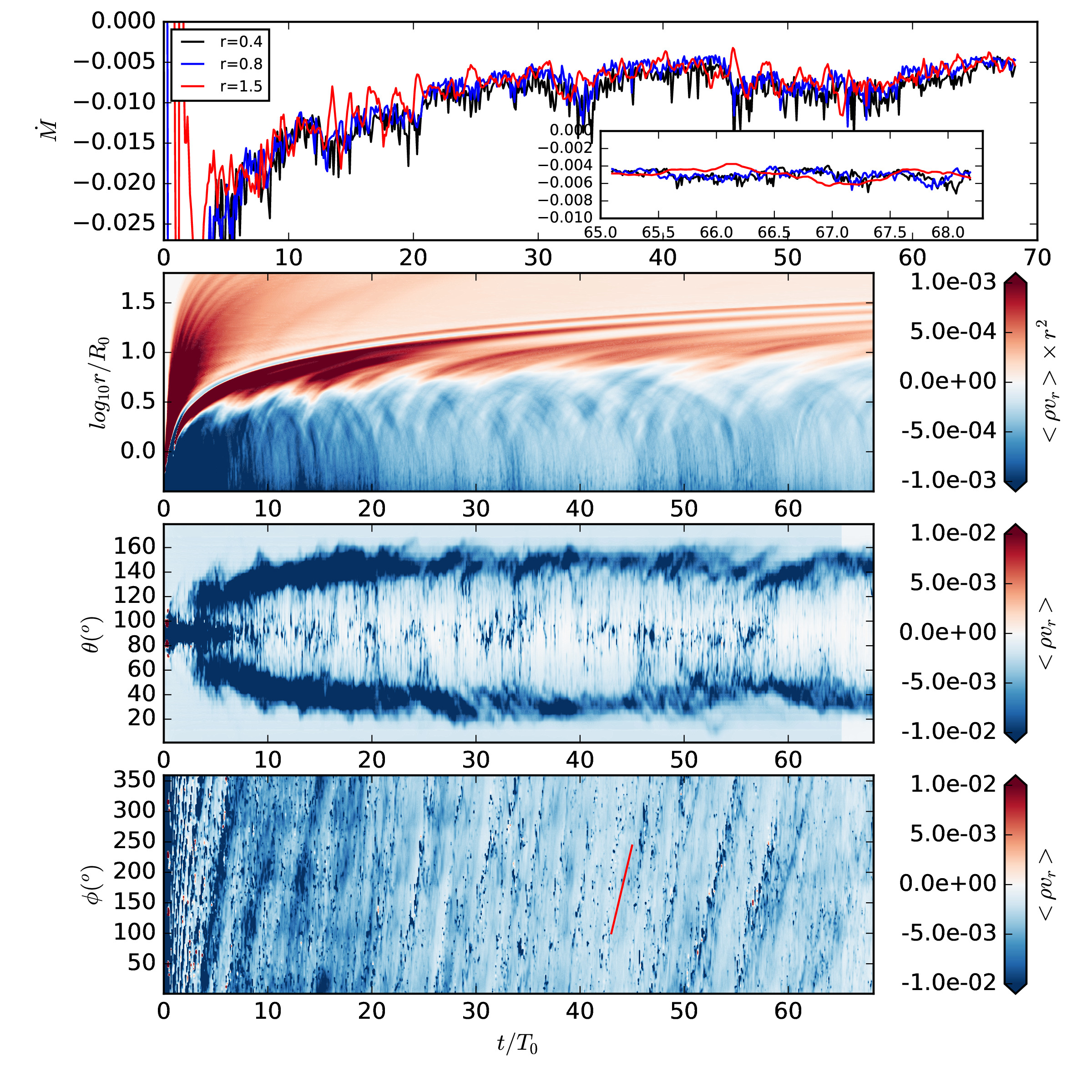}
\figcaption{Top panel: evolution of the mass accretion rate at $r=0.4$, $0.8$, and $1.5$ with time. The insert zooms into the last 3 orbits. The bottom three panels are
space time diagram for $\rho v_r$ along $r$, $\theta$, and $\phi$ directions. The $\dot{M}$ and $\rho v_r$ in the top two panels are integrated and
averaged over the sphere respectively. The $\rho v_r$ in the $t$-$\theta$ and $t$-$\phi$ panels are averaged along the $\phi$ and $\theta$ directions at $r=0.4$.  The red line in the bottom panel shows the azimuthal movement for a hot spot if its orbital frequency is  20\% of the Keplerian frequency at the
magnetospheric truncation radius.   \label{fig:spacetime}} 
\end{figure}

Since most accretion is concentrated in a few accretion columns and these columns appear and disappear dynamically due to the interchange instability, it is natural to ask if such an accretion is steady. We integrate
the total accretion rate over the sphere at $r=0.4$, 0.8, and 1.5 respectively, and plot the accretion rates with respect to time in the top panel of Figure \ref{fig:spacetime}.  The accretion rates at all three radii are almost the same, indicating a constant accretion rate from the disc to the star. There is little time lag among the accretion rates at all three radii, even when the accretion rate changes by a factor of 2 over $\sim$10 orbits at the beginning of the simulation. This simultaneous change of the accretion rates at all three radii is due to the fast radial inflow from the disc surface accretion and magnetospheric accretion.  

\begin{figure}[t!]
\includegraphics[width=3.5in]{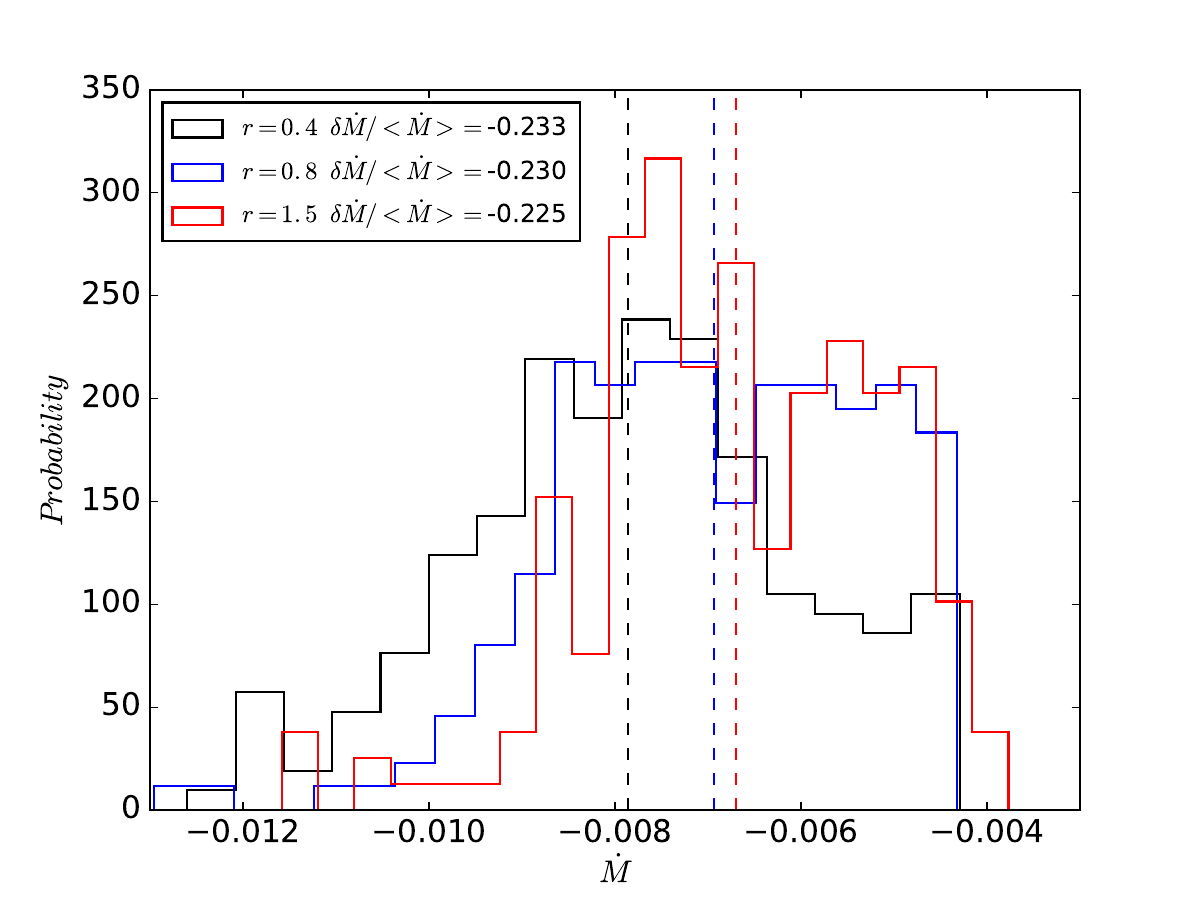}
\figcaption{ The probability distribution function of the accretion rates at r=0.4, 0.8, and 1.5 (black, blue, and red curves). They are derived using 341 snapshots uniformly spanned from
half of the total time ($t$=34.1) to the end of the simulation. The integration of the probability over all the $\dot{M}$ is 1.  The vertical lines are the averaged accretion rates. The standard deviation of the accretion rates is also given in the upper left corner.   \label{fig:mdotp}}
\end{figure}

After 30 orbits, the accretion rates become almost constant.   From the insert of Figure \ref{fig:spacetime}, it is evident that the accretion rate in the disc (e.g. $r=1.5$) is smoother than 
those in the magnetosphere (e.g. $r=0.4$ and $r=0.8$). These short time-scale fluctuations within the magnetosphere are probably due to the filaments produced by the interchange instability. On the other hand, the amplitudes of the $\dot{M}$ variability are similar at these three radii, as shown in Figure \ref{fig:mdotp}. The averaged rates are within 15\% of each other and the ratio between the standard deviation of the accretion rates and the mean accretion rates is also close to each other ($\delta\dot{M}/\langle\dot{M}\rangle\sim 23\%$). This relatively steady accretion is consistent with TW Hya's steady accretion over the past 20 years \citep{Herczeg2023}.

The $\dot{M}$ distribution along the $r$, $\theta$, and $\phi$ directions are shown in the spacetime plots in Figure \ref{fig:spacetime}. In the radial direction, there is a region of steady accretion which grows with time, beyond which there is a transition region that also moves outwards (with positive $\rho v_{r}$). In the $\theta$ direction, most accretion onto the star occurs around 30 and 150 degrees. The $\phi$ panel shows that the accretion is not axisymmetric. This is expected due to the filaments within the magnetosphere. However, most interestingly, the accretion hot spot rotates around the central star at 20\% of the Keplerian frequency at the magnetospheric truncation radius. 

\begin{figure}[t!]
\includegraphics[width=3.5in]{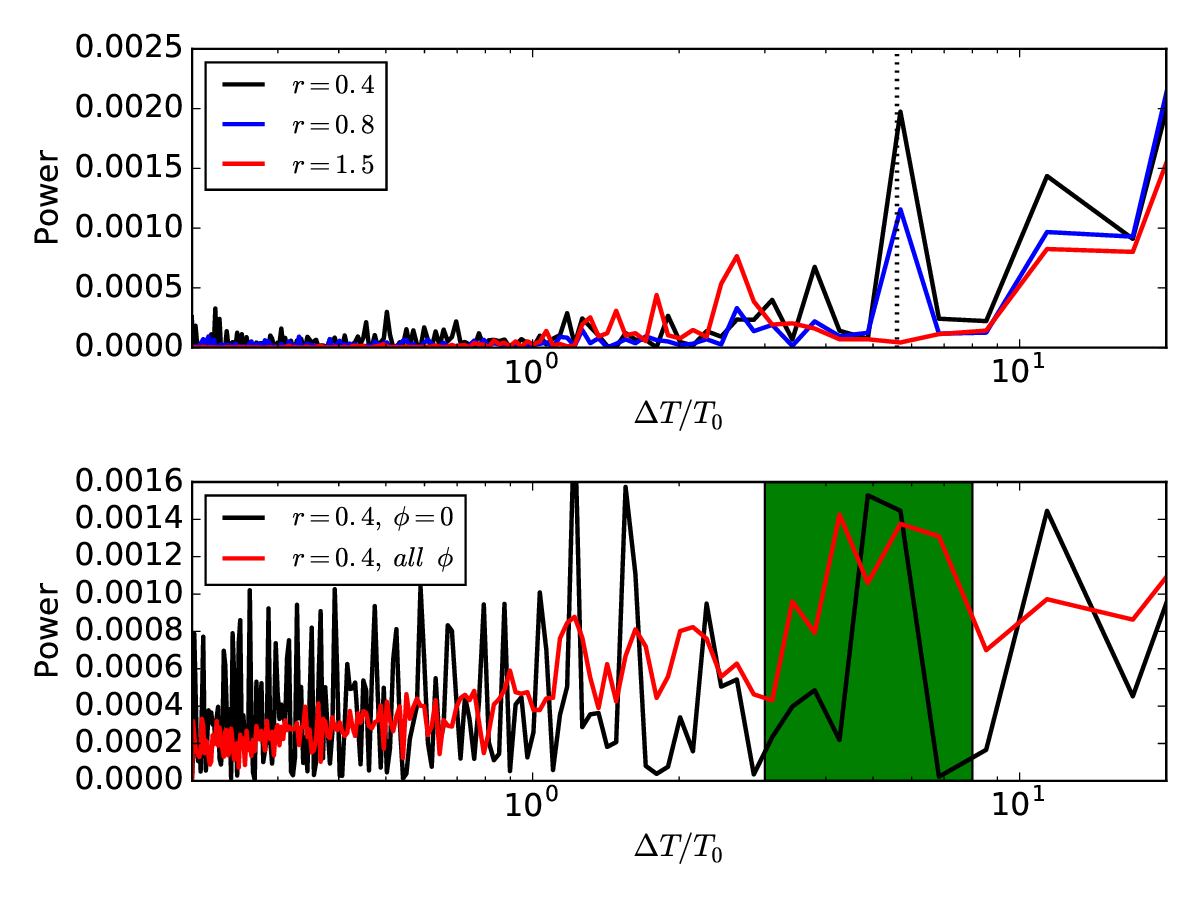}
\figcaption{Top panel: the periodogram for the integrated mass accretion rate at $r=0.4$, $0.8$, and $1.5$.
Bottom panel: the periodogram for the accretion rate at $r=0.4$ and $\phi=0$ (black curve) and the averaged curve
of all the periodograms for the accretion rate at $r=0.4$ in 256 $\phi$ directions (red curve). The mass accretion rate data are from the top and bottom panels of Figure \ref{fig:spacetime}, and we use 341 snapshots from  34.1 to 68.2 $T_0$.The red  curve in the bottom panel has been stretched vertically by a factor of 1.5 to highlight its features. \label{fig:period}}
\end{figure}

This 5$T_0$ periodicity is also shown in the periodogram for the accretion rate at one particular $\phi$ angle. The bottom panel of Figure \ref{fig:period} shows that the periodogram for $\phi=0$ has a peak around 5$T_0$. On the other hand, there are many other peaks due to the statistical noise. Thus, we averaged all the periodograms in different $\phi$ directions to lower the noise, and the averaged curve is the red curve. We still see a bump around  5$T_0$, confirming the trend in the space-time diagram at the bottom panel of Figure \ref{fig:spacetime}. This 5$T_0$ modulation is linked to the orbital motion of the magnetic bubble in Section \ref{sec:asym}. The middle panels in Figure \ref{fig:outflow} show that the magnetic bubble develops around $R_T$ but it extends all the way to the central star. A dense filament can be seen at the edge of the bubble, connecting to the star. The filament is most apparent in the middle panels from 54 to 61 $T_0$ and this accreting filament can also be seen in the bottom panel of Figure \ref{fig:spacetime} during the same period of time (the bubble and filament actually start around 50 $T_0$ shown in the movie after Figure \ref{fig:outflow}). In terms of astronomical observations, we anticipate that such accretion modulation would be observable in inclined systems when the accretion hot spot undergoes orbital motion around the star at this particular frequency.

We have also calculated the periodogram for the integrated accretion rate over the sphere at $r=0.4$, $0.8$, and $1.5$ (top panel of Figure \ref{fig:period}). At both $r=0.4$ and $0.8$, there is also a peak at $\sim$5.5 $T_0$ which could also be related to the modulation and duration of the magnetic bubble. On the other hand, the power in the periodogram increases with $\Delta T$, and there could be more peaks at larger $\Delta T$ which can only be studied with long timescale simulations.

\subsection{The Spin-up Torque}
\label{sec:torque}

\begin{figure}[t!]
\includegraphics[trim=15mm 30mm 15mm 5mm, clip, width=3.5in]{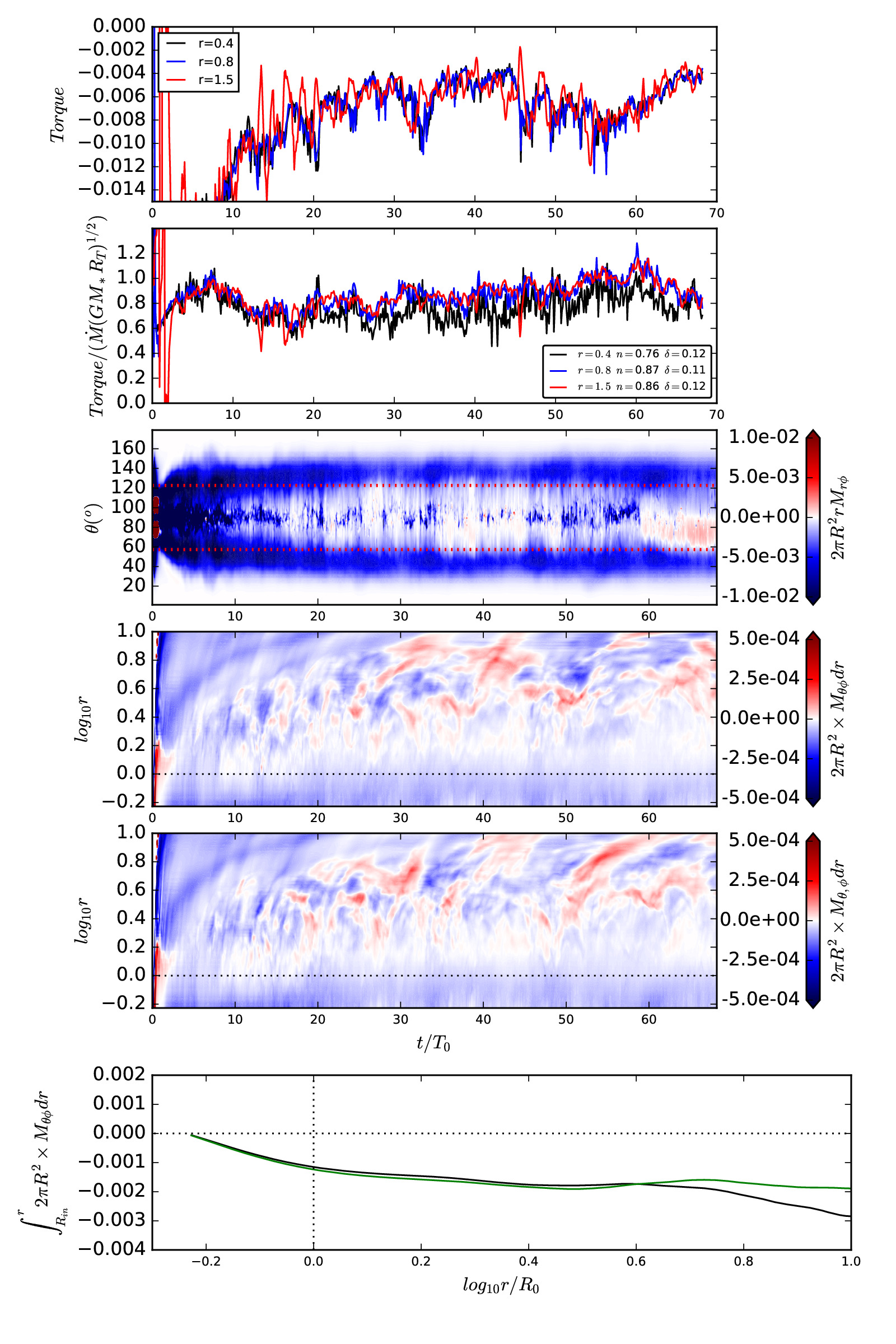}
\figcaption{Top two panels: total torque and the $n$ parameter at $r$=0.4, 0.8, and 1.5 with time. The average and the standard deviation of $n$ after 10 orbits are given in the legend of the second panel. Third panel: the torque ($\langle M_{r\phi}\rangle 2\pi r^3 {\rm sin}^2\theta$) at $r=0.6$ along the $\theta$ direction. Fourth and fifth panels: $\langle M_{r\phi}\rangle 2\pi r^2 {\rm sin}^2\theta\Delta r$ along the $r$ direction at $\theta$=1 and $\pi$-1 ($S_2$ and $S_3$ areas in Figure \ref{fig:torqueillu}).  Bottom panel: integrated torques along $S_2$ (black curve) and $S_3$ (green curve) averaged from $t=50$ $T_0$ to the end of the simulation.  \label{fig:torque}}
\end{figure} 

\begin{figure}[t!]
\includegraphics[trim=0mm 5mm 0mm 0mm, clip, width=3.5in]{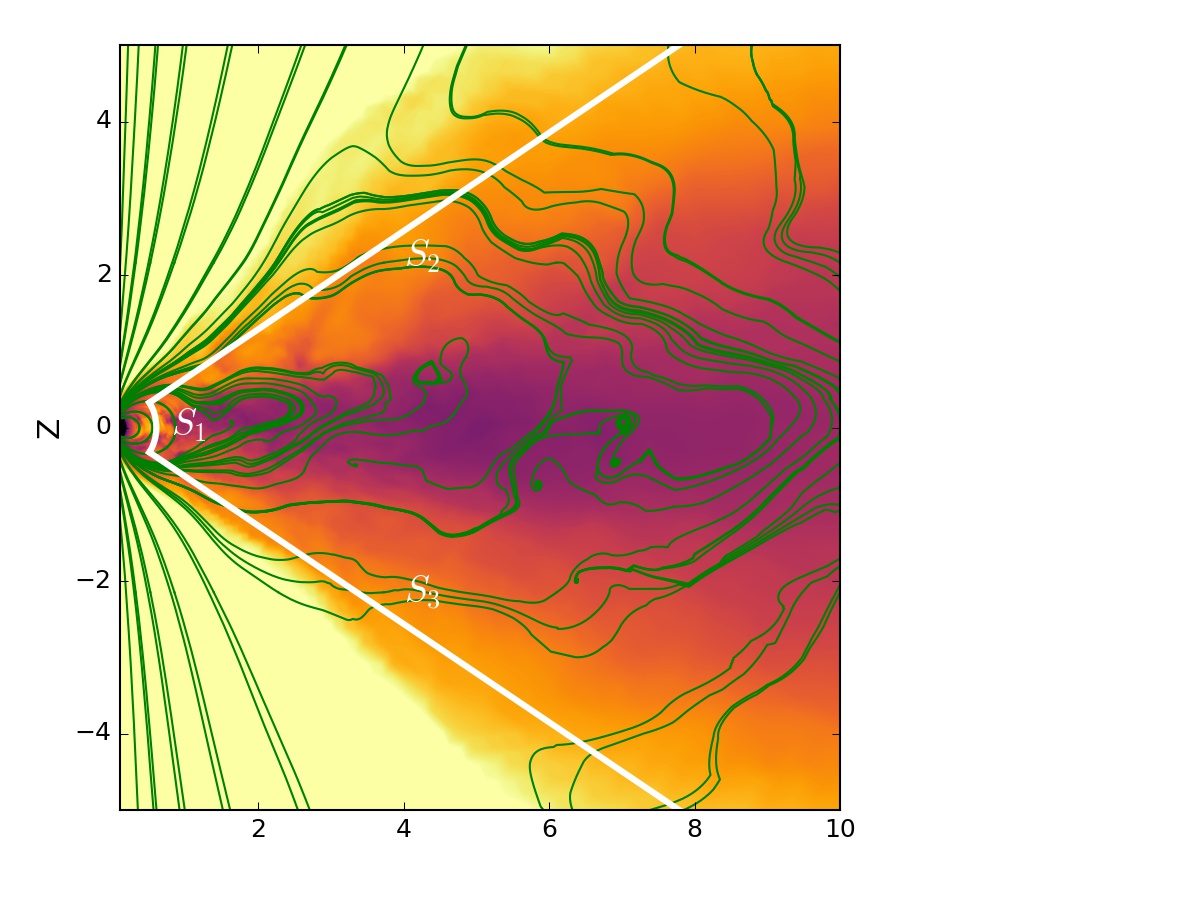}
\figcaption{The azimuthally averaged density and poloidal magnetic streamlines at the end of the simulation. The white lines label the surfaces where the torques are calculated in Figure \ref{fig:torque}.  \label{fig:torqueillu}}
\end{figure}

\begin{figure*}[t!]
\includegraphics[trim=20mm 10mm 30mm 10mm, clip, width=7in]{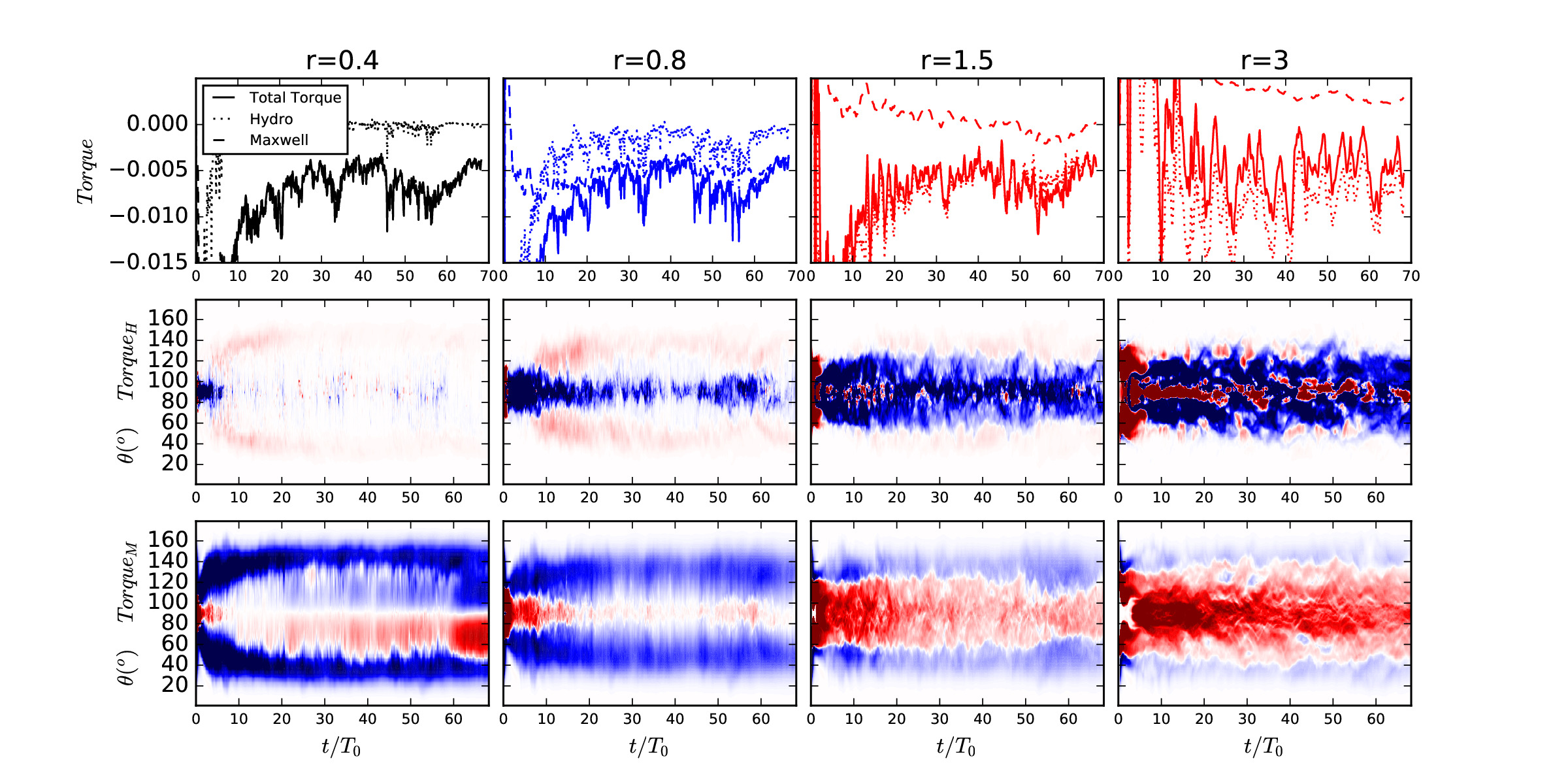}
\figcaption{Similar to Figure \ref{fig:torque} but for different stress components. Top panels: the integrated torque over spheres at different radii (left to right panels) with time. The middle and bottom panels: the hydrodynamical and magnetic components of the torque along the $\theta$ direction with time. The normalization and colorbar are the same as those in the third panel in Figure \ref{fig:torque}. \label{fig:torquesep}}
\end{figure*}

\begin{figure*}[t!]
\includegraphics[trim=20mm 15mm 30mm 10mm, clip, width=7in]{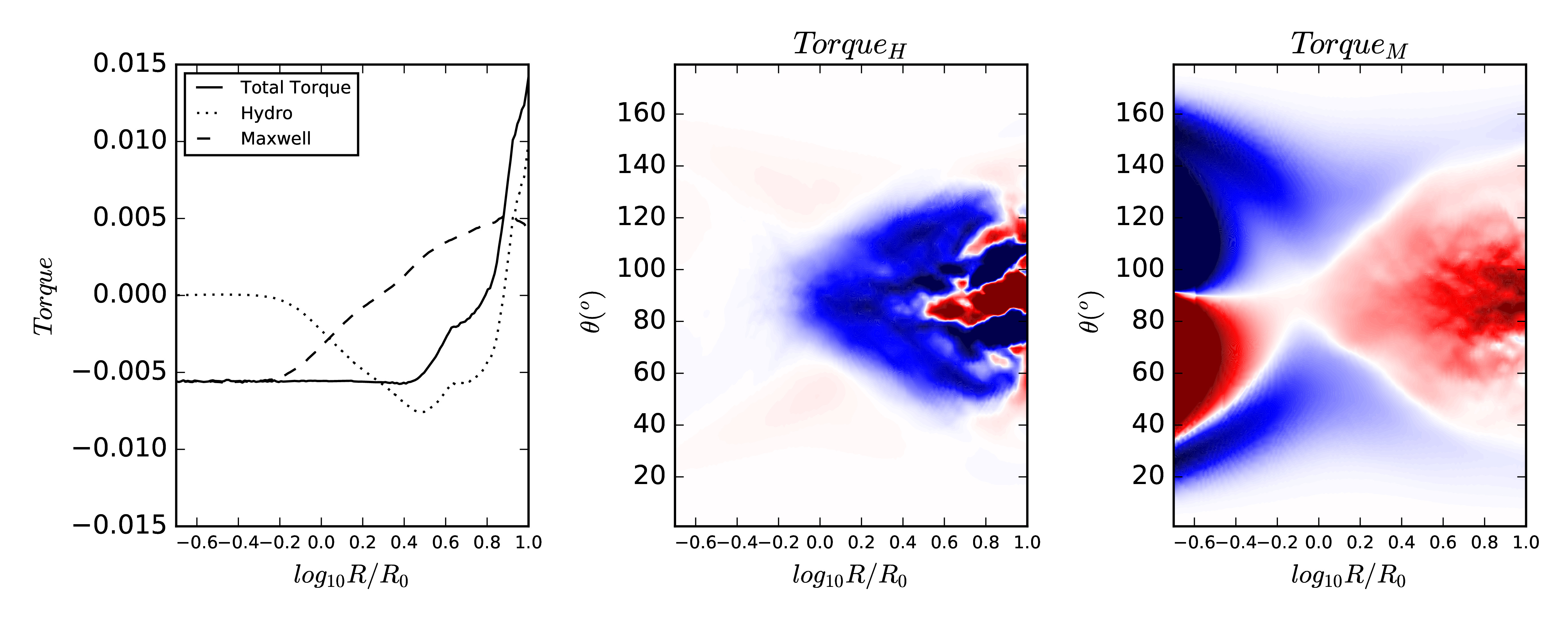}
\figcaption{The left panel: the integrated torque over spheres at different radii. The middle and right panels: the hydrodynamical and magnetic components of the torque along the $\theta$ direction with radii. The normalization and colorbar are the same as those in the third panel in Figure \ref{fig:torque}. The torques have been averaged using 100 snapshots over the last 10 orbits. \label{fig:torqueseptr}}
\end{figure*}

The star is spun up by its coupling with the disc through magnetic fields. The total angular momentum of the star and the disc is conserved. If we integrate Equation \ref{eq:ame} over the whole volume from $r=r_{in}$ to $r=r_{out}$, we can derive the region's angular momentum change 
\begin{align}
\frac{\partial \int_{r_{in}}^{r_{out}}R\rho v_{\phi} dv}{\partial t} = &\int_{r_{in}} R_{in}(\rho v_{r}v_{\phi}-B_r B_{\phi})dS \nonumber\\
-  &\int_{r_{out}} R_{out}(\rho v_{r}v_{\phi}-B_r B_{\phi})dS \,,
\end{align}
where the first and second terms on the right-hand side are the integrals over the sphere at $r_{in}$ and $r_{out}$. { When the integrated stress at $r$ is positive, the spherical region within it loses angular momentum. } If the disc region extends from $r_{in}$ to a very large $r_{out}$ where the density and velocity are close to zero,  the second integral on the right-hand side becomes zero. Since  angular momentum can be changed by torque, the first term on the right-hand side can be considered as the torque between the disc and the star. The $B_r B_{\phi}$ term is the magnetic stress/torque and the $\rho v_r v_{\phi}$ term is the hydrodynamical stress/torque that includes both the turbulent stress/torque ($\rho v_r \delta v_{\phi}$)  and angular momentum carried by the accreting material ($\rho v_r \langle v_{\phi}\rangle$). We integrate the total torque over the sphere at $r=0.4$, 0.8, and 1.5, and plot them with time in Figure \ref{fig:torque}. All three curves overlap with each other, suggesting that the disc has reached a constant angular momentum flow within $r=1.5$. If we divide the torque by $\dot{M}(GM_* R_T)^{1/2}$, we derive the $n$ parameter in Equation \ref{eq:Tstar}. The measured mean value of $n$ is $\sim$0.8 with the standard deviation of $\sim$0.1, shown in the second panel of Figure \ref{fig:torque}. The distribution of the torque along the $\theta$ direction at $r=0.6$ is shown in the third panel of Figure \ref{fig:torque}. We can see that most of the torque is exerted at $\theta\sim 45^o$ and $135^o$, corresponding to regions where the highest accretion rates are observed at $r=0.6$.

{ The torque between the star and the disc sets the constant in Equation \ref{eq:torqueconst}, and this constant is essential for the disc evolution. To understand how different stress/torque terms contribute to the total torque, we plot the Maxwell stress/torque and the hydrodynamical stress/torque in Figure \ref{fig:torquesep}. Close to the stellar surface (e.g. $r$=0.4), the magnetic stress dominates and is exerted at high latitudes  where most of accretion occurs. This indicates that the accretion disc twists the magnetic fields that connect to the star, and these field lines torque the star while channeling the accretion flow. While the magnetic stress spins up the star (bottom panels), the small hydrodynamical stress actually tries to spin down the star (middle panels). This is because the infalling gas rotates in the opposite direction from the disc (\S \ref{sec:surface}) and the star thus accretes gas with negative angular momentum. Around the magnetospheric truncation radius (e.g. $r$=0.8), the hydrodynamical stress from the penetrating filaments at the disc midplane becomes more apparent, while the total stress remains the same as the stress at $r$=0.4. At $r=1.5$, the magnetic stress is $\sim$ 0, and the positive magnetic stress in the disc region is balanced by the negative magnetic stress higher above. This means that the hydrodynamical stress equals the constant.  For the disc region (e.g. $r=3$), both hydrodynamical and magnetic stresses become larger than the constant (see the discussion after Equation \ref{eq:torqueconst}), and the magnetic stress drives the accretion inwards. { The stress profiles along the radial direction is shown in Figure \ref{fig:torqueseptr}.} Overall, the magnetic fields within the magnetosphere transfer angular momentum to the star, while the magnetic fields in the disc transfers angular momentum outwards. The magnetic stress changes sign at $r\sim1.5$, slightly outside the magnetospheric truncation radius. }

The fact that the $n$ parameter is $\sim$1 suggests that most of the coupling between the magnetosphere and the disc occurs around $R\sim R_{T}$. To confirm this and understand how different disc regions contribute to the star's spin-up, we calculate the stress at the interface between the disc region and the magnetosphere, shown as $S_1$, $S_2$, and $S_3$ in Figure \ref{fig:torqueillu}. The total torque in the disc region can be separated into
\begin{align}
\frac{\partial \int R\rho v_{\phi} dv}{\partial t} = &\int_{S_1} R_{in}(\rho v_{r}v_{\phi}-B_r B_{\phi})dS \nonumber \\
+ &\int_{S_2} R(\rho v_{\theta}v_{\phi}-B_{\theta} B_{\phi})dS \nonumber \\
- &\int_{S_3} R(\rho v_{\theta}v_{\phi}-B_{\theta} B_{\phi})dS \,,
\end{align}
where $S_1$ is the shell from $\theta=1$ to $\pi-1$ at $r_{in}=0.6$, while $S_2$ and $S_3$ are surfaces of a cone at $\theta=1$ and  $\theta=\pi-1$ from $r_{in}=0.6$ outwards. $S_2$ and $S_3$ are chosen to enclose the disc region, including the surface accretion region. The space-time diagrams for these three terms are shown in the third to fifth panels of Figure \ref{fig:torque}. Since most of the torque at $r=0.6 $ is exerted beyond the $\theta$=[1, $\pi$-1] region (the third panel), the integrated torque at $S_1$ is small compared with the torque at $S_2$ and $S_3$. The fourth and fifth panels show that most of the torque at $S_2$ and $S_3$ is from the region within $r\sim1$. { The white band extending from $log_{10}r\sim $0 to 0.4 suggests that the torque density is $\sim$0 beyond $r\sim$1.} This is also confirmed in the bottom panel showing the integrated torques at $S_2$ and $S_3$. The total torque averaged over the last 50 $T_0$ is -0.0065, and the torque at $S_1$ during the same period of time is -0.0020. The bottom panel shows that the torques at $S_2$ and $S_3$ are also $\sim$-0.0020. At $r=1$, the integrated torques at $S_{2}$ and $S_3$ are both -0.0012.
Thus, 70\% of the total torque is exerted at the disc surface within $r=1$, and 90\% of the torque is exerted within $r=3$.

{ Our torque results are noticeably  different from recent work by \cite{Takasao2022} who finds significant hydrodynamical stress contribution close to the star and the $n$ parameter is significantly smaller than 1. Especially for their model C, whose large corotation radius is similar to our non-rotator setup, its $n$ value is $\sim$0. Two factors could contribute to the difference. The first is that our truncation radius is 10 times the stellar radius while their truncation radius is 2 times the stellar radius due to their much weaker stellar field. As shown in Figure \ref{fig:torquesep}, magnetic stress is more important deeper into the magnetosphere. The second difference is the strong outflow in \cite{Takasao2022}, which is absent in our simulations. The difference in the outflow rate could be due to the disc fields applied in \cite{Takasao2022} the different adopted coronal density/density floor around the star. The impact of the coronal density/density floor is an important issue which needs to be thoroughly examined in future.  }

\subsection{Field Transport}
\label{sec:field}
\begin{figure}[t!]
\includegraphics[trim=0mm 0mm 0mm 0mm, clip, width=3.5in]{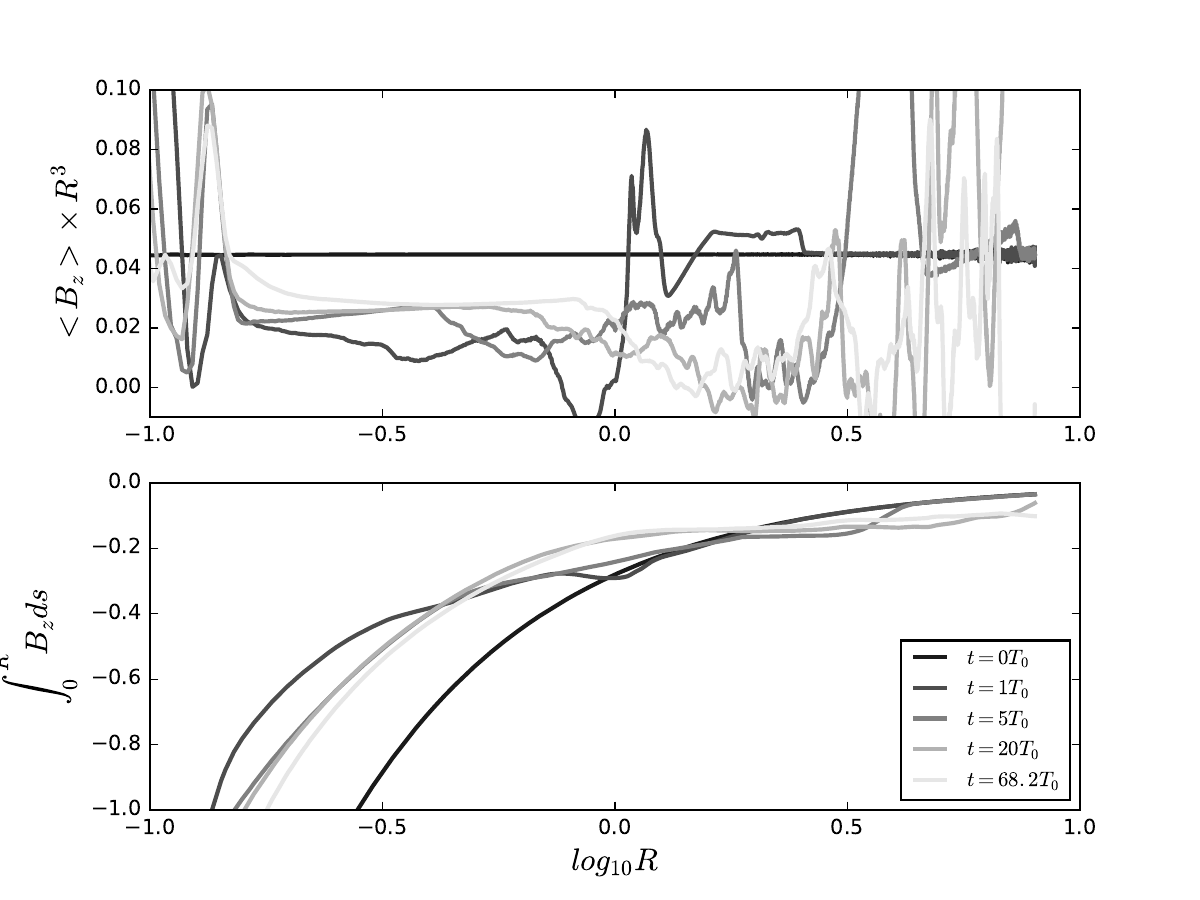}
\figcaption{ The azimuthally averaged vertical magnetic fields (upper panel) and the radially integrated magnetic flux (lower panel) at the midplane. 
Dark to light colored curves show $t=$0, 1, 5, 20, 68.2 $T_0$ respectively. \label{fig:midbz}}
\end{figure}

The transport of net magnetic flux in discs directly controls the disc's long-term evolution. However, studying magnetic field transport in discs can be challenging, often influenced by inner boundary conditions of the simulation. Fortunately, our simulation setup has two advantages allowing us to study field transport. First, our setup incorporates the central star within the simulation domain, and the  stellar magnetic fields are the only source of magnetic fields in the problem. Thus, there is no need for special boundary conditions at the stellar surface.  Second, Athena++ conserves the total magnetic flux in the whole domain to machine precision, except for flux losses at the boundaries. 

To examine flux transport, we monitor the evolution of magnetic flux integrated outward from the central star.  Given our Cartesian grid setup, we integrate the flux over the area of a circle at the disc midplane. The integrated flux within different sized circles around the central star is shown at the bottom panel of Figure \ref{fig:midbz}. Shortly after the simulation starts, the dipole magnetic fields begin moving outward, reducing the dipole field strength within $r\sim 1$. This decrease of dipole fields is likely due to field inflation and the reconnection at the beginning of the simulation (Figure \ref{fig:evolve}).  The outward moving fields are piled up at the disc region (the high $B_z$ values that are above the initial field strengths in the upper panel of Figure \ref{fig:midbz}).  With outer regions becoming MRI active, the fields are transported further outwards. At the end of our simulation, the whole region within $R\sim5$ has weaker fields than the initial condition. The disc region beyond $R\sim$1 seems to have the most significant field reduction compared with the magnetosphere region within $R\sim$1. Overall MRI turbulence in the disc seems to be efficient at transporting the fields outwards. The outward moving fields may accelerate the operation of MRI at the outer discs.  

{ Such outward field transport is different from field transport in simulations with net vertical fields. Previous net vertical field simulations find that either the fields are transported to the star \citep{ZhuStone2018, Jacquemin2021} at the mass accretion timescale \citep{Jacquemin2021} or maintains a quasi-steady state \citep{Mishra2020M}. Such difference indicates that field transport also depends on the initial field distribution besides the accretion disc properties. To achieve a long-term equilibrium field configuration, it is necessary to conduct simulations that run for significantly longer timescales. The field transport could also be affected by non-ideal MHD effects at the outer disc or the multipole components of the fields \citep{Russo2015}.}

\subsection{Implications for Planet Formation}
\label{sec:planet}

Our simulation can be scaled to realistic astronomical systems that undergo magnetospheric accretion. When considering a disc that is threaded by the stellar dipole fields and it has the same temperature slope ($q=-1/2$ in Equation \ref{eq:qvalue}) as in our simulation, there are only two dimensionless free parameters to define the system: the disc aspect ratio at the magnetospheric truncation radius ($h(R_T)$), and the ratio between $R_T$ and $R_*$. Since material undergoes free fall toward the central star within the magnetosphere, the region immediately surrounding the star is unlikely to have a significant impact on the dynamics within the disc, except through thermal feedback. Thus, the only important free parameter is $h(R_T)$. Although we have only studied the thick disc case with  $h(R_T)=0.1$ here, we will explore thinner discs, which will be more applicable to protoplanetary discs,  in future works. Nevertheless, we still use our current simulation results to study protoplanetary discs and will justify some parameter choices later.

We consider a typical protoplanetary disc with an accretion rate of $\dot{M}=10^{-8}\msunyr$ around a 2 $r_\odot$, 0.5 $M_\odot$ star having a 1kG dipole magnetic field. The magnetic truncation radius (Equation \ref{eq:RT})
is thus
\begin{align}
R_T=&14.4 \left(\frac{B_*}{1kG}\right)^{4/7} \left(\frac{M_*}{0.5 M_\odot}\right)^{-1/7}  \nonumber \\
&\left(\frac{\dot{M}}{10^{-8}\msunyr}\right)^{-2/7}\left(\frac{r_*}{2r_{\odot}}\right) ^{12/7} r_\odot\,.\label{eq:RTS}
\end{align}
$R_T/r_*=7.2$ which is quite close to our $R_T/r_*=10$ in the simulation \footnote{{ For a star with weaker magnetic fields, $r_*/R_T$ is larger.
To scale our simulation for such a star, we could assume that the stellar surface is at the given $r_*/R_T$ in the current simulation. }}.
To represent this fiducial system, the length unit in our simulation $R_0$ is  
14.4 $r_\odot$ or 0.067 au since $R_0\sim R_T$. The time unit ($1/\Omega_0$) is thus 0.0039 years.
If we equate $\dot{M}=-0.005$ in our code unit with $\dot{M}=10^{-8}\msunyr$, the mass unit is  $7.8\times 10^{-9}\msun$.
The surface density unit is then 15.5 g/cm$^2$. Thus, the disc surface density $\sim$ 0.003 $R$ from $R=1$ to $R=10$ (Figure \ref{fig:onedradial}) is equivalent to a protoplanetary disc with the surface density of 
\begin{equation}
\Sigma=0.7 \times (R/{\rm au})\; {\rm g/cm}^2 \;\;\;at\;\;\; r<0.7\; au \,.\label{eq:sigmapd}
\end{equation}
The increase of $\Sigma$ with R is due to the fast decrease of $\alpha$ with R. With this surface density, the total gas mass within R is
\begin{equation}
M(R)=\int_0^R2\pi R \Sigma dR=0.055 (R/au)^3 M_{\oplus} \,.
\end{equation}
Assuming the dust-to-gas mass ratio is 1 to 100, the total dust mass is 100 times smaller.
This low value of $\Sigma$ is caused by the large $\alpha$ value within the disc resulting from surface accretion. Figure \ref{fig:onedradial} shows that $\alpha_{int}\sim 1$ at $R\sim$3, which is scaled to
\begin{equation}
\alpha_{int}\sim 0.1\times (R/au)^{-1.5}\,,\label{eq:alphapro}
\end{equation}
in the MRI active region of the protoplanetary disc. Using $\dot{M}=3\pi\nu\Sigma$, we can estimate that $\Sigma\sim$2 g/cm$^2$ at $R\sim$0.2 $au$ assuming a more realistic $h/r\sim$0.05. This surface density is $\sim$10 times larger than Equation \ref{eq:sigmapd}. But even so,
the mass is orders of magnitude smaller than what would be required to explain the discovered exoplanets  within 1 au. Furthermore, dust may evaporate in this region. Thus, exoplanets may not be able to form
in the MRI active inner disc. On the other hand, they could form at the inner edge of the dead-zone and later migrate inwards. Dust growth in this inner disc region is crucial for both planet formation and explaining dipper stars \citep{Li2022}.

To estimate the location of the inner edge of the dead zone, it's necessary to calculate the temperature distribution within the disc. MRI becomes active when the disc temperature exceeds $\sim$ 1000 K.
An accurate estimate requires us to know how accretion energy is dissipated in the disc.
Since we have little knowledge on this, we simply estimate the lower limit of the disc temperature using the irradiation equilibrium temperature.
Assuming that the disc absorbs a fraction ($\epsilon$) of the total stellar luminosity ($L_*$), the equilibrium temperature is then
\begin{equation}
T_{irr}=394 \left(\frac{R}{au}\right)^{-1/2}\left(\frac{\epsilon L_*}{L_{\odot}}\right)^{1/4}K\,.
\end{equation}
Since MRI becomes active when $T\gtrsim 1000$K, the inner edge of the deadzone is 0.07 au if $\epsilon L_*=0.2 L_{\odot}$. On the other hand, if viscous heating is included, the inner deadzone edge can be 0.2 au \citep{DAlessio1998}. For Herbig Ae-Be stars, this radius can be even larger, reaching to 1 au \citep{Dullemond2010}.
Within this radius where the disc couples efficiently with the stellar magnetic fields and maintains a low surface density, the formation of the exoplanets within 0.1 au (10 day period) through in-situ formation is challenging. Instead, these exoplanets are likely to form at the outer discs and migrate inwards.
 
We can estimate the planet migration timescale in the MRI active disc  using the derived disc surface density in Equation \ref{eq:sigmapd}. The type I migration timescale \citep{Baruteau2014} for a planet around a 0.5 $M_{\odot}$ star is then
\begin{align}
t_{I, mig}&=\Omega^{-1}h^2q^{-1}\left(\frac{\Sigma R^2}{M_*}\right)^{-1}\nonumber\\
&=1.1\times10^{10}\left(\frac{R}{0.1\; au}\right)^{-1.5}\left(\frac{q}{10^{-5}}\right)^{-1}\left(\frac{h}{0.05}\right)^2 yr \,,\label{eq:typeI}
\end{align} 
where $q\equiv M_p/M_*$.
When this mass ratio is higher than $\alpha^{1/2}h^{5/2}$ \citep{Zhu2013}, a gap will be induced in the disc and the planet undergoes Type II migration.
The Type II migration rate is \citep{Ivanov1999,Dempsey2020} 
\begin{equation}
t_{II, mig}=\tau_{visc}\frac{M_p}{\Sigma R^2}=\Omega^{-1}h^{-2}\alpha^{-1}q\left(\frac{\Sigma R^2}{M_*}\right)^{-1}\,. \label{eq:typeII}
\end{equation}

\begin{figure}[t!]
\includegraphics[trim=0mm 0mm 0mm 0mm, clip, width=3.5in]{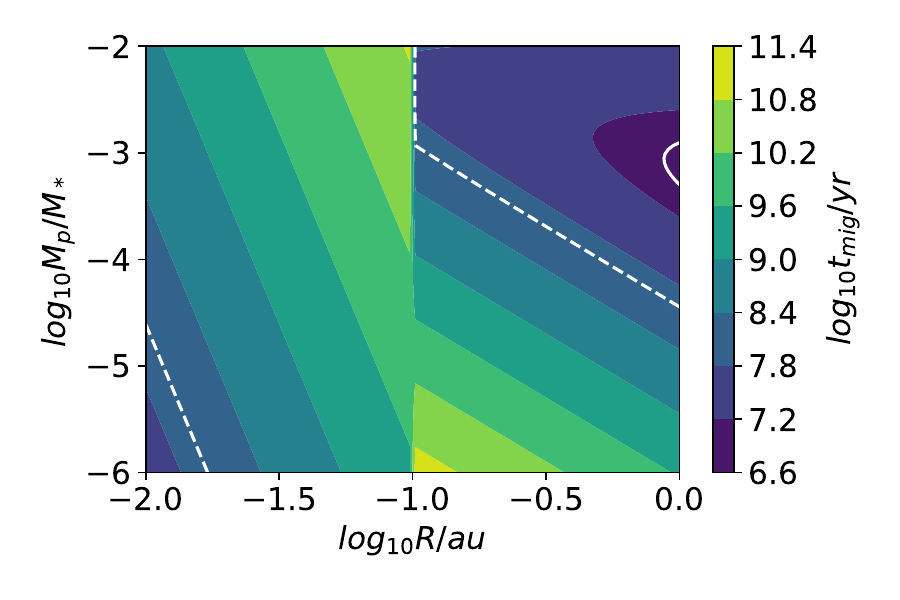}
\figcaption{{ The migration timescale for a planet at the inner disc. We assume that the magnetospheric truncation radius is at 0.1 au. Outside this radius, disc-driven migration (Equation \ref{eq:mig}) is important. We adopt the disc's surface density from Equation \ref{eq:sigmapd}, $h=0.05$, and $\alpha$ from Equation \ref{eq:alphapro}. Within 0.1 au, aerodynamic drag is important (Equation \ref{eq:migaero}).
The solid and dashed white contours label the migration timescale of $10^7$ and $10^{8}$ years. The central star is a 0.5 $M_{\odot}$ star.}  \label{fig:migt}}
\end{figure}

We could combine both Type I and Type II migration rates (Equations \ref{eq:typeI}, \ref{eq:typeII}), and incorporate the effects of the gap-opening planet mass into a single equation that represents the overall planet migration rate
\begin{equation}
t_{mig}=\Omega^{-1}h^2q^{-1}\left(\frac{\Sigma R^2}{M_*}\right)^{-1}(1+h K)\,, \label{eq:mig}
\end{equation}
where $K\equiv q^2/(\alpha h^5)$. When $K\lesssim1/h$, it reduces to the Type I rate,  as shown in Figure 12 of \cite{Dempsey2020}. 
Unlike  Type I migration, the Type II migration timescale is longer for a more massive planet. Thus, a planet that can marginally induce gaps ($hK\sim$1) migrates fastest in the disc. With the high $\alpha$ value in our simulations, Jupiter mass planets marginally induce gaps in the disc.
Figure \ref{fig:migt} shows the planet's migration timescale calculated with our disc structure from Equation \ref{eq:sigmapd}. For this calculation, we ignore the fact that the planet will undergo stochastic migration due to MRI turbulence. The planet's stochastic migration in a disc that undergoes magnetospheric accretion will be presented in a future publication. Nevertheless, if we only consider Type I and Type II migration, the planet migration timescale is much longer than
the disc's lifetime, except for giant planets at  $\sim$1 au. Most planets cannot undergo disc migration within the inner MRI active disc, except early times when the disc's accretion rate and surface density is a lot higher. Planets are likely to stall or become trapped at the inner edge of the deadzone. Thus, it is the inner deadzone edge, instead of the magnetospheric truncation radius, that determines the planet's final position before the protoplanetary disc dissipates.
This could have important implications for the distribution of exoplanets.

{
On the other hand, if a planet manages to migrate through the inner MRI turbulent disc (via either disc migration or planet-planet scattering) and gets into the magnetosphere, the planet will be subject to strong aerodynamic drag which accelerates its migration to the central star.
For a Keplerian orbiting object in a Keplerian rotating disc, the relative motion between the planet and the local disc flow is  small, and  the interactions between the planet and the disc are mostly through resonance interactions (e.g. Lindblad and Corotation resonances). However, if the relative motion between the planet and background flow becomes significant, dynamical friction and aerodynamic drag start to play a more important role. One example where these effects manifest is the interaction between an inclined planet and a Keplerian rotating disc \citep{Rein2012, Arzamasskiy2018}. As the material in the magnetosphere corotates with the star, it rotates significantly slower than the Keplerian speed. In our simulation having a non-rotating star, the material inside the magnetosphere has nearly zero azimuthal velocity (Figure \ref{fig:mid1dfluc}). Thus, the relative speed between the planet and the magnetosphere is the local Keplerian speed. As a result, the planet within the magnetosphere experiences strong head-wind and migrates inwards. 

Although both aerodynamic drag and dynamical friction could be important when the relative motion between the object and the background flow is nonzero, aerodynamic drag plays a more important role for a planet in the magnetosphere. The ratio between the aerodynamic drag force and the dynamical friction force is roughly the square of the ratio between the object's size and its Bondi radius ($R_{Bondi}=GM_p/v_{rel}^2$) (e.g., \citealt{Rein2012,Wang2023}). The Bondi radius, for a planet that is within 10 solar radius distance to the star, is at least one order of magnitude smaller than the planet size assuming $v_{rel}\sim v_K$. If we only consider the aerodynamic drag force
\begin{equation}
\mathbf{f_{aero}}=-\pi s_{p}^2 \rho v_{rel}\mathbf{v_{rel}}\,,
\end{equation}
where $s_p$ is the planet's radius, we can estimate the migration timescale 
\begin{equation}
t_{mig}=\frac{M_p v_{rel}}{f_{aero}}=\frac{4 s_p \rho_p}{3 v_{rel} \rho}\,, \label{eq:migaero}
\end{equation}
where $\rho_p$ is the material density of the planet. The background density within the magnetosphere can be estimated by assuming that the accretion is from the spherical infall at the free-fall speed
\begin{equation}
\rho=\frac{\dot{M}}{4\pi R^2 v_{ff}}\,.\label{eq:magrho}
\end{equation}
We verify that,  at a distance of 5 stellar radii, this density is only a factor of 2 smaller than the midplane density found in our simulation. The presence of intruding filaments resulting from interchange instability ensures that the gas density at the midplane within the magnetosphere remains non-negligible and approaches values estimated from the spherical infall. { Our simulation here ignores the coronal heating within the magnetosphere which will change the temperature and density within the magnetosphere. However, the aerodynamic drag is mainly determined by the density, which is estimated from the mass conservation  (Equation \ref{eq:magrho}) and less affected by the coronal heating. }
Again, using the typical stellar parameters and the disc accretion rate as before, we calculate this timescale within 0.1 au, shown in Figure \ref{fig:migt}. The dashed contours label the parameter space for planets with  a migration timescale of $10^8$ years. Since both the disc-driven and aerodynamic-driven migration timescales are inversely proportional to the disc's surface density, these contours can also be interpreted as the parameter space of planets with a migration timescale of
 $10^7$ years in a $\dot{M}=10^{-7} M_{\odot}$yr$^{-1}$ disc.

Although the aerodynamic drag seems to accelerate the planet's migration within the magnetosphere, the migration timescale is still longer than the disc's lifetime for most parameter spaces. Thus, we would expect that any planet that ends in this region will stay in this region. However, their orbital configurations (e.g. eccentricity and inclination) might evolve due to the planet-disc or planet-magnetosphere interactions. Finally, we caution that we have ignored any electromagnetic effect (including dipole-dipole and dipole-conductor interactions, \citealt{Bromley2022}) on the planet migration. 

}
\section{Conclusion}

\begin{figure*}[t!]
\includegraphics[trim=0mm 0mm 0mm 0mm, clip, width=7.in]{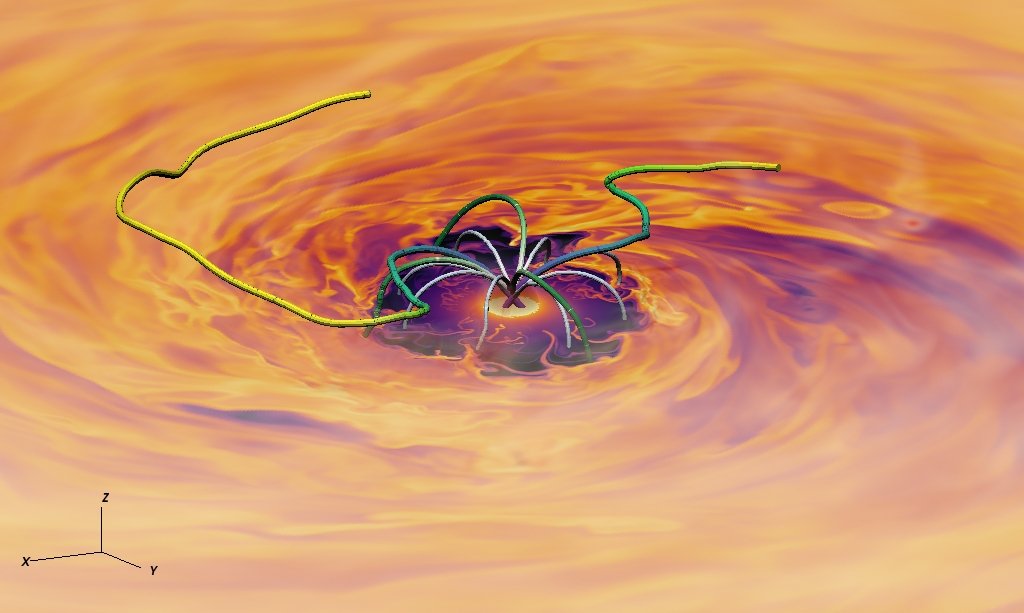}
\figcaption{ The color contours for density at the disc midplane, and the magnetic field lines within the magnetosphere and the disc. 
  \label{fig:summary}}
\end{figure*}
We have carried out high-resolution long-timescale MHD simulations to study magnetospheric accretion onto a non-rotating star. Adopting a Cartesian grid with mesh-refinement allows us to resolve both the disc and the polar accretion region equally well and reduces the computational cost significantly. We run the simulation for 68 orbits at the magnetospheric truncation radius ($R_T$), which is equivalent to 2157 Keplerian orbits at the stellar surface. A steady accretion is reached within $R\sim$6 $R_T$. 

Figure \ref{fig:summary} summarizes some of our key results. Surrounding the star, the flow within the magnetosphere is highly dynamic and filamentary. Within the magnetospheric truncation radius $R_T$, the filamentary flow is in the force-free limit, moving along the magnetic field lines.  The formation of these filamentary structures is driven by the interchange instability at $R_T$ where the density increases with radial distance.
The developed filaments (``fingers'') could penetrate deep into the magnetosphere. The density within the filaments could be more than 3 orders of magnitude higher than the background density.  As these filaments move in, they are lifted from the midplane and move along the dipole magnetic field lines. Eventually, most material accretes at  30$^o$ from the magnetic poles and falls to the star at close to the free-fall speed. More than 50\% (90\%) of accretion occurs within accretion columns covering 5\% (20\%) of the stellar surface area. Thus, we consider the filling factor to be $\sim$5-20\%. Multiple accretion columns could develop simultaneously, forming an onion like structure with multiple isolated layers. Despite the filamentary structures, the total accretion rate onto the star is relatively steady with 23\% standard deviation. Material falling onto the star has negative azimuthal velocity, since it follows the stellar magnetic field lines that are pitched forward by disc dragging. The stress from the magnetic fields spins up the star.  
The ratio between the spin-up torque and $\dot{M}(GM_*R_T)^{1/2}$ is $\sim$ 0.8, independent of the long-term accretion rate change. This constant torque will affect the disc's long term evolution. 

Many properties of the simulated flow  within the magnetosphere are consistent with observations, including hot spots at high altitudes, free-fall velocities, low filling factors, and multiple accretion layers. On the other hand, recent observations  by \cite{Thanathibodee2023} find that low accretors with 
$\dot{M}<2\times 10^{-10}\msunyr$ have magnetospheres with 
sizes $\sim$ 2 $-$ 5 $R_*$, which is significantly smaller than the theory prediction ($\sim$7 $R_{*}$ from Equation \ref{eq:RTS}).  Possible explanations include a weaker dipole magnetic field \citep{Long2008}, a quadrupole/multipole magnetic field, or a more complicated thermal structure of the magnetosphere. Future theoretical and observational studies in these directions are desired.

Outside the magnetosphere, we have the highly magnetized disc.  Although, at the disc midplane, the transition radius between the magnetosphere and the disc agrees well with the traditional magnetospheric truncation radius, these two regions are less distinct above the midplane. The disc surface accretion smoothly joins the magnetospheric accretion. 
If we use  the Alf\'ven surface or the $E_k/E_m\sim1$ surface to separate these two regions, the transition radius becomes larger when it is higher up in the disc. The azimuthal velocity and azimuthal magnetic field also reverse their signs at the transition radius.

The disc region outside $R_T$ is also highly variable due to the strong net vertical magnetic fields.
Magnetic reconnection and interchange instability could occasionally reorganize magnetic fields around the truncation radius, leading to a large-scale density void that orbits around the central star at sub-Keplerian speed, similar to the structures in MADs around blackholes. It takes $\sim$5 $T_0$ for the density void to finish one orbit. The density void extends all the way to the stellar surface, leading to hot spots that orbit at the same frequency as the void. 
The periodogram of disc accretion also shows a peak at $\sim$20\% of the Keplerian frequency at $R_T$, which corresponds to the orbital motion of the hot spot and the lifetime of the bubble. We have also observed outflows that originate from the bubble. But the mass loss rate is quite low.
Overall, both smaller-scale filaments and larger-scale magnetic bubbles are characteristics of a disc that is magnetically disrupted by strong fields. 

Further away into the disc, a magnetically supported surface region plays a crucial role in disc accretion, which is distinctly different from the traditional model.  Keplerian differential rotation stretches the radial component of the dipole magnetic fields to generate strong azimuthal fields, which lead to a low-density region up to $z\sim$R. The resulting strong $R-\phi$ stress 
makes this surface region accrete inwards at supersonic speeds, which is similar to the surface accretion of an accretion disc with net vertical magnetic fields. However, little disc wind is launched above this region, since the magnetic fields there are connected to the non-rotating star instead of the Keplerian rotating disc. 
Both the net vertical magnetic fields and the disc $\alpha$ decrease sharply with radii, which leads to a disc with a surface density proportional to $R$ within $R\sim10 R_T$.  

{ Our disc structure is significantly different from previous ``X-wind model'' \citep{Shu1994} and the unsteady field inflation model \citep{LyndenBell1994, Lovelace1995, Uzdensky2002}, which are built upon $R$-$z$ 2-D models. During the early stage in the simulation when the disc is largely axisymmetric, we observe strong field inflation as in previous works. However, when various 3-D instabilities start to operate, including the magnetic interchange instability around $R_T$ and the MRI in the outer disc, the field lines diffuse across disc material, acting like a large anomalous resistivity. Thus, open fields lines return to the dipole configuration due to the anomalous resistivity. 
With a large resistivity, the slippage of field lines  balances the azimuthal shear, allowing the quasi-steady accretion.}

After scaling our simulations to protostars with $\dot{M}=10^{-8}\msunyr$, we find that the inner MRI active disc has a very low surface density ($<$1 g cm$^{-2}$) due to the efficient surface accretion. The timescale for Type-I/II planet migration is longer than the disc lifetime, suggesting that planets (especially low mass planets) are not able to migrate in the inner MRI active region and likely to be stalled at the inner edge of the deadzone. If the planets could move into the magnetosphere, aerodynamic drag can accelerate the planet's migration, although the migration timescale is still long. 

\section*{ACKNOWLEDGEMENTS}
All  simulations are carried out using with NASA Pleiades supercomputer. Z. Z. acknowledges support from NASA award 80NSSC22K1413. J.M.S. acknowledges support from the Schmidt Futures Fund to the IAS. ZZ thanks Bart Ripperda, Catherine Dougados, Dong Lai, Yihan Wang, Douglas N.C. Lin for discussions and suggestions.

\section*{DATA AVAILABILITY}
The data underlying this article will be shared on reasonable request to the corresponding author.



\appendix

\section{Conserved Quantities Within the Magnetosphere}
\label{sec:constantsofintegral}

\begin{figure*}[t!]
\includegraphics[trim=0mm 15mm 15mm 15mm, clip, width=6in]{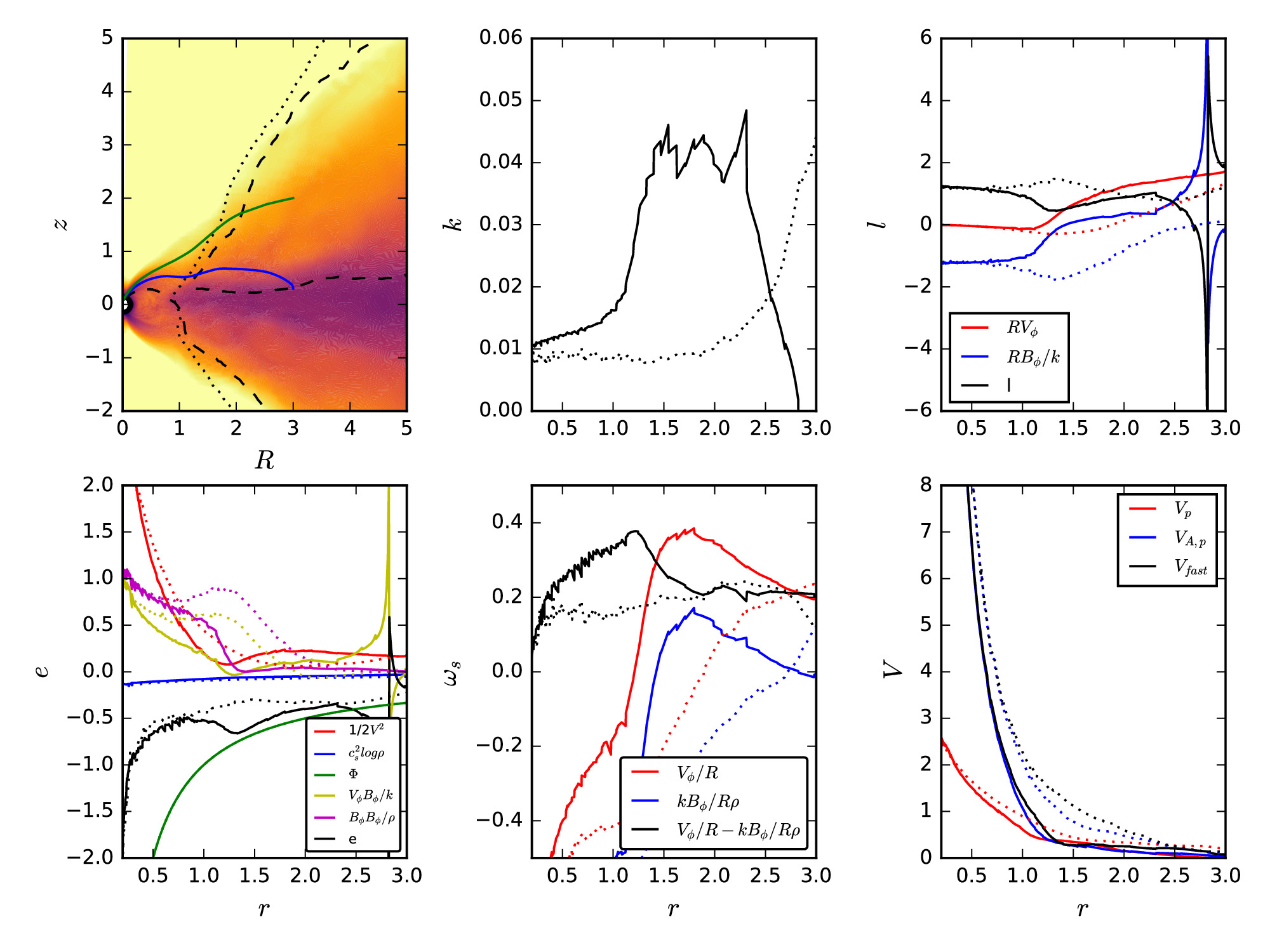}
\figcaption{ Various conserved quantities ($k$, $l$, $e$, $\omega$) and characteristic speeds (poloidal speed, Alfv\'en speed using polar B fields, and fast magnetosonic speed)
 along two fluid streamlines shown as the blue and green curves in the upper left panel. The dotted and dashed curves in the upper left panel represent where $E_k=E_B$ and $B_\phi=0$.
 The solid curves in the rest panels show quantities for the streamline starting at $R$=3 and $z=0.3$, while the dotted curves are for the streamline starting at $R=3$ and $z=2$. 
All the primitive variables have been averaged over both the azimuthal direction and time (using 20 snapshots from t=66.2 to 68.2 $T_{0}$).  The streamlines are also calculated with the
averaged velocity. \label{fig:stream}} 
\end{figure*}

Within the magnetosphere, magnetic fields are dominated by large-scale smooth fields ($\langle B \rangle^2/\langle B^2\rangle\sim$1 in the $\langle B \rangle^2/\langle B^2\rangle$ panel of Figure \ref{fig:magconst}). Although the flow is filamentary, the flow's azimuthally averaged properties may still follow the four conserved quantities ($k$, $l$, $\omega_s$, $e$, Equations \ref{eq:kintegral} to \ref{eq:eintegral}), which are plotted at the $R-z$ plane in Figure \ref{fig:magconst} and along two streamlines in Figure \ref{fig:stream}. Figure \ref{fig:stream} shows that
the conserved quantities are roughly constant along fluid streamlines in the magnetically dominated region. The black solid and dotted curves in the $k$, $l$, $\omega_s$, and $e$ panels show the four quantities along the blue and green curves in the upper left panel, and these quantities approach constant values within $r\sim 1$ and 2 respectively.  For these streamlines, $r\sim 1$ and 2 coincide with the $E_k=E_B$ boundary. The $k$ value decreases significantly when the streamline is moving from the disc region to the magnetosphere region. This is also demonstrated in the $k$ panel of Figure \ref{fig:magconst} where the $E_k>E_B$ region has a higher $k$ value. This higher mass loading factor for each streamline is related to the turbulent magnetic fields (or the break of the 2-D assumption) as shown in the $\langle B \rangle^2/\langle B^2\rangle$ panel of Figure \ref{fig:magconst}. The $k$, $\omega_s$ and $l$ panels in Figure \ref{fig:magconst} all show sharp changes of values at the region where $\langle B \rangle^2/\langle B^2\rangle$ suddenly becomes small. Different components of the conserved quantities are also shown in Figure \ref{fig:stream}. For  $l$, we can see that the azimuthal magnetic fields dominate over angular momentum transport within the magnetosphere. For  $e$  at small $r$, the two dominant terms are the potential term and the kinetic energy term, which confirms the statement after Equation \ref{eq:eintegral} that the inflow should have free-fall speed. For the black dotted curve,  $\omega_s$ is  $\sim$0.2 before it drops towards zero within $r\sim$0.5. Finally, the lower right panel of Figure  \ref{fig:stream} shows that the region having conserved quantities are sub-Alfv\'enic. The flow becomes sub-Alfv\'enic inside $r\sim$ 1.3 and 2.5 for these two streamlines, again corresponding to where  $E_k\sim E_B$. { We note that the conserved quantities are not strictly constant within the magnetically dominated regions, especially for $e$ and $\omega_s$ along the streamline starting at $R=3$ and $z$=0.3. This is probably due to the turbulent structure. On the other hand, considering that the conserved quantities only vary by a factor of 2-3 while some components of the conserved quantities vary by one order of magnitude in the magnetically dominated region, we consider that}  the { traditional 2-D axisymmetric steady-state model roughly captures the azimuthally-averaged flow structure inside the magnetic-dominated region. However, this only reflects some conversion between various components (e.g. the magnetic, motion, gravity components), and blindly extending the 2-D theory to describe the averaged 3-D flow structure can be misleading.  With properly azimuthally averaging the fluid and induction equations, a better set of conserved quantities may be possible to describe the averaged flow, which is beyond the scope of this work.}

\section{Disc Structure and Angular Momentum Transport}
\label{sec:diskaccretion}

\begin{figure*}[t!]
\includegraphics[trim=0mm 5mm 0mm 0mm, clip, width=6.in]{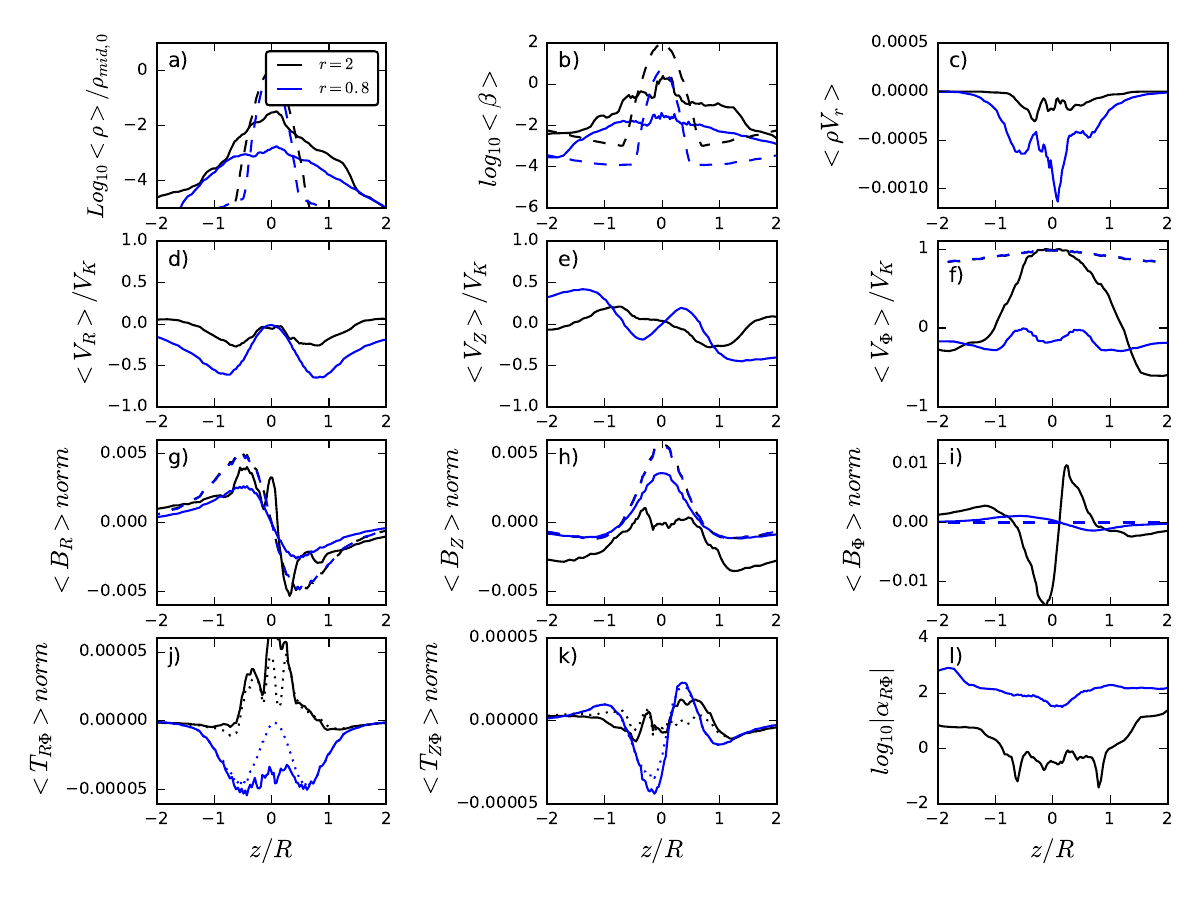}
\figcaption{Various quantities along the vertical direction at R=0.8 and 2. All quantities are averaged over both the azimuthal direction and time (20 snapshots from 66.2 and 68.2 $T_0$).  The dashed curves represent the initial condition. 
In the bottom two rows, the quantities at $R=0.8$ are multiplied by $(0.8/2)^{-3}$ to compare with the quantities at $R=2$. With this normalization, the initial magnetic fields at these two radii share the same profiles. The dotted curves in the $\langle T_{R\phi}\rangle$ and $\langle T_{Z\phi}\rangle$ panels are calculated with the space and time averaged velocities and magnetic fields. For the Reynolds stress component, the azimuthally averaged $v_{\phi}$ has been subtracted.
\label{fig:vertical1d}}
\end{figure*}

To examine the disc structure in more detail, Figure \ref{fig:vertical1d} shows the profiles of various quantities along the $z$ direction at $R=0.8$ and 2. Since the magnetosphere is within $R\sim$1, the profiles at $R=0.8$ and 2 respectively represent the magnetosphere and disc structure. With stronger magnetic fields, the density drops more at the inner radii. But the stronger magnetic fields also push the material to higher elevations. Compared with the initial condition, the density profile has a smoother transition from the midplane to the surface (Panel a of Figure \ref{fig:vertical1d}). While the magnetosphere and the disc surface are magnetically dominated ($\langle \beta \rangle\sim 10^{-4}-10^{-1}$), the disc midplane is still matter dominated (Panel b). Within the magnetosphere, the radial velocity is approaching the Keplerian speed (also the free-fall speed). Within the disc surface region, the radial speed becomes supersonic at $z\sim R$ (Panel d, with $c_{s}\sim$0.1$v_k$). Since the disc density decreases towards a higher $z$, the mass flux rate at $R=2$ peaks at $z\sim$0.3$R$ or $3H$ (Panel c). The azimuthal velocity becomes negative within the magnetosphere, while the disc midplane is still rotationally supported (Panel f). The poloidal magnetic fields still maintain the dipole configuration within both the magnetosphere and the disc region (Panel g and h). However, azimuthal fields are significantly amplified in the disc region (esp. at the disc surface), and they can be several times stronger than the poloidal components (Panel i). Since $B_{\phi}$ reverses its direction within the magnetosphere, $\langle T_{R\phi}\rangle$ also changes the sign between the magnetosphere and the disc region (Panel j). While $T_{z\phi}$ is comparable to $T_{R\phi}$ within the magnetosphere at $R=0.8$, $T_{z\phi}$ is much weaker than $T_{R\phi}$ at the disc region ($R=2$) since $B_z$  is much smaller than $B_R$ there. Thus, most of angular momentum transport in the disc region, including both the midplane and the surface region, is still through $T_{R\phi}$. { The $T_{R\phi}$ stress mostly comes from the turbulent stress at the midplane, while it is from the net fields at the surface accretion region ($z/R\sim$ 1) (by comparing dotted and solid curves in Panel j).}

\begin{figure*}[t!]
\includegraphics[trim=0mm 5mm 0mm 0mm, clip, width=6.5in]{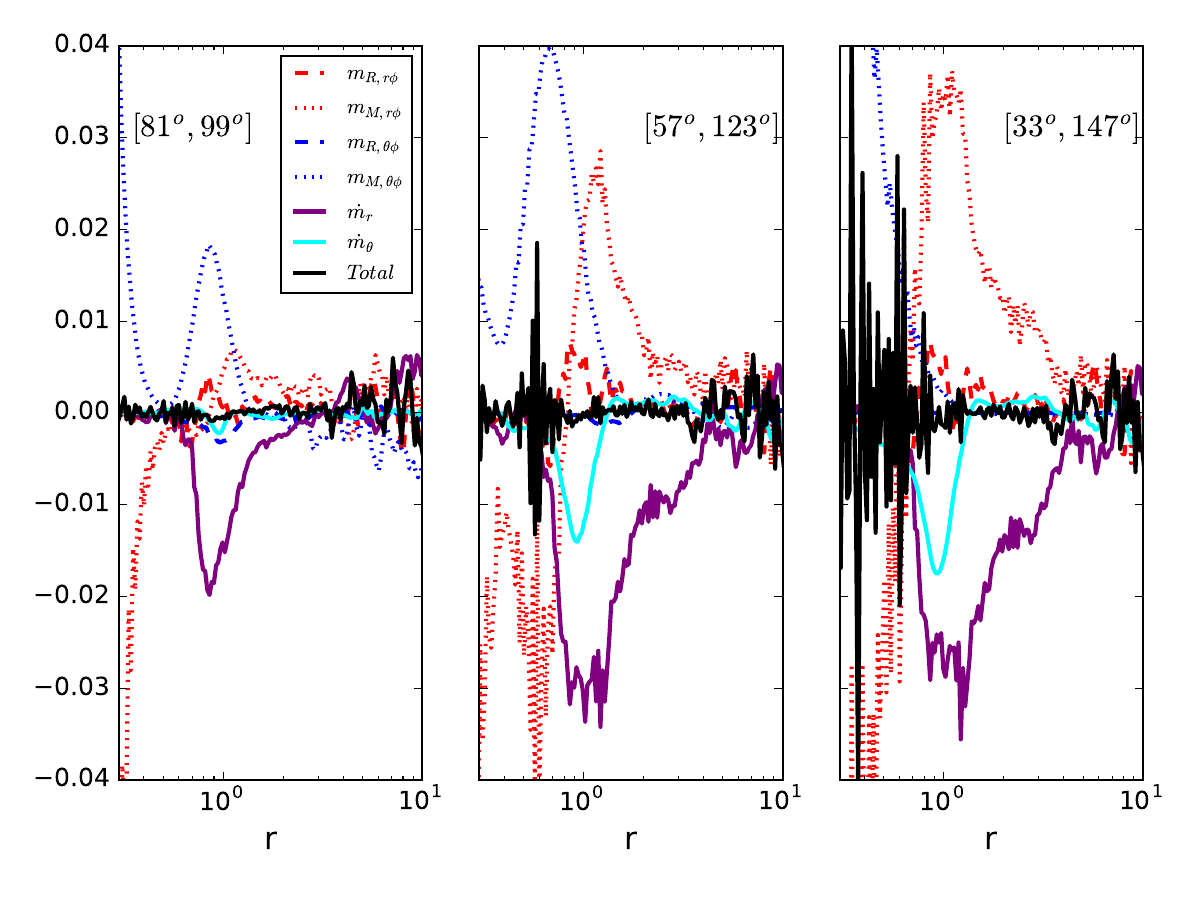}
\figcaption{Various torque terms in Equation \ref{eq:budget} integrated over different $\theta$ limits (around disc midplane, surface accretion boundary, outflow boundary, from the left to right panels).   \label{fig:torque1d}}
\end{figure*}

To understand the roles played by $r$-$\phi$ and $\theta$-$\phi$ stress terms, we write the angular momentum conservation equation in a spherical-polar coordinate system: 
\small
\begin{align}
\frac{\partial \langle r {\rm sin}\theta\rho\delta v_{\phi}\rangle}{\partial t}&=-\frac{{\rm sin}\theta}{r^2}\frac{\partial ( r^3  \langle \rho v_{r}\delta v_{\phi}-B_{r}B_{\phi}\rangle)}{\partial r}-{\rm sin}\theta\langle\rho v_{r}\rangle\frac{\partial r\overline{v_{\phi}}}{\partial r}\nonumber\\
&-\frac{1}{{\rm sin}\theta}\frac{\partial ({\rm sin}^2\theta\langle  \rho v_{\theta}\delta v_{\phi}-B_{\theta}B_{\phi}\rangle)}{\partial \theta}-
\langle\rho v_{\theta}\rangle\frac{\partial ({\rm sin}\theta \overline{v_{\phi}})}{\partial \theta}
\end{align}
\normalsize
where $\delta v_{\phi}=v_{\phi}-\overline{v_{\phi}}$ and $\overline{v_{\phi}}$ is the azimuthally averaged $v_{\phi}$. If we integrate the equation over the poloidal and azimuthal directions weighted by area ($r^2{\rm sin}\theta$), we have
\small
\begin{align}
\frac{\partial \int r^3 {\rm sin}^2\theta\langle \rho\delta v_{\phi}\rangle d\theta}{\partial t}=&-\int{\rm sin}^2\theta\frac{\partial ( r^3  \langle \rho v_{r}\delta v_{\phi}-B_{r}B_{\phi}\rangle)}{\partial r}d\theta-\int r^2{\rm sin}^2\theta\langle\rho v_{r}\rangle\frac{\partial r\overline{v_{\phi}}}{\partial r}d\theta\nonumber\\
&-\int r^2\frac{\partial ({\rm sin}^2\theta\langle  \rho v_{\theta}\delta v_{\phi}-B_{\theta}B_{\phi}\rangle)}{\partial \theta}d\theta-\int r^2 {\rm sin}\theta\langle\rho v_{\theta}\rangle\frac{\partial ({\rm sin}\theta \overline{v_{\phi}})}{\partial \theta}d\theta \label{eq:angsph}
\end{align}
\normalsize
or simply
\begin{equation}
\frac{\partial \int r^3{\rm sin}^2\theta\langle   \rho\delta v_{\phi}\rangle d\theta}{\partial t}=-m_{R,r\phi}-m_{M,r\phi}-\dot{m_{r}}-m_{R,\theta\phi}-m_{M,\theta\phi}-\dot{m_{\theta}}\,.\label{eq:budget}
\end{equation}
The left term is the change of angular momentum. In a quasi-steady state, it is zero. The first two terms on the right side of Equation \ref{eq:budget} are the radial gradients of the Reynolds and Maxwell $r$-$\phi$ stresses. The stresses could be due to MRI turbulence 
(e.g. at the disc midplane) or large-scale organized magnetic fields (e.g. at the disc surface). 
The third term on the right is the radial advection term which represents angular momentum carried by the radial accretion flow. The last three terms on the right are similar to the first three terms except that they describe $\theta$-$\phi$ stresses and advection in the $\theta$ direction (e.g. flow in the $\theta$ direction). 

We calculate these terms from our simulation by averaging 310 snapshots from 65.1 to 68.2 $T_0$. We integrate these terms over the $\theta$ direction for different sized wedges, shown in Figure \ref{fig:torque1d}. We can see that the disc has reached a quasi-steady state within $R\sim$6, where the black curves have much smaller values than other curves. The plots also show that the Maxwell stress (dotted curves) dominates over the Reynolds stress (dashed curves). Within 1.5 disc scale heights above and below the midplane (the leftmost panel), the disc is matter dominated ($\langle\beta\rangle\gtrsim 1$) beyond $R\sim 1.5$. The radial mass flux is quite low at this region around the midplane. Instead of driving the radial inflow (the $\dot{m_r}$ term), the $r$-$\phi$ stress is balanced by the $\theta$-$\phi$ stress at $R\gtrsim$2. Thus, the magnetic breaking from the midplane to the surface transports angular momentum into the midplane, which reduces the midplane's radial inflow. Only within the magnetically supported surface region at $r\lesssim$1.5, the $\theta$-$\phi$ stress removes the flow's angular momentum, leading to disc accretion.  Within the highly magnetized region ($r\lesssim$0.6), the $r$-$\phi$ and $\theta$-$\phi$ magnetic stresses balance each other again, indicating that magnetic fields there follow the force-free solution. If we include the disc surface in the integration (middle and right panels), we can see that the radial inflow now becomes significant beyond $R\sim$1.5 and is mainly driven by the $r$-$\phi$ stress. The advection in the $\theta$ direction also carries part of angular momentum away, but it is much smaller than the $r$-$\phi$ stress, except around the magnetosphere boundary  at $r\sim$1. The $\theta$-$\phi$ stress is also much weaker than the $r$-$\phi$ stress in the disc region. It only becomes important within the magnetosphere. Overall, in the disc region beyond the magnetosphere, the flow and magnetic structure are very similar to the net vertical field simulations in  \cite{ZhuStone2018}, with the $r$-$\phi$ stress leading to the overall radial inflow and the $\theta$-$\phi$ stress connecting the surface to the midplane. 

On the other hand, there are significant differences between this simulation and that in \cite{ZhuStone2018}, mainly due to 
the dipole field configuration.   Very high above the disc surface ($z\gtrsim R$), magnetic fields are pointing down towards the central star ($\langle B_{\phi}\rangle$ panels in Figure \ref{fig:compare}). 
$B_{\phi}$ there thus becomes negative due to the differential motion between the non-rotating star and the Keplerian disc. Material that follows the field lines moving inwards have negative $v_{\phi}$,
as explained in \S \ref{sec:magstr}. For the net vertical field simulations, the fields at $z\gtrsim R$ are pointing upwards. The anchor points at $z\sim R$ drag the field lines in the direction of disc rotation.
This generates negative $B_{\phi}$ and launches disc wind like ``beads on the wire'' \citep{BlandfordPayne1982}. At $B_{\phi}=0$, which is close to the wind launching point, the flow's azimuthal angular frequency $\Omega$ equals the conserved constant $\omega_s$ (Equation \ref{eq:omegaintegral}). When the wind is moving outwards along the field lines, $\Omega$ decreases as $B_{\phi}$ becomes more negative. However, this decrease of $\Omega$ is slower than the decrease of the Keplerian frequency with $R$ ($R^{-3/2}$), so that the wind flow appears to be super-Keplerian compared with the local flow at $R$ (the $v_{\phi}$ panel in the bottom row of Figure \ref{fig:compare}). Overall, for the material high above the disc surface, the flow has opposite  $r$ and $\phi$ directions between the magnetospheric accretion disc and the  disc with net vertical fields.

\end{document}